\documentclass[12pt]{article}
\usepackage{tikz}
\usepackage{caption}
\usepackage{subcaption}
\usetikzlibrary{positioning}

\usepackage[utf8]{inputenc}
\usepackage{textcomp}
\usepackage{chetnew}
\usepackage{tikz}
\usetikzlibrary{decorations.pathreplacing}
\usepackage{amssymb,amsfonts,mathrsfs,dsfont,yfonts,bbm}\usetikzlibrary{calc}
\usepackage{xcolor}
\usepackage{braket}

\renewcommand{\tilde}{\widetilde}
\newcommand{\CO}{\mathcal{O}}
\newcommand{\CS}{\mathcal{S}}
\newcommand{\CT}{\mathcal{T}}
\newcommand{\CI}{\mathcal{I}}

\newcommand{\CN}{\mathcal{N}}

\newcommand{\AD}{AD}

\usepackage{amsmath}	

\begin{document}

\preprint{QMUL-PH-19-16}

\title{Peculiar Index Relations, 2D TQFT, \\[2.5mm] and Universality of SUSY Enhancement}

\author{Matthew Buican$^{\diamondsuit,1}$, Linfeng Li$^{\clubsuit,1}$, and Takahiro Nishinaka$^{\heartsuit,2}$}

\affiliation{\smallskip $^{1}$CRST and School of Physics and Astronomy\\
Queen Mary University of London, London E1 4NS, UK\\ $^{2}$Department of
Physical Sciences, College of Science and Engineering\\ Ritsumeikan
University, Shiga 525-8577,
Japan\emails{$^{\diamondsuit}$m.buican@qmul.ac.uk, $^{\clubsuit}$linfeng.li@qmul.ac.uk, $^{\heartsuit}$nishinak@fc.ritsumei.ac.jp}}

\abstract{We study certain exactly marginal gaugings involving arbitrary numbers of Argyres-Douglas (AD) theories and show that the resulting Schur indices are related to those of certain Lagrangian theories of class $\mathcal{S}$ via simple transformations. By writing these quantities in the language of 2D topological quantum field theory (TQFT), we easily read off the $S$-duality action on the flavor symmetries of the AD quivers and also find expressions for the Schur indices of various classes of exotic AD theories appearing in different decoupling limits. The TQFT expressions for these latter theories are related by simple transformations to the corresponding quantities for certain well-known isolated theories with regular punctures (e.g., the Minahan-Nemeschansky $E_6$ theory and various generalizations). We then reinterpret the TQFT expressions for the indices of our AD theories in terms of the topology of the corresponding 3D mirror quivers, and we show that our isolated AD theories generically admit renormalization group (RG) flows to interacting superconformal field theories (SCFTs) with thirty-two (Poincar\'e plus special) supercharges. Motivated by these examples, we argue that, in a sense we make precise, the existence of RG flows to interacting SCFTs with thirty-two supercharges is generic in a far larger class of 4D $\CN=2$ SCFTs arising from compactifications of the 6D $(2,0)$ theory on surfaces with irregular singularities.}

\date{July 2019}

\setcounter{tocdepth}{2}

\maketitle
\toc

\section{Introduction}
\label{sec:Intro}
In this paper, we begin by focusing on a particularly simple---yet surprisingly rich---class of strongly interacting 4D $\CN=2$ SCFTs called the $D_2(SU(N))$ theories, with $N=2n+1$ an odd integer \cite{Cecotti:2012jx}. These theories are often imagined as arising in type IIB string theory\footnote{Although note that the simplest example, $D_2(SU(3))$, was originally constructed in \cite{Argyres:1995xn}.} at local Calabi-Yau singularities and are part of a larger class of theories called the $D_p(G)$ theories, where $G$ is the ADE flavor symmetry of the SCFT. However, using the methods of \cite{Xie:2012hs}, we will primarily think of these theories as coming from twisted compactifications of the 6D $(2,0)$ theory on Riemann surfaces with an irregular puncture.\footnote{Depending on the realization, the twisted compactification may or may not be accompanied by an extra regular singularity.} 

While the strongly coupled $D_2(SU(2n+1))$ SCFTs are of Argyres-Douglas (AD) type\footnote{In other words, they have $\CN=2$ chiral operators (i.e., operators annihilated by the anti-chiral half of $\CN=2$ superspace sometimes called \lq\lq Coulomb branch" operators) of non-integer scaling dimension.} and therefore lack $\CN=2$ Lagrangians, they behave in various surprising ways like collections of free hypermultiplets:

\begin{itemize}
\item{The role of the $D_2(SU(3))$ theory in the $S$-duality studied in \cite{Buican:2014hfa,Buican:2017fiq,Buican:2018ddk} is reminiscent of the role played by some of the hypermultiplets in the $S$-duality of $\CN=2$ $SU(3)$ Supersymmetric Quantum Chromodynamics (SQCD) with $N_f=6$ flavors \cite{Argyres:2007cn}.}
\item{The so-called \lq\lq Schur" limits of the 4D $\CN=2$ superconformal indices of the $D_2(SU(2n+1))$ theories are related to the Schur indices of free hypermultiplets by a simple rescaling of the superconformal fugacity and a specialization of the flavor fugacities \cite{Xie:2016evu,Song:2017oew}.}
\item{The (partially refined) Schur indices of the $D_2(SU(2n+1))$ theories can be computed via theories of free non-unitary hypermultiplets with wrong statistics in 4D \cite{Buican:2017rya,Buican:2019huq}.}
\end{itemize}

Given these parallels, it is interesting to ask if at least some of these close relations with Lagrangian theories persist upon conformally gauging subgroups of the flavor symmetry of the $D_2(SU(2n+1))$ theories. As we will see below, the answer to this question is a resounding, \lq\lq yes." In particular, we will show that the Schur indices of an infinite set of theories gotten by gauging various diagonal flavor symmetries of collections of arbitrarily large numbers of $D_2(SU(2n+1))$ theories and hypermultiplets are related to the Schur indices of certain Lagrangian theories of class $\CS$ \cite{Gaiotto:2009we} by simple transformations. Rephrasing these relations in the language of 2D TQFT allows us to efficiently study the action of $S$-duality on the flavor symmetries of the $D_2(SU(2n+1))$ quiver gauge theories (see \cite{Xie:2017vaf,Xie:2017aqx,Xie:2019zlb} for recent discussions of other $S$-duality properties of these theories).

Beyond the action on flavor symmetries, one of the most interesting aspects of $\CN=2$ $S$-duality is the emergence of exotic isolated theories at cusps in the space of exactly marginal gauge couplings. For example, Argyres and Seiberg found the exotic $E_6$ Minahan-Nemeschansky theory in $SU(3)$ SQCD with $N_f=6$ emerging at a dual cusp with a weakly coupled $SU(2)\subset E_6$ gauge group \cite{Argyres:2007cn}. This construction was then vastly generalized to find new classes of isolated non-Lagrangian $\CN=2$ SCFTs (e.g, see \cite{Gaiotto:2009we,Chacaltana:2010ks}).

\begin{table}
\begin{center}
\nonumber
     \begin{tabular}{| c| c |c |}
\hline  {\rm AD theory} & ${\rm Class\ \CS\ fixture\ analog}$ & ${\rm Flow\ to\ 32\  supercharges}$ \cr\hline \hline 
         $D_2(SU(2n+1))$ &$Y_{\rm simple},\ Y_{\rm full},\ Y_{\rm full}$ ;\ {\rm(free)} & {\rm no} \cr \hline 
              $ R_{0,p}^{2,AD}$ & $Y_{2}^{(1)},\ Y_{\rm full},\ Y_{\rm full};\ {\rm(interacting)}$ & {\rm yes}\cr\hline 
               $T_{(
m_1,m_2,m_3)}^{2,AD}$ & $Y_{m_1}^{(1)},\ Y_{m_2}^{(1)},\ Y_{m_3}^{(1)};\ {\rm(interacting)}$ & {\rm yes}  \cr\hline 
      \end{tabular}
\caption{Three important classes of isolated SCFTs we study in this paper are in the leftmost column (note that we assume, without loss of generality, that $m_3\ge m_2\ge m_1$; these quantities obey further constraints discussed in the main text). The middle column indicates the corresponding regular puncture class $\CS$ fixture (specified by a triple of Young diagrams) in the sense described in Sec. \ref{sec:S-duality}, where $Y_k^{(\ell)}$ is the Young diagram shown in Fig.~\ref{fig:Young}. The parenthetical comment in this column indicates whether the class $\CS$ fixture is interacting or not. The final column indicates if the theory admits an RG flow, of the type described in the main text, to an interacting SCFT with thirty-two (Poincar\'e plus special) supercharges. The above AD relatives of interacting class $\CS$ fixtures always admit such flows while relatives of free fixtures do not. All the above theories can be realized as type $III$ in the nomenclature of \cite{Xie:2012hs}. In Sec. \ref{flows32}, we vastly generalize these results.}
\end{center}
\end{table}\label{theories}

As we will see, the TQFT relations we find between the AD quivers and their Lagrangian cousins lead to an interesting new expression for the Schur index of the exotic AD analog of the $E_6$ theory, the so-called \lq\lq $\CT_{X}$" SCFT, arising via the $S$-duality studied in \cite{Buican:2014hfa,Buican:2017fiq,Buican:2018ddk}. Moreover, we are able to find the Schur indices for infinitely many generalizations of the $\CT_{X}$ theory arising via various AD generalizations of $S$-dualities involving only regular punctures. For example, we find indices for AD analogs of the $R_{0,p}$ theories (with $p\in\mathbb{Z}_{\ge0,{\rm odd}}$) arising via the $S$-dualities studied in \cite{Chacaltana:2010ks}. We call these theories $R_{0,p}^{2,AD}$ SCFTs. In all cases, the AD index expressions we find are related to those of their regular puncture relatives (e.g., see \cite{Gadde:2011ik}) by simple transformations on the fugacities. We term these types of AD theories \lq\lq AD fixtures" in reference to the terminology for the corresponding isolated theories arising from three-punctured spheres in class $\CS$ (e.g., see the terminology in \cite{Chacaltana:2010ks}). In this context, one may also think of the $D_2(SU(2n+1))$ theories as AD relatives of free regular puncture fixtures. On the other hand, the $R_{0,n}^{2,AD}$ SCFTs (and other theories we construct below) are AD relatives of interacting regular puncture fixtures (see  Table. \ref{theories}).

However, the TQFT index expressions we find for these isolated exotic theories are rather illuminating in their own right. For example, unlike the usual expressions for regular puncture theories, the AD indices feature products over TQFT wave functions that are not independent. We then interpret this lack of independence in terms of the topology of the corresponding quivers of the 3D mirrors associated with the AD theories \cite{Xie:2012hs}. As we will see, the quiver topology of our AD relatives of interacting fixtures is characterized by a loop of non-abelian gauge nodes in the 3D mirror. This loop has interesting physical consequences: it guarantees that one can take these isolated AD theories, compactify them on $S^1$, and flow (up to free decoupled matter fields) to interacting theories with thirty-two (Poincar\'e plus special) superchages (thereby generalizing the examples in \cite{Buican:2018ddk}).\footnote{Like their free AD fixture counterparts, the $D_2(SU(2n+1))$ fixtures do not admit RG flows via vevs and relevant deformations to interacting theories with thirty-two supercharges (note that we do not consider turning on additional gauge couplings in these flows).} We believe that these latter fixed points uplift to 4D $\CN=4$ theories, but we leave a detailed study of this correspondence to future work.\footnote{See also \cite{Argyres:2016xmc} for examples of $\CN=2\to\CN=4$ enhancement (in the case of theories with integer dimensional Coulomb branch operators).}

Based on the generic existence of RG flows with enhancement to thirty-two supercharges in the exotic isolated AD theories we study,\footnote{In fact, this enhancement can also occur in AD quivers. Indeed, these theories also have indices with non-independent wave functions, and some of the general results we prove below apply to these theories as well. The fact that we gauge some symmetries to build these theories means that the 3D mirror interpretation of their indices is more subtle.} we ask more generally when such flows can occur. As we will see, the existence of these types of flows is in fact generic in the space of 4D $\CN=2$ SCFTs (with known 3D Lagrangian mirrors) obtained by compactifying the 6D $(2,0)$ theory on a Riemann surface with an irregular singularity (we may or may not add an additional regular singularity).\footnote{In this sense, the word \lq\lq exotic" for our isolated AD theories is inappropriate. Indeed, although flows to thirty-two supercharges of the type we describe are not common among the AD theories often studied in the literature, we will see that this is because such theories are actually rather special.} Combined with the results of \cite{Maruyoshi:2016tqk,Maruyoshi:2016aim,Agarwal:2016pjo,Benvenuti:2017lle,Agarwal:2017roi,Giacomelli:2017ckh,Agarwal:2018oxb,Benvenuti:2017bpg,Giacomelli:2018ziv}, our work here and in \cite{Buican:2018ddk} suggests that AD theories naturally live along RG flows with accidental SUSY.\footnote{Although note that here and in \cite{Buican:2018ddk} we imagine that the accidental SUSY enhancement arises along RG flows emanating from the AD theories in the UV. On the other hand, in \cite{Maruyoshi:2016tqk,Maruyoshi:2016aim,Agarwal:2016pjo,Benvenuti:2017lle,Agarwal:2017roi,Giacomelli:2017ckh,Agarwal:2018oxb,Benvenuti:2017bpg,Giacomelli:2018ziv} the accidental SUSY enhancement mainly arises for flows ending on AD theories in the IR.} We discuss further implications of these ideas in the conclusions.

The outline of the rest of the paper is as follows. In the next section we give more details regarding the $D_2(SU(N))$ theories, the resulting quiver gauge theories, and the index relations between these quivers and certain Lagrangian theories of class $\CS$. We then move on to construct the 2D TQFT expressions for our indices and study $S$-duality using these expressions. We conclude this section by computing indices for various exotic type $III$ AD fixtures that arise via $S$-duality and relating them to indices of better-known theories consisting purely of regular punctures. In the following section, we analyze the implications of these expressions for the quivers of the corresponding 3D mirrors. We then move on to a discussion of the resulting RG flows with accidental supersymmetry enhancement to thirty-two supercharges and conclude by proving a theorem on the universality of such flows in the class of theories arising from compactification of the $(2,0)$ theory on surfaces with irregular punctures and known 3D mirrors.

Note that throughout our discussion below, we will use the following shorthand to refer to the $D_2(SU(N))$ theories in order to ease notational burden:
\begin{equation}\label{ADNlabel}
AD_N\equiv D_2(SU(N))~,  \ \ \ N\in\mathbb{Z}_{\ge0,\rm odd}~.
\end{equation}

\section{Conformal gauging of $AD_N\equiv D_2(SU(N))$ theories with $N\in\mathbb{Z}_{\ge0,\rm odd}$}
\label{sec:Dp}
In this section we introduce relevant technical aspects of the $AD_N\equiv D_2(SU(N))$ SCFTs (with $N$ odd) and the quiver theories built by conformally gauging them. In particular, we first construct an intermediate building block, $\mathcal{T}^{(\ell)}_{n_1,n_2}$, and then construct the main quiver theories of interest, $\mathcal{T}^{(\ell)}_{n_1,n,n_2}$. We then move on to construct Schur indices for these quivers and relate them to Schur indices of certain Lagrangian theories.

\subsection{More details of the AD quiver building blocks}
The $AD_N$ theories are a class of isolated strongly coupled 4D $\CN=2$ SCFTs. Their Coulomb branch chiral rings are generated by operators of dimensions $\frac{N}{2}-i$ for $0\leq i\leq\lceil\frac{N}{2}\rceil-2$ \cite{Cecotti:2012jx}. Since $N$ is odd, these theories have $\CN=2$ chiral primaries (i.e., \lq\lq Coulomb branch" operators) of non-integer dimension and are therefore of AD type.\footnote{In particular, $\AD_3$ is identical to the $H_2$ Argyres-Douglas theory \cite{Argyres:1995xn} and is sometimes also called the $(A_1,D_4)$ theory \cite{Cecotti:2010fi}.} The conformal anomalies of $\AD_N$ are given by $a_{\AD_N} = \frac{7}{96}(N^2-1)$ and $c_{\AD_N} = \frac{1}{12}(N^2-1)$. Most importantly for us in what follows, the flavor symmetry of $\AD_N$ is $SU(N)$, and the corresponding flavor central charge is given by
\begin{align}
k_{SU(N)} = N~,
\label{eq:k}
\end{align}
where a fundamental hypermultiplet of $SU(N)$ contributes as $k_{SU(N)}=2$. 

\begin{figure}
 \begin{center}
\begin{tikzpicture}[place/.style={circle,draw=blue!50,fill=blue!20,thick,inner sep=0pt,minimum size=8mm},transition2/.style={rectangle,draw=black!50,fill=red!20,thick,inner sep=0pt,minimum size=8mm},transition3/.style={rectangle,draw=black!70,thick,inner sep=0pt,minimum size=8mm},auto]

 \node[place] (1) at (-3.9,0) [shape=circle] {\footnotesize\;$n+\ell$\;};
 \node[transition2] (2) at (-5.6,0) {\;$\AD_{2n+\ell}$\;} edge (1);
 \node[transition2] (3) at (-2.2,0) {\;$\AD_{2n+3\ell}$\;} edge (1); 
  \end{tikzpicture}
\vskip .5cm
\caption{The quiver diagram of two conformally gauged $\AD_{N}$ SCFTs. The left box stands for an $AD_{2n+\ell}$ theory, the right box stands for an $AD_{2n+3\ell}$ theory, and the middle circle stands for an $SU(n+\ell)$ vector multiplet diagonally gauging the two AD theories. Here $n$ is an integer, and $\ell$ is an odd integer. This is the simplest example of a conformally gauged AD building block for the more complicated class of quivers we will focus on (see Fig. \ref{fig:quiver1}).}
\label{fig:ex1}
 \end{center}
\end{figure}
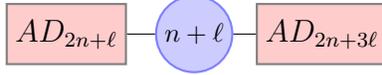

From the isolated $AD_N$ theories, we can construct an intermediate building block for the theories we are interested in as follows. Consider $\AD_{2n+\ell}$ and $\AD_{2n+3\ell}$ for a positive integer $n$ and an {\it odd} positive integer $\ell$ (so that $2n+\ell$ and $2n+3\ell$ are odd). These theories have $SU(2n+\ell)$ and $SU(2n+3\ell)$ flavor symmetries respectively. We can couple an $SU(n+\ell)$ vector multiplet to these SCFTs by gauging a diagonal $SU(n+\ell)$ flavor
symmetry. The flavor central charge \eqref{eq:k} implies that this gauging is exactly marginal. The resulting theory is an $\mathcal{N}=2$ SCFT described by the quiver diagram in Fig.~\ref{fig:ex1} and has $U(n) \times U(n+2\ell)$ flavor symmetry.

Given this flavor symmetry, we can further gauge an $SU(n+2\ell)\subset U(n+2\ell)$ subgroup. This gauging is exactly marginal when the $SU(n+2\ell)$ vector multiplet is coupled to an additional $\AD_{2n+5\ell}$ theory in such a way that the residual flavor symmetry of the $\AD_{2n+5\ell}$ sector is $U(n+3\ell)$. The resulting theory now has $U(n)\times U(1) \times U(n+3\ell)$ flavor symmetry.

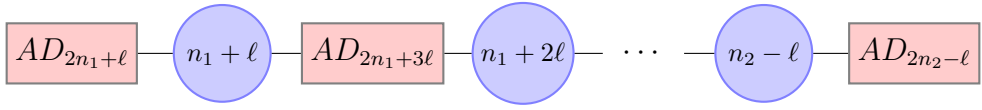
\begin{figure}
 \begin{center}
\begin{tikzpicture}[place/.style={circle,draw=blue!50,fill=blue!20,thick,inner sep=0pt,minimum size=8mm},transition2/.style={rectangle,draw=black!50,fill=red!20,thick,inner sep=0pt,minimum size=8mm},transition3/.style={rectangle,draw=black!70,thick,inner sep=0pt,minimum size=8mm},auto]

 \node[transition2] (1) at (0,0) {\;$\AD_{2n_1+\ell}$\;};
 \node[place] (2) at (2,0) [shape=circle] {\footnotesize\,\;$n_1+\ell$\,\;} edge (1);
 \node[transition2] (3) at (4,0) {\;$\AD_{2n_1+3\ell}$\;} edge (2); 
\node[place] (4) at (6,0) {\footnotesize\;$n_1+2\ell$\;} edge (3);
\node (5) at (7.6,0) {\;$\cdots$\;} edge (4);
\node[place] (6) at (9.2,0) {\footnotesize\;\,$n_2-\ell$\,\;} edge (5);
\node[transition2] (7) at (11.2,0) {\;$\AD_{2n_2-\ell}$\;} edge (6);

  \end{tikzpicture}
\vskip .5cm
\caption{The quiver diagram of the $\CT^{(\ell)}_{n_1,\,n_2}$
  building block for the larger quiver we will consider in Fig. \ref{fig:quiver1} and focus on in the next section. Here $\ell$ is a positive odd integer, and $n_1$ and $n_2$
  are two integers such that $(n_2-n_1)/\ell$ is a positive integer. The
  gauge group of the quiver is $SU(n_1+\ell)\times SU(n_1+2\ell)\times
  \cdots \times SU(n_2-\ell)$. The flavor symmetry is $U(n_1)\times
  U(1)^{\frac{n_2-n_1}{\ell}-2}\times U(n_2)$.}
\label{fig:quiver0}
 \end{center}
\end{figure}

By continuing this procedure, we obtain a series of conformal linear quiver theories whose matter sector is comprised of various $\AD_N$ theories. The quiver diagram for these theories is shown in Fig. \ref{fig:quiver0}, where the gauge group is $SU(n_1+\ell)\times SU(n_1+2\ell)\times \cdots \times SU(n_2-\ell)$ for a positive odd integer, $\ell$, and two integers, $n_1$ and $n_2$, such that $(n_2-n_1)/\ell$ is a positive integer. We denote this theory by $\CT_{n_1,\,n_2}^{(\ell)}$, and it has $U(n_1)\times U(1)^{\frac{n_2-n_1}{\ell}-2}\times U(n_2)$ flavor symmetry.\footnote{Note that, when we write $\mathcal{T}^{(\ell)}_{n_1,\,n_2}$, we always have $n_1<n_2$ so that $(n_2-n_1)/\ell$ is a positive integer.} From the quiver diagram, we see that the flavor central charge of the $SU(n_1)$ and $SU(n_2)$ subgroups are $2n_1+\ell$ and $2n_2-\ell$, respectively.

\subsection{The main quiver theories of interest: the $\CT^{(\ell)}_{n_1,\,n,\, n_2}$ SCFTs}

\begin{figure}
 \begin{center}
\begin{tikzpicture}[place/.style={circle,draw=blue!50,fill=blue!20,thick,inner sep=0pt,minimum size=10mm},transition2/.style={rectangle,draw=black!50,fill=red!20,thick,inner sep=0pt,minimum size=8mm},transition3/.style={rectangle,draw=black!70,thick,inner sep=0pt,minimum size=8mm},auto]

\node[place] (0) at (0,0) {\;$n$\;};
\node[transition2] (-1) at (-1.5,0) {\;$\CT^{(\ell)}_{n_1,n}$\;}
 edge (0);
\node[transition2] (1) at (1.5,0) {\;$\CT^{(\ell)}_{n_2,n}$\;}
 edge (0);
\node[transition3] (2) at (0,1.3) {\;$\ell$\;} edge (0);

  \end{tikzpicture}
\caption{The diagram for the main quiver theory of interest: the  $\CT^{(\ell)}_{n_1,\,n,\,n_2}$ theory. The middle $SU(n)$ diagonally gauges the $SU(n)$ flavor subgroups of $\CT^{(\ell)}_{n_1,n},\,\CT^{(\ell)}_{n_2,n}$, and $\ell\ge1$ fundamental hypermultiplets (recall that $\ell\in\mathbb{Z}_{\rm odd}$). This theory has $U(n_1)\times U(n_2)\times U(\ell)\times  U(1)^{\frac{2n-n_1-n_2}{\ell}-2}$ flavor symmetry.}
\label{fig:quiver1}
 \end{center}
\end{figure}
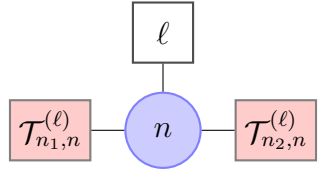

Now we come to the main quiver theories of interest that are built from the above SCFTs and also from fundamental hypermultiplets. To be more explicit, let us take $\CT^{(\ell)}_{n_1,n}$, $\CT^{(\ell)}_{n_2,n}$, and $\ell$ fundamental hypermultiplets of $SU(n)$.\footnote{Note that $n,n_1,n_2$, and $\ell$ are positive integers such that $\ell$ is odd, and $(n-n_1)/\ell$ and $(n-n_2)/\ell$ are positive integers. } By the discussion in the previous subsection, if we gauge a diagonal $SU(n)$ flavor subgroup of these theories, the beta function vanishes:
\begin{align}
 \beta = (2n-\ell) + (2n-\ell) + 2\ell - 4n = 0~,
\end{align}
where $\CT^{(\ell)}_{n_1,n}$ and $\CT^{(\ell)}_{n_2,n}$ both contribute $2n-\ell$, the $\ell$ fundamental hypermultiplets contribute $2\ell$, and the $SU(n)$ vector multiplet contributes $-4n$. The resulting theory is an $\mathcal{N}=2$ SCFT described by the quiver diagram in Fig.~\ref{fig:quiver1} and has $U(n_1)\times U(n_2)\times U(\ell)\times U(1)^{\frac{2n-n_1-n_2}{\ell}-2}$ flavor symmetry. We denote this theory by $\CT^{(\ell)}_{n_1,\,n,\,n_2}$, where the middle $n$ in the subscript stands for the largest rank of the simple components of the gauge group.

\subsection{Schur index}

In this subsection, we construct the Schur indices of the $\CT^{(\ell)}_{n_1,n,n_2}$ SCFTs from the various building blocks described previously. As we will see, these quantities turn out to be closely related to the indices of certain
Lagrangian theories of class $\CS$.

To understand these statements, first recall that the Schur index of a general $\mathcal{N}=2$ SCFT, $\CT$, is defined as  \cite{Gadde:2011ik, Gadde:2011uv}\footnote{The Schur index is a particular limit of a more general superconformal index \cite{Kinney:2005ej, Romelsberger:2005eg}.}
\begin{align}
\CI_{\CT}(q;{\bf x}) \equiv \text{Tr}_{\mathcal{H}}(-1)^F
 q^{E-R}\prod_{i=1}^{\text{rank}\, G_F}({\bf x}_i)^{f_i}~,
\end{align}
where $\mathcal{H}$ is the Hilbert space of local operators of $\CT$, $E$ is the scaling dimension, $R$ is the Cartan generator of $SU(2)_R$ normalized so that the fundamental representation has eigenvalues $\pm \frac{1}{2}$, $G_F$ is the flavor symmetry of the theory, and $f_i$ is the $i^{\rm th}$ Cartan generator of $G_F$ (i.e., the $i^{\rm th}$ flavor charge).

In the case of $AD_N$, the Schur index was conjectured to be \cite{Xie:2016evu} (see also the mathematical results in \cite{kac2017remark,Creutzig:2017qyf})\footnote{The $N=3$ case is also discussed in \cite{Buican:2015ina,Buican:2015hsa,Cordova:2015nma,Buican:2015tda}, and the formula in \eqref{eq:AD} agrees with the formula found in these references.}
\begin{align}
 \CI_{\AD_N}(q;{\bf x}) =
 P.E.\left[\frac{q}{1-q^2}\chi_\text{adj}^{SU(N)}({\bf x})\right]~,
\label{eq:AD}
\end{align}
where ${\bf x} = (x_1,\cdots,x_N)$ subject to $\prod_{i=1}^Nx_i = 1$ is the fugacity for the $SU(N)$ flavor symmetry, $\chi_{\rm adj}^{SU(N)}({\bf x})$ is the character of the adjoint representation, and $P.E.$ is the \lq\lq plethystic exponential." This latter quantity is defined as
\begin{equation}
P.E.[f(x_1,\cdots,x_M)]\equiv\exp\left(\sum_{p=1}^{\infty}f(x_1^p,\cdots,x_M^p)\right)~.
\end{equation}

Let us focus on the case $N=2n_1+\ell$ for a positive integer $n_1$ and an odd positive integer $\ell$, since these theories enter the quivers we are interested in. In order to make contact with the index of the $\CT_{n_1,n_2}^{(\ell)}$ SCFT, it is useful to consider the splitting of the $SU(2n_1+\ell)$ fugacity, ${\bf x}$, into those for the $SU(n_1)\times SU(n_1+\ell)\times U(1)\subset SU(2n_1+\ell)$ subgroup. In particular, ${\bf x}$ splits into ${\bf y}=(y_1,\cdots,y_{n_1}),\, {\bf z} =(z_1,\cdots,z_{n_1+\ell})$, and $a$ such that $\prod_{i=1}^{n_1}y_i = \prod_{i=1}^{n_1+\ell}z_i = 1$.\footnote{The precise relation between ${\bf x}$ and $({\bf y},{\bf z},a)$ is given by
\begin{align}
 a =
\left(\prod_{i=1}^{n_1}x_i\right)^{\frac{1}{n_1}}~,\quad
 y_i = x_i/a \;\;\text{ for }\;\; i =1,\cdots,n_1~,\quad z_i = x_i
 a\;\;\text{for}\;\; i=n_1+1,\cdots,2n_1+\ell~. 
\end{align} 
}
In terms of these variables, the Schur index \eqref{eq:AD} for $N=2n_1+\ell$ is\footnote{For $n_1=1$, we instead have
\begin{align}
\CI_{\AD_{\ell+2}}(q;{\bf z},a) =
 P.E.\left[\frac{q}{1-q^2}\left(1+\chi_\text{adj}^{SU(\ell+1)}({\bf z}) + a
 \chi_\text{afund}^{SU(\ell+1)}({\bf z}) +
 a^{-1}\chi_\text{fund}^{SU(\ell+1)}({\bf z}) \right)\right]~.
\end{align}
}
\begin{align}
 \CI_{\AD_{2n_1+\ell}}(q;{\bf y},{\bf z},a) &=
 P.E.\left[\frac{q}{1-q^2}\left(1+\chi_\text{adj}^{SU(n_1)}({\bf y}) +
 \chi_\text{adj}^{SU(n_1+\ell)}({\bf z})  \right)\right]\times
 \mathcal{I}_\text{bfund}^{n_1\times (n_1+\ell)}(q^2;{\bf y},{\bf z},a)~.
\label{eq:AD2}
\end{align}
where $\chi_{R}^{SU(N)}$ is the character of an $SU(N)$ representation
$R$, ``$\text{adj}$'' stands for the adjoint representation, and
$\mathcal{I}_\text{bfund}^{N\times M}(q,{\bf y},{\bf z},a)$ is the Schur index of a
bifundamental hypermultiplet of $SU(N)\times SU(M)$
\begin{align}
\mathcal{I}_\text{bfund}^{N\times M}(q;{\bf y},{\bf z},a) &\equiv  P.E.\left[\frac{q^{\frac{1}{2}}}{1-q}\left(
 a\,\chi_\text{fund}^{SU(N)}({\bf
 y})\chi_\text{afund}^{SU(M)}({\bf z}) +
 a^{-1}\chi_\text{afund}^{SU(N)}({\bf
 y})\chi_\text{fund}^{SU(M)}({\bf z})\right)\right]~,
\end{align}
with ``$\text{fund}$'' and ``$\text{afund}$'' being fundamental and anti-fundamental representations, respectively. Note that the last factor of \eqref{eq:AD2} is identical to the Schur index of a bifundamental hypermultiplet of $SU(n_1)\times SU(n_1+\ell)$ with $q$ replaced by $q^2$. This expression will be important in our discussions below.

\subsubsection{The index of the $\CT^{(\ell)}_{n_1,n_2}$ building block}
Let us now evaluate the Schur indices of the $\CT^{(\ell)}_{n_1,n_2}$ quiver building blocks we will eventually use to construct the Schur indices of the quivers of ultimate interest. Since the $\CT^{(\ell)}_{n_1,n_2}$ SCFTs are obtained by conformally gauging $\AD_N$ theories, their indices are evaluated as integrals of products of the indices associated with each sector of the quivers.

To describe this gauging, let ${\bf z}_0,\,{\bf z}_{\frac{n_2-n_1}{\ell}}$, and $\vec{a} \equiv (a_1,\cdots,a_{\frac{n_2-n_1}{\ell}})$  be fugacities for $SU(n_1)$, $SU(n_2)$, and $U(1)^{\frac{n_2-n_1}{\ell}}$ subgroups of the $\CT^{(\ell)}_{n_1,n_2}$ flavor symmetry, respectively. Then the quiver diagram in Fig.~\ref{fig:quiver0} implies that  
\begin{align}
&\CI_{\CT^{(\ell)}_{n_1,n_2}}(q;{\bf
 z}_0,\vec{a},{\bf z}_{\frac{n_2-n_1}{\ell}}) 
\nonumber\\
&=\int
 \left(\prod_{i=1}^{\frac{n_2-n_1}{\ell}-1}\!\!\!d\mu_i({\bf
 z}_i)\; \CI_\text{vec}^{SU(n_1 + i\ell)}(q;{\bf z}_i)\right)\left(\prod_{i=0}^{\frac{n_2-n_1}{\ell}-1}\CI_{\AD_{2n_1+(2i+1)\ell}}(q;{\bf
 z}_i,{\bf z}_{i+1},a_{i+1})\right)~,
\label{eq:Tnn1}
\end{align}
where the integral is taken over $SU(n_1+\ell)\times SU(n_1+2\ell)\times \cdots \times SU(n_2-\ell)$, $d\mu_i$ is the Haar measure on $SU(n_1+i\ell)$, ${\bf z}_i$ for $1\leq i\leq \frac{n_2-n_1}{\ell}-1$ is the $SU(n_1+i\ell)$ fugacity associated with $d\mu_i$, and 
\begin{align}
 \CI^{SU(N)}_\text{vec}(q;{\bf z}) \equiv
 P.E.\left[\frac{-2q}{1-q}\chi_\text{adj}^{SU(N)}({\bf z})\right]~,
\end{align}
is the index contribution from an $SU(N)$ vector multiplet. 

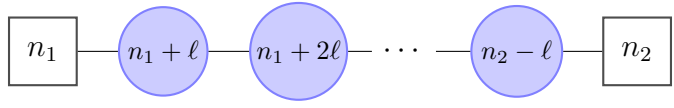
\begin{figure}
 \begin{center}
\begin{tikzpicture}[place/.style={circle,draw=blue!50,fill=blue!20,thick,inner sep=0pt,minimum size=9mm},transition2/.style={rectangle,draw=black!50,fill=red!20,thick,inner sep=0pt,minimum size=8mm},transition3/.style={rectangle,draw=black!70,thick,inner sep=0pt,minimum size=9mm},auto]

\node[transition3] (0) at (-4.6,0) {\;$n_1$\;};
 \node[place] (1) at (-3,0) [shape = circle] {\footnotesize\;$n_1+\ell$\;} edge (0);
\node[place] (2) at (-1.2,0) {\,{\footnotesize $n_1+2\ell$}\,} edge (1);
\node (3) at (.2,0) {$\cdots$} edge (2);
\node[place] (4) at (1.7,0) {\;{\footnotesize $n_2-\ell$}\;} edge (3);
\node[transition3] (5) at (3.3,0) {\;$n_2$\;} edge (4);

\end{tikzpicture}
\caption{The quiver diagram of the Lagrangian theory we call $\mathcal{L}^{(\ell)}_{n_1,n_2}$. Each edge connecting two nodes stands for a bifundamental hypermultiplet, and each box labeled by \lq\lq $n$" stands for $n$ fundamental hypermultiplets. The flavor symmetry of $\mathcal{L}^{(\ell)}_{n_1,n_2}$ is generically the same as that of  $\CT^{(\ell)}_{n_1,n_2}$.}
\label{fig:quiver2}
 \end{center}
\end{figure}

Note that, up to adjoint-valued pre-factors (whose role we will clarify below) and a $q\to q^2$ fugacity rescaling,  the Schur indices of the $\AD_{2n_1+\ell}$ SCFTs in \eqref{eq:AD2} are just the indices of bifundamental hypermultiplets. As a result, the indices of the $\CT^{(\ell)}_{n_1,n_2}$ SCFTs will also have a close connection with those of
Lagrangian theories.
Indeed, using the identities \eqref{eq:rewrite-vec} and \eqref{eq:AD2}, one can
rewrite \eqref{eq:Tnn1} as
\begin{align}
 \CI_{\CT^{(\ell)}_{n_1,n_2}}(q;{\bf z}_0,\vec{a},{\bf
 z}_{\frac{n_2-n_1}{\ell}}) &=  \frac{1}{(q;q^2)^{\frac{n_2-n_1}{\ell}}}\, P.E.\left[\frac{q}{1-q^2}\left(
 \chi_\text{adj}^{SU(n_1)}({\bf z}_0) + \chi_\text{adj}^{SU(n_2)}({\bf
 z}_{\frac{n_2-n_1}{\ell}})\right)\right] 
\nonumber\\[1mm]
&\qquad \times \CI_{\mathcal{L}_{n_1,n_2}^{(\ell)}}(q^2;{\bf
 z}_0,\vec{a},{\bf z}_{\frac{n_2-n_1}{\ell}})~,
\label{eq:connection1}
\end{align}
where 
\begin{align}
&  \mathcal{I}_{\mathcal{L}_{n_1,n_2}^{(\ell)}}(q,{\bf z}_0,\vec{a},{\bf
 z}_{\frac{n_2-n_1}{\ell}}) 
\nonumber\\
&\equiv  \int\left(\prod_{i=1}^{\frac{n_2-n_1}{\ell}-1} d\mu_i({\bf
 z}_i) \CI_\text{vec}^{SU(n_1+i\ell)}(q;{\bf z}_i) \right)
\prod_{i=0}^{\frac{n_2-n_1}{\ell}-1}\CI_\text{bfund}^{(n_1+i\ell)\times
 (n_1+(i+1) \ell)}(q;{\bf
 z}_i,{\bf
 z}_{i+1},a_{i+1})~,
\label{eq:Ln1n2}
\end{align}
is the Schur index of the Lagrangian theory described by the quiver in Fig.~\ref{fig:quiver2}. Note that this quiver has the same gauge group as in Fig.~\ref{fig:quiver0}, but its matter sector is composed purely of fundamental and bifundamental hypermultiplets.\footnote{The $\mathcal{L}^{(\ell)}_{n_1,n_2}$ theory has the same flavor symmetry as $\CT^{(\ell)}_{n_1,n_2}$ unless there is an accidental enhancement. Therefore, its Schur index is a function of the same set of fugacities as $\CI_{\CT^{(\ell)}_{n_1,n_2}}$.} The expression \eqref{eq:connection1} shows that the Schur index of $\mathcal{T}^{(\ell)}_{n_1,n_2}$ has a close connection with that of $\mathcal{L}^{(\ell)}_{n_1,n_2}$ (we need only multiply by adjoint-valued prefactors and rescale $q\to q^2$).

Let us briefly comment on the plethystic exponential pre-factor in front of $\CI_{\mathcal{L}^{(\ell)}_{n_1,n_2}}$ on
the RHS of \eqref{eq:connection1}. This term is inherited from the AD theories at the ends of the quiver and is independent of the abelian flavor fugacities, $\vec{a}$. On the other hand, this pre-factor does depend on the fugacities, ${\bf x}$ and ${\bf y}$, for the non-abelian flavor subgroup. The role of this dependence can be understood by noting that $\CI_{\CT^{(\ell)}_{n_1,n_2}}$ and $\CI_{\mathcal{L}^{(\ell)}_{n_1,n_2}}$ satisfy recursive relations. Indeed,
$\CI_{\CT^{(\ell)}_{n_1,n_2}}$ satisfies
\begin{align}
 \CI_{\CT^{(\ell)}_{n_1,n_2}}(q;{\bf
 x},\vec{a},{\bf y}) &= \int_{SU(n_2-i\ell)}  d\mu({\bf z})\,
 \CI_{\CT^{(\ell)}_{n_1,n_2-i\ell}}(q;{\bf
 x},\vec{b},{\bf z})\,\CI_\text{vec}^{SU(n_2-i\ell)}(q;{\bf
 z})\,\CI_{\CT^{(\ell)}_{n_2-i\ell,n_2}}(q;{\bf z},{\bf
 y},\vec{c})~,
\end{align}
where $1\leq i\leq \frac{n_2-n_1}{\ell}$, and $\vec{a} = (b_1,\cdots,b_{\frac{n_2-n_1}{\ell}-i},c_1,\cdots, c_i)$. There is a similar recursive relation for $\CI_{\mathcal{L}^{(\ell)}_{n_1,n_2}}$, where all $\CI_{\CT^{(\ell)}_{n,m}}$ are replaced with $\CI_{\mathcal{L}^{(\ell)}_{n,m}}$. These two recursive relations are consistent with \eqref{eq:connection1} if the $P.E.$ factor is present in the relation \eqref{eq:connection1}.\footnote{The flavor-independent part of the pre-factor multiplying $\CI_{\mathcal{L}^{(\ell)}_{n_1,n_2}}$ in \eqref{eq:connection1}, $(q;q^2)^{-\left(n_2-n_1\over\ell\right)}$, is present in order to make up the difference in $a-c$ between the Lagrangian and non-Lagrangian theories in the Cardy limit of the index.}

\subsubsection{The indices of the $\CT^{(\ell)}_{n_1,n,n_2}$ quivers}
Let us now assemble our previous results and compute the Schur indices of the quivers we will ultimately be interested in for our discussion below---the $\CT^{(\ell)}_{n_1,n,n_2}$ SCFTs. To begin, we let $({\bf x}_1,a_1),\,({\bf x}_2,b_1)$, and $({\bf y},c)$ denote the fugacities for the flavor $U(n_1),\,U(n_2)$, and $U(\ell)$ subgroups, respectively. We also let $(a_2,\cdots, a_{\frac{n-n_1}{\ell}})$ and $(b_2,\cdots,b_{\frac{n-n_2}{\ell}})$ represent the fugacities for the residual $U(1)^{\frac{2n-n_1-n_2}{\ell}-2}$ flavor subgroup. From its quiver description in Fig. \ref{fig:quiver1}, we see that the Schur index of $\CT^{(\ell)}_{n_1,n,n_2}$ can be evaluated as
\begin{align}
 \CI_{\CT^{(\ell)}_{n_1,n,n_2}}(q;{\bf
 x}_{1},\vec{a},({\bf y},c),\vec{b},{\bf x}_2) &= \int_{SU(n)} d\mu({\bf z}) \,\CI_\text{vec}^{SU(n)}(q;{\bf
 z})\CI_\text{bifund}^{\,\ell\times n}(q;{\bf y},{\bf z},c)
\nonumber\\
&\qquad \qquad \times 
 \CI_{\CT^{(\ell)}_{n_1,n}}(q;{\bf x}_1,\vec{a},{\bf
 z})\CI_{\CT^{(\ell)}_{n_2,n}}(q;{\bf
 x}_2,\vec{b},{\bf z})~,
\label{eq:Tnnn1}
\end{align}
where $\vec{a}\equiv (a_1,\cdots,a_{\frac{n-n_1}{\ell}})$ and $\vec{b}
\equiv (b_1,\cdots,b_{\frac{n-n_2}{\ell}})$.

\begin{figure}
 \begin{center}
\begin{tikzpicture}[place/.style={circle,draw=blue!50,fill=blue!20,thick,inner sep=0pt,minimum size=10mm},transition2/.style={rectangle,draw=black!50,fill=red!20,thick,inner sep=0pt,minimum size=8mm},transition3/.style={rectangle,draw=black!70,thick,inner sep=0pt,minimum size=8mm},auto]

\node[transition3] (a-1) at (-1.5,-2.5) {\;{\small $\mathcal{L}^{(\ell)}_{n_1,n}$}\;};
\node[place] (a0) at (0,-2.5) {\;$n$\;} edge (a-1);
\node[transition3] (a+1) at (1.5,-2.5) {\;{\small $\mathcal{L}^{(\ell)}_{n_2,n}$}\;} edge (a0);

\node[transition3] (a+5) at (0,-1.2) {\;$2\ell$\;} edge (a0);

\draw (8,-2) ellipse (3 and 1.2);

\node at (5.8,-2) {\small $Y_{n_1}^{(\ell)}$};
\draw[fill] (6.8,-1.5) circle (.04);
\draw[fill] (7.2,-1.5) circle (.04);
\draw[fill] (7.6,-1.5) circle (.04);
\node at (8.4,-1.5) {$\cdots$};
\draw[fill] (9.2,-1.5) circle (.04);
\node at (10.2,-2) {\small $Y_{n_2}^{(\ell)}$};

  \end{tikzpicture}
\vskip .3cm

\caption{The left quiver is a weak coupling description of the Lagrangian theory $\mathcal{L}^{(\ell)}_{n_1,n,n_2}$. The gauge group is the same as that in Fig.~\ref{fig:quiver1}, but the matter sector is composed purely of fundamental and bifundamental hypermultiplets. The rank of the flavor symmetry group of $\mathcal{L}^{(\ell)}_{n_1,n,n_2}$ is larger than that of $\CT^{(\ell)}_{n_1,n,n_2}$ by $\ell$. The $\mathcal{L}^{(\ell)}_{n_1,n,n_2}$ theory is obtained by compactifying the 6D $(2,0)$ $A_{n-1}$ theory on the punctured sphere shown in the right picture. The sphere has $\frac{2n-n_1-n_2}{\ell}$ simple punctures (represented by black points) and two additional regular punctures associated with $Y_{n_1}^{(\ell)}$ and $Y_{n_2}^{(\ell)}$. The complex structure moduli space of this punctured sphere is identified as the conformal manifold of $\mathcal{L}^{(\ell)}_{n_1,n,n_2}$.}
\label{fig:quiver3}
 \end{center}
\end{figure}
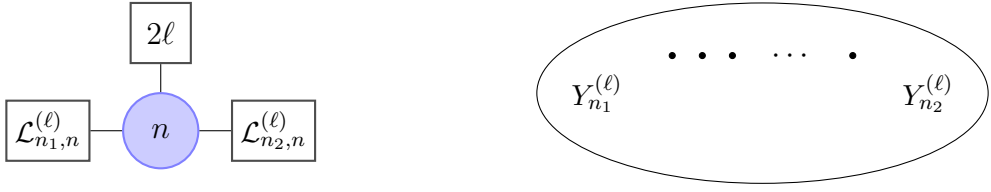

As in the case of $\CT^{(\ell)}_{n_1,n_2}$, this Schur index is also related to the index of a quiver gauge theory with a Lagrangian description. Indeed, using \eqref{eq:rewrite-vec}, \eqref{eq:rewrite-bfund} and
\eqref{eq:connection1}, one can rewrite 
\eqref{eq:Tnnn1} as\footnote{In the case of $n_i=1$, the factor $\chi_\text{adj}^{SU(n_i)}$ is replaced with $0$. In the case of $n_i=0$, it is replaced by $-1$.}
\begin{align}
\CI_{\CT^{(\ell)}_{n_1,n,n_2}}(q;{\bf x}_1,\vec{a},({\bf
 y},c),\vec{b},{\bf x}_2) 
&=
 \frac{1}{(q;q^2)^{\frac{2n-n_1-n_2}{\ell}}}P.E.\left[\frac{q}{1-q^2}\left(\chi_\text{adj}^{SU(n_1)}
({\bf x}_1) + \chi_\text{adj}^{SU(n_2)}({\bf x}_2) \right)\right]
\nonumber\\[2mm]
&\qquad \times  \CI_{\mathcal{L}^{(\ell)}_{n_1,n,n_2}}(q^2;{\bf
 x}_1,\vec{a},({\bf y},cq^{\frac{1}{2}}),({\bf
 y},cq^{-\frac{1}{2}}),\vec{b},{\bf x}_2)~,
\label{eq:connection2}
\end{align}
where
\begin{align}
 &\CI_{\mathcal{L}^{(\ell)}_{n_1,n,n_2}}(q;{\bf x}_1,\vec{a},({\bf
 y}_1,c_1),({\bf y}_2,c_2),\vec{b},{\bf x}_2) 
\nonumber\\
&\equiv \int d\mu({\bf
 z})\,\CI_\text{vec}^{SU(n)}(q;{\bf z}) \, \CI_{\mathcal{L}^{(\ell)}_{n_1,n}}(q;{\bf
 x}_1,\vec{a},{\bf
 z})\, 
\CI_{\mathcal{L}^{(\ell)}_{n_2,n}}(q;{\bf
 x}_2,\vec{b},{\bf z})
\,
\prod_{i=1}^2\CI^{\ell\times n}_\text{bifund}(q;{\bf y}_i,{\bf
 z},c_i)~,
\label{eq:Lnnn1}
\end{align}
is the Schur index of a Lagrangian theory described by the quiver diagram in Fig.~\ref{fig:quiver3}. We call this quiver gauge theory $\mathcal{L}^{(\ell)}_{n_1,n,n_2}$.

Note that the flavor symmetry of $\mathcal{L}^{(\ell)}_{n_1,n,n_2}$ is $U(n_1)\times U(n_2) \times U(2\ell)\times U(1)^{\frac{2n-n_1-n_2}{\ell}-2}$. In \eqref{eq:Lnnn1}, $({\bf x}_1,a_1)$ and $({\bf x}_2,b_1)$ are fugacities for the $U(n_1)$ and $U(n_2)$ flavor subgroups respectively, while $({\bf y}_1,{\bf y}_2,c_1,c_2)$ are fugacities for the $U(2\ell)$ flavor subgroup. Note that the flavor symmetry of $\mathcal{L}_{n_1,n,n_2}^{(\ell)}$ is {\it not} the same as the flavor symmetry of $\CT^{(\ell)}_{n_1,n,n_2}$. Indeed, the rank of the flavor symmetry of $\mathcal{L}^{(\ell)}_{n_1,n,n_2}$ is larger than that of $\CT^{(\ell)}_{n_1,n,n_2}$ by $\ell$.  Therefore, in the relation \eqref{eq:connection2}, $2\ell$ fugacities for the $U(2\ell)$ flavor subgroup of $\mathcal{L}^{(\ell)}_{n_1,n,n_2}$ are restricted to $\ell$ fugacities $({\bf y},c)$. Finally, note that the $P.E.$ factor depending on ${\bf x}_1$ and ${\bf x}_2$ plays the same role as in the case of $\CT^{(\ell)}_{n_1,n_2}$.

\section{TQFT expressions for the Schur indices and S-duality}
\label{sec:S-duality}

In this section we begin by focusing on the $\CT^{(\ell)}_{n_1,n,n_2}$ SCFTs and studying the resulting $S$-dualities via the connection with $\mathcal{L}^{(\ell)}_{n_1,n,n_2}$ discussed in the previous section. In particular, this connection leads us to simple TQFT expressions for the Schur indices of the $\CT^{(\ell)}_{n_1,n,n_2}$ SCFTs and makes it straightfoward to read off the action of $S$-duality on the corresponding abelian flavor symmetries.\footnote{The generalization of this discussion to $\CT^{(\ell)}_{n_1,n_2}$ is straightforward but involves extra  decoupled hypermultiplets.} Moreover, as we will see, the TQFT approach gives rise to interesting new expressions for indices of certain exotic AD building blocks that appear at certain cusps in the conformal manifolds of the  $\CT^{(\ell)}_{n_1,n,n_2}$ theories.

One useful aspect of the Lagrangian quiver theory, $\mathcal{L}^{(\ell)}_{n_1,n,n_2}$, is that it can be obtained by compactifying the 6D (2,0) $A_{n-1}$ theory on a sphere with $\frac{2n-n_1-n_2}{\ell}$ simple punctures and two additional regular punctures associated with $Y_{n_1}^{(\ell)}$ and $Y_{n_2}^{(\ell)}$ (see Fig.~\ref{fig:quiver3}) \cite{Gaiotto:2009we}. This fact implies that the superconformal index of $\mathcal{L}^{(\ell)}_{n_1,n,n_2}$ can be computed via a TQFT on the sphere \cite{Gadde:2009kb}. In this context, its Schur index, $\CI_{\mathcal{L}_{n_1,n,n_2}^{(\ell)}}$, is written as a correlation function of $q$-deformed Yang-Mills ($q$-YM) theory \cite{Gadde:2011ik}. Moreover, since the compactificaiton of the 6D $(2,0)$ theory involves only regular punctures, the TQFT expression for $\CI_{\mathcal{L}_{n_1,n,n_2}^{(\ell)}}$ is particularly simple.

On the other hand, AD theories arise from the compactifications of the $(2,0)$ theory with one irregular puncture and, depending on the case, at most one additional regular puncture. The resulting TQFT index expressions tend to be considerably more elaborate \cite{Buican:2015ina,Buican:2017uka}.

\begin{figure}
 
 \begin{center}
\begin{tikzpicture}[place/.style={circle,draw=blue!50,fill=blue!20,thick,inner sep=0pt,minimum size=8mm},transition2/.style={rectangle,draw=black!50,fill=red!20,thick,inner sep=0pt,minimum size=8mm},transition3/.style={rectangle,draw=black!70,thick,inner sep=0pt,minimum size=8mm},auto]

\draw (0,0) rectangle (1.45,2);
\draw (.25,0) -- (.25,2);
\draw (.5,0) -- (.5,2);
\draw (1.2,0) -- (1.2,2);
\node at (.9,1) {$\cdots$}; 

\draw (-1.5,0) rectangle (0,.25);
\draw (-1.25,0) -- (-1.25,.25);
\draw (-1,0) -- (-1,.25);
\draw (-.25,0) -- (-.25,.25);
\node at (-.6,.125) {$\cdots$}; 

\draw (0,.25) -- (.5,.25);
\draw (0,.5) -- (.5,.5);

\draw (0,2-.25) -- (.5,2-.25);

\draw (1.45,.25) -- (1.45-.25,.25);
\draw (1.45,.5) -- (1.45-.25,.5);

\draw (1.45,2-.25) -- (1.45-.25,2-.25);

\node at (.125,1.25) {$\vdots$};

\node at (.25+.125,1.25) {$\vdots$};

\node at (1.45-.125,1.25) {$\vdots$};

\draw [decorate,decoration={brace,amplitude=5pt,mirror,raise=4pt}]
(1.45,0) -- (1.45,2) node[midway,right=.3]{$\frac{n-k}{\ell}$};

\draw [decorate,decoration={brace,amplitude=5pt, raise=4pt}]
(0,2) -- (1.45,2) node[midway, above=.3]{$\ell$};

\draw [decorate,decoration={brace,amplitude=5pt, mirror, raise=4pt}]
(-1.5,0) -- (0,0) node[midway, below=.3]{$k$};

\draw (6-1,0) rectangle (11.5-1,1);
\draw (7.1-1,0) -- (7.1-1,1);
\draw (8.2-1,0) -- (8.2-1,2);
\draw (9.3-1,0) -- (9.3-1,2);
\draw (10.4-1,0) -- (10.4-1,2);
\draw (8.2-1,2) -- (11.5-1,2) -- (11.5-1,1);

\node at (6.55-1,.5) {\scriptsize $f^{-2}x_1$};
\node at (7.65-1,.5) {\scriptsize $f^{-2}x_2$};
\node at (8.78-1,.5) {\scriptsize $q^{-\frac{1}{2}}\!fy_1$};
\node at (9.88-1,.5) {\scriptsize $q^{-\frac{1}{2}}\!fy_2$};
\node at (10.98-1,.5) {\scriptsize $q^{-\frac{1}{2}}\!fy_3$};
\node at (8.78-1,1.5) {\scriptsize $q^{\frac{1}{2}}\!fy_1$};
\node at (9.88-1,1.5) {\scriptsize $q^{\frac{1}{2}}\!fy_2$};
\node at (10.98-1,1.5) {\scriptsize $q^{\frac{1}{2}}\!fy_3$};

\end{tikzpicture}
\caption{The left picture shows the Young diagram $Y_{k}^{(\ell)}$ with $n$ boxes. Here $k$ and $\ell$ are non-negative integers such that $\frac{n-k}{\ell}$ is a positive integer. There are $k$ columns of height one and $\ell$ columns of height $\frac{n-k}{\ell}$. We also use the shorthand notation $Y_\text{simple} \equiv Y_1^{(1)}$ and $Y_\text{full}\equiv Y_{0}^{(n)} = Y_{n-1}^{(1)}$ in the main text. The right picture shows how the $SU(n)$ fugacity ${\bf w}$ in \eqref{eq:wave1} is related to the $SU(k)\times SU(\ell)\times U(1)$ fugacities $({\bf x},{\bf y},f)$ (in the particular case of $n=8,k=2$, and $\ell=3$), where $w_1,\cdots,w_n$ are assigned to the boxes.}
\label{fig:Young}
 \end{center}
\end{figure}
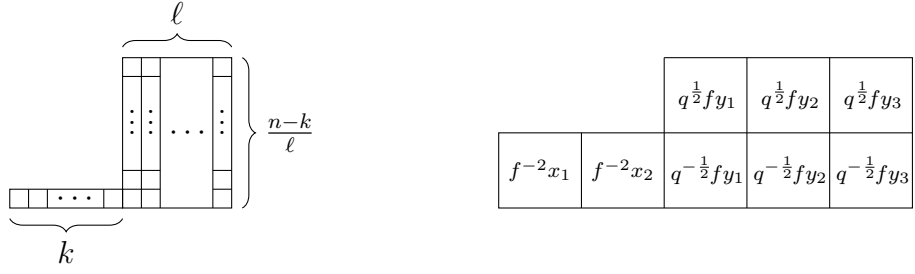

However, the simple TQFT expression for $\CI_{\mathcal{L}^{(\ell)}_{n_1,n,n_2}}$ and the relation \eqref{eq:connection2} imply that the Schur index of the non-Lagrangian quiver theory $\CT^{(\ell)}_{n_1,n,n_2}$ also has a simple TQFT expression. Indeed, applying the transformation in \eqref{eq:connection2} to the $q$-YM expression for $\CI_{\mathcal{L}^{(\ell)}_{n_1,n,n_2}}$, we obtain
\begin{align}
& \CI_{\CT^{(\ell)}_{n_1,n,n_2}}(q;{\bf x}_1,\vec{a}, ({\bf
 y},c),\vec{b},{\bf x}_2) 
\nonumber\\
&=  \frac{1}{(q;q^2)^{\frac{2n-n_1-n_2}{\ell}}}P.E.\left[\frac{q}{1-q^2}\left(\chi_\text{adj}^{SU(n_1)}
({\bf x}_1) + \chi_\text{adj}^{SU(n_2)}({\bf x}_2) \right)\right]
\nonumber\\
&\times \!\!\! \sum_{R: \text{ irreps of
 }\mathfrak{su}(n)}\!\!\! \frac{f_R^{Y_{n_1}^{(\ell)}}\!(q^2;{\bf x}_1,{\bf y},e_0)\Big(\prod_{i=1}^{\frac{n-n_1}{\ell}}
 f_R^{Y_\text{simple}}(q^2;e_i)\Big)\Big(\prod_{j=1}^{\frac{n-n_2}{\ell}}f_R^{Y_\text{simple}}(q^2;f_j)\Big)f_R^{Y_{n_2}^{(\ell)}}\!(q^2;{\bf
 x}_2,{\bf y}^*\!,f_0)}{\big(C_R(q^2)\big)^{\frac{2n-n_1-n_2}{\ell}}}~,
\label{eq:qYM1}
\end{align}
where ${\bf y}^* \equiv (y_1^{-1}, \cdots,y_\ell^{-1})$ and
\begin{align}
C_R(q) &\equiv
 \frac{\prod_{\ell=1}^{n-1}(1-q^\ell)^{n-\ell}}{(q;q)_\infty^{n-1}}\chi_R^{SU(n)}(q^{\frac{n-1}{2}},q^{\frac{n-3}{2}},\cdots,
 q^{-\frac{n-1}{2}})~.
\end{align}
The parameters $e_i$ and $f_i$ are functions of $\vec{a},\ \vec{b},\ c$, and $q$ satisfying
\begin{align}
 (e_0)^n \equiv
 q^{\frac{n_1}{2}}c^{n_1}\prod_{i=1}^{\frac{n-n_1}{\ell}}(a_i)^{-n_1}~&,
 \qquad (f_0)^n =
 q^{\frac{n_2}{2}}c^{-n_2}\prod_{j=1}^{\frac{n-n_2}{\ell}}(b_j)^{n_2}~,
\label{eq:def-ef0}
\\
 \left(e_i\right)^n = q^{\frac{\ell}{2}}c^\ell(a_{i})^{n_1+(i-1)\ell}
 \prod_{k=i+1}^{\frac{n-n_1}{\ell}}(a_{k})^{-\ell}~&,\qquad
 \left(f_j\right)^{n} =q^{\frac{\ell}{2}}c^{-\ell} (b_{j})^{-n_2-(j-1)\ell}
 \prod_{k=j+1}^{\frac{n-n_2}{\ell}}(b_{k})^{\ell}~,
\label{eq:def-efi}
\end{align}
for $1\leq i\leq \frac{n-n_1}{\ell}$ and $1\leq j \leq \frac{n-n_2}{\ell}.$\footnote{Note that not all $e_i$ and $f_j$ are independent. Indeed, we see that there is one constraint on them:
\begin{align}
\left(\prod_{i=0}^{\frac{n-n_1}{\ell}}e_i\right)\left(\prod_{j=0}^{\frac{n-n_2}{\ell}}f_j\right)
 = q~.
\end{align}
}
The ``wave function'' $f_R^{Y}$ depends on the Young diagram $Y$, and
$Y^{(\ell)}_k$ is the $n$-box Young diagram with $k$ columns of height
one and $\ell$ columns of height $\frac{n-k}{\ell}$ (see Fig.~\ref{fig:Young}). We use the short-hand notation $Y_\text{simple} \equiv Y_1^{(1)}$ and $Y_\text{full}\equiv Y_{n-1}^{(1)} = Y_{0}^{(n)}$. The wave function $f_R^{Y}$ for $Y=Y_k^{(\ell)}$ is given by \cite{Gadde:2011ik, Beem:2014rza}
\begin{align}
 f_R^{Y_{k}^{(\ell)}}(q;{\bf x},{\bf y},f) &\equiv K^{Y_{k}^{(\ell)}}(q;{\bf
 x},{\bf y},f) \chi^{SU(n)}_R({\bf w})~,
\label{eq:wave1}
\\
K^{Y_k^{(\ell)}}(q;{\bf x},{\bf y},f) &\equiv
 P.E.\left[\frac{q}{1-q}\chi_\text{adj}^{SU(k)}({\bf x}) +
 \frac{q^{\frac{1}{2}\left(\frac{n-k}{\ell}+1\right)}}{1-q}f^{-\frac{n}{k}}\chi_\text{fund}^{SU(k)}({\bf
 x})\chi_\text{afund}^{SU(\ell)}({\bf y})\right.
\nonumber\\
&\left. \qquad +
 \frac{q^{\frac{1}{2}\left(\frac{n-k}{\ell}+1\right)}}{1-q}f^{\frac{n}{k}}\chi_\text{afund}^{SU(k)}({\bf
 x})\chi_\text{fund}^{SU(\ell)}({\bf y})+
 \frac{q(1-q^{\frac{n-k}{\ell}})}{(1-q)^2}\chi_\text{adj}^{U(\ell)}({\bf
 y})\right]~,
\label{eq:K-factor}
\end{align}
where ${\bf w}$ is an $SU(n)$ fugacity such that $w_i \equiv f^{-\frac{n-k}{k}}x_i$ for $1\leq i\leq k$ and  $w_{k
+\frac{n-k}{\ell}(i-1) + j} \equiv q^{\frac{1}{2}\left(\frac{n-k}{\ell}+1\right)-j}fy_i$ for $1\leq i\leq \ell$ and $1\leq j \leq \frac{n-k}{\ell}$ (see Fig.~\ref{fig:Young}), and $\chi_\text{adj}^{U(\ell)}({\bf y})= \chi_\text{adj}^{SU(\ell)}({\bf y}) + 1$.\footnote{For $Y_\text{simple}$ and $Y_\text{full}$, this expression reduces to
\begin{align}
f_R^{Y_\text{simple}}(q;f) &=
 P.E.\left[\frac{q^{\frac{n}{2}}}{1-q}(f^n+f^{-n})\right]\frac{\prod_{\ell=1}^{n-2}(1-q^\ell)^{n-\ell-1}}{(q;q)_\infty^{n-1}} \chi_R^{SU(n)}(fq^{\frac{n-2}{2}},
\cdots, fq^{-\frac{n-2}{2}},f^{1-n})~,
\\[1mm]
  f_R^{Y_{\text{full}}}(q;{\bf x}) &=
 P.E.\left[\frac{q}{1-q}\chi_\mathrm{adj}^{SU(n)}({\bf
 x})\right]\chi_R^{SU(n)}(\bf x)~.
\end{align}
}
Note here that the wave function factors in \eqref{eq:qYM1} are {\it not} directly given by \eqref{eq:wave1} but involve the rescaling $q\to q^2$.

Note also that the expression \eqref{eq:qYM1} for the Schur index of $\CT^{(\ell)}_{n_1,n,n_2}$ is invariant under the
permutations of $(e_1,\cdots,e_{\frac{n-n_1}{\ell}},f_1,\cdots,f_{\frac{n-n_2}{\ell}})$. It turns out that such permutations are realized by reparameterizing  $a_i,b_j$ and $c$. Indeed, $e_i \longleftrightarrow e_{i+1}$ is realized by
\begin{align}
 a_i \to (a_i)^{\frac{\ell}{n_2+i\ell}}
 (a_{i+1})^{\frac{n_2+(i+1)\ell}{n_2+i\ell}}~,\qquad a_{i+1}\to
 (a_i)^{\frac{n_2+(i-1)\ell}{n_2+i\ell}}(a_{i+1})^{-\frac{\ell}{n_2+i\ell}}~,
\label{eq:S-gen-1}
\end{align}
with the other fugacities kept fixed. Similarly, $f_i\longleftrightarrow f_{i+1}$ is realized by a transformation of $b_i$
and $b_{i+1}$. Finally, $e_{\frac{n-n_1}{\ell}}\longleftrightarrow f_{\frac{n-n_2}{\ell}}$ is realized by 
\begin{align}
a_{\frac{n-n_1}{\ell}}\to
 (a_{\frac{n-n_1}{\ell}})^{\frac{\ell}{n}}&(b_{\frac{n-n_2}{\ell}})^{\frac{\ell}{n}-1}c^{-\frac{2\ell}{n}}~,\qquad
 b_{\frac{n-n_2}{\ell}}\to (a_{\frac{n-n_1}{\ell}})^{\frac{\ell}{n}-1}(b_{\frac{n-n_2}{\ell}})^{\frac{\ell}{n}}c^{-\frac{2\ell}{n}}~,
\nonumber\\[1mm]
c &\to
 (a_{\frac{n-n_1}{\ell}})^{\frac{\ell}{n}-1}(b_{\frac{n-n_2}{\ell}})^{\frac{\ell}{n}-1}c^{1-\frac{2\ell}{n}}~,
\label{eq:S-gen-2}
\end{align}
with the other fugacities kept fixed.  Note that all these transformations keep $e_0$ and $f_0$ invariant, and therefore preserve the wave functions $f_R^{Y_{n_1}^{(\ell)}}(q^2;{\bf x}_1,{\bf y},e_0)$ and $f_{R}^{Y_{n_2}^{(\ell)}}(q^2;{\bf x}_2,{\bf y^*},f_0)$.  This discussion shows that the Schur index of $\CT^{(\ell)}_{n_1,n,n_2}$ is invariant under the action of
$S_{\frac{2n-n_1-n_2}{\ell}}$. As discussed below, this invariance can be regarded as a natural generalization of an $S_{2n}$ symmetry of the index of $\CT^{(1)}_{1,n,1} = (A_{2n-1},A_{2n-1})$, which was identified in \cite{Buican:2017uka} as the action of the S-duality group (see also \cite{Creutzig:2018lbc}). It is therefore natural to interpret the above $S_{\frac{2n-n_1-n_2}{\ell}}$ invariance as a consequence of the S-duality invariance of $\CT^{(\ell)}_{n_1,n,n_2}$. 

In the next section, we carefully study two special cases, $\CT^{(n)}_{0,n,0}$ and $\CT^{(1)}_{1,n,1}$, and show that, from various S-dual descriptions of these theories, one can read off the Schur indices of various infinite series of exotic type $III$ AD theories that decouple at cusps in the space of gauge couplings.

\section{$S$-duality and indices for exotic AD fixtures}
In this section we perform a more thorough analysis of two sets of examples of the $S$-dualities discussed in the previous section. In particular, we construct indices for exotic AD fixtures that arise in certain decoupling limits of the $\CT^{(n)}_{0,n,0}$ and $\CT^{(1)}_{1,n,1}$ SCFTs. The first set of examples gives rise to theories that generalize the $\CT_X$ theory discussed in \cite{Buican:2018ddk} and are AD analogs of the $R_{0,n}$ theories studied in \cite{Chacaltana:2010ks}. Some of the theories in the second set of examples are  AD analogs of other regular puncture fixtures (although, we will see there are some interesting subtleties in this analysis).

\subsection{S-duality of the $\CT^{(n)}_{0,n,0}$ SCFTs and AD analogs for $R_{0,n}$ theories}
\label{subsec:Tnl}

Let us first focus on the $\CT^{(n)}_{0,n,0}$ theory, where $n\geq 3$ is an odd positive integer. The quiver diagrams of $\CT^{(n)}_{0,n,0}$ and $\mathcal{L}^{(n)}_{0,n,0}$ are shown in Fig.~\ref{fig:quiver1} and Fig.~\ref{fig:quiver3}, respectively. For $\ell = n$ and $n_1=n_2=0$, the TQFT expression \eqref{eq:qYM1} reduces to
\begin{align}
 \CI_{\CT_{0,n,0}^{(n)}}(q;{\bf y},c) = \sum_{R:\text{ irreps
 of }\mathfrak{su}(n)}\frac{f_R^{Y_\text{full}}(q^2;{\bf
 y})f_R^{Y_\text{simple}}(q^2;q^{\frac{1}{2}}c)f_R^{Y_\text{simple}}(q^2;q^{\frac{1}{2}}c^{-1})f_R^{Y_\text{full}}(q^2;{\bf
 y}^*)}{\big(C_R(q^2)\big)^2}~,
\label{eq:qYM-n0n0-1}
\end{align}
where ${\bf y}$ and $c$ are fugacities for $SU(n)\subset U(n)$ and $U(1)\subset U(n)$ subgroups of the flavor $U(n)$ symmetry, respectively.\footnote{Note that, for $n_1=n_2=0$, the first line of the RHS in \eqref{eq:qYM1} reduces to $1$. Moreover, the Young diagrams $Y_{n_1}^{(\ell)}$ and $Y_{n_2}^{(\ell)}$ reduce to $Y_\text{full}\equiv Y_0^{(n)} = Y_{n-1}^{(1)}$. We also note that $e_0 = f_0=1$ in this case.} Note that, unlike in the case of regular puncture theories, the two full puncture wave functions are not independent of each other since they have conjugate fugacities (the same statement applies for the simple puncture wave functions). We will discuss some implications of this fact in the context of the isolated theories that emerge from cusps in the $\CT_{0,n,0}^{(n)}$ gauge coupling space.\footnote{In fact, some of these implications apply to the gauged theories as well. However, the 3D mirror analysis is more complicated in this case.}

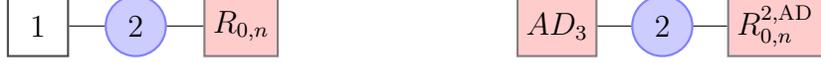
\begin{figure}
 \begin{center}
\begin{tikzpicture}[place/.style={circle,draw=blue!50,fill=blue!20,thick,inner sep=0pt,minimum size=8mm},transition2/.style={rectangle,draw=black!50,fill=red!20,thick,inner sep=0pt,minimum size=8mm},transition3/.style={rectangle,draw=black!70,thick,inner sep=0pt,minimum size=8mm},auto]

\node[place] (1) at (0,0) {$2$};
\node[transition2] at (1.4,0) {\;$R_{0,n}$\;} edge (1);
\node[transition3] at (-1.3,0) {\;$1$\;} edge (1);

\node[place] (2) at (7,0) {$2$};
\node[transition2] at (8.5,0) {\;$R^{2,\text{AD}}_{0,n}$\;} edge (2);
\node[transition2] at (5.6,0) {\;$\AD_3$\;} edge (2);

  \end{tikzpicture}
\vskip.3cm
\caption{S-dual descriptions for $\mathcal{L}^{(n)}_{0,n,0}$ (left) and $\CT^{(n)}_{0,n,0}$ (right). In the left quiver, an $SU(2)$ gauge group is coupled to a fundamental hypermultiplet and an isolated SCFT called $R_{0,n}$. In the right quiver, an $SU(2)$ gauge group is coupled to $\AD_3$ (playing the role of the hypermultiplet) and an exotic fixture we call  $R_{0,n}^{2,\text{AD}}$ (this latter theory is a type $III$ theory in the nomenclature of \cite{Xie:2012hs}).}
\label{fig:quiver-n0n0-2}
 \end{center}
\end{figure}

\begin{figure}
 \begin{center}
\begin{tikzpicture}[place/.style={circle,draw=blue!50,fill=blue!20,thick,inner sep=0pt,minimum size=8mm},transition2/.style={rectangle,draw=black!50,fill=red!20,thick,inner sep=0pt,minimum size=8mm},transition3/.style={rectangle,draw=black!70,thick,inner sep=0pt,minimum size=8mm},auto]

\draw (0,0) arc (180:25:.8);
\draw (0,0) arc (180:335:.8);

\draw (5,0) arc (0:161:1);
\draw (5,0) arc (0:-161:1);

\draw[bend right=20] (1.2,.4) to (3.5,.4);
\draw[bend left=20] (1.2,-.4) to (3.5,-.4);

\draw[fill] (.7, .4) circle (.04);
\draw[fill] (.7, -.4) circle (.04);

\node at (6-1.8,.5) {\footnotesize $Y_\text{full}$};
\node at (6-1.8,-.5) {\footnotesize $Y_\text{full}$};

  \end{tikzpicture}
\vskip.3cm
\caption{The pants decomposition for the punctured sphere corresponding to the S-dual description of $\mathcal{L}^{(n)}_{0,n,0}$ shown on the left of Fig.~\ref{fig:quiver-n0n0-2}. The left and right spheres correspond to a fundamental hypermultiplet and $R_{0,n}$ respectively, while the middle cylinder corresponds to the $SU(2)$ vector multiplet.}
\label{fig:decomposition1}
 \end{center}
\end{figure}
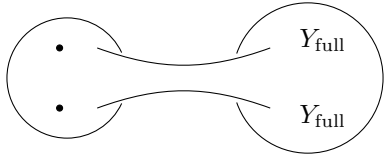

Let us now discuss the different $S$-duality frames of the $\CT_{0,n,0}^{(n)}$ SCFTs and the exotic fixtures that appear at certain cusps in the gauge coupling constant space. In order to proceed, it is useful to first review the corresponding story for the $\mathcal{L}^{(n)}_{0,n,0}$ theories. To that end, recall that the $\mathcal{L}^{(n)}_{0,n,0}$ theory has another S-dual description in terms of the quiver diagram on the left of Fig.~\ref{fig:quiver-n0n0-2}, where the $SU(2)$ gauge group is coupled to a fundamental hypermultiplet and an isolated SCFT / fixture called $R_{0,n}$ \cite{Chacaltana:2010ks}. The flavor symmetry of $R_{0,n}$ is generically $SU(2)\times SU(2n)$, which is enhanced to $E_6$ in the case $n=3$.\footnote{Since its Coulomb branch operators are all of integral dimension, the $R_{0,n}$ theory is {\it not} an AD theory.} The gauge coupling, $\tau'$, of the dual description is related to the coupling, $\tau$, of the original description by $\tau' = \frac{1}{1-\tau}$. In terms of the punctured sphere on the right of Fig.~\ref{fig:quiver3}, this description corresponds to the pants decomposition shown in Fig.~\ref{fig:decomposition1}. This dual description implies that the Schur index of $\mathcal{L}^{(n)}_{0,n,0}$ can also be expressed as
\begin{align}
\CI_{\mathcal{L}_{0,n,0}^{(n)}}(q;({\bf w}_1,c_1),({\bf w}_2,c_2)) &=  \int_{SU(2)}d\mu({\bf z})\,\CI_\text{fund}^{SU(2)}(q;{\bf
 z},s)\CI_\text{vec}^{SU(2)}(q;{\bf
 z})\CI_{R_{0,n}}(q;{\bf z},r,{\bf w}_1,{\bf w}_2^*)~,
\label{eq:Ln0n0}
\end{align}
where $({\bf w}_i,c_i)$ are $U(n)$ fugacities as in \eqref{eq:Lnnn1}, ${\bf z}=(z,z^{-1})$ is an $SU(2)$ fugacity,  and $s \equiv (c_1c_2)^{\frac{n}{2}}$ and $r\equiv c_1/c_2$ are $U(1)$ fugacities. The last factor in \eqref{eq:Ln0n0} is the Schur
index of $R_{0,n}$ given by \cite{Gadde:2011ik}
\begin{align}
 \CI_{R_{0,n}}(q;{\bf z},r,{\bf w}_1,{\bf w}_2^*) =
 \sum_{R:\text{ irrep of
 }\mathfrak{su}(n)}\frac{f_R^{Y_2^{(1)}}\!\!(q;{\bf z},r)\,f_R^{Y_\text{full}}(q;{\bf
 w}_1)\,f_R^{Y_\text{full}}(q;{\bf w}_2^*)}{C_R(q)}~,
\label{eq:R0n}
   \end{align}
where only an $SU(2)\times U(1) \times SU(n)^2$ subgroup of the flavor symmetry is manifest.

As we discuss in appendix \ref{app:Tn0n0}, the $\CT^{(n)}_{0,n,0}$ theory has a similar S-dual description, which is described by the quiver shown on the right of Fig. \ref{fig:quiver-n0n0-2}. The gauge group is again $SU(2)$, which is now coupled to an $\AD_3$ theory (acting as an AD generalization of hypermultiplets) and a type $III$ AD theory in the language of \cite{Xie:2012hs}. This type $III$ AD theory is labeled by three Young diagrams $Y_1= Y_2= [n-1,n-1,2]$ and $Y_3=[2,\cdots,2,1,1]$ with $2n$ boxes and generically has $SU(2)\times SU(n)$ flavor symmetry (the $n=3$ case has $SU(3)\times SU(2)\times SU(2)$ flavor symmetry). We denote this type $III$ theory by $R^{2,\text{AD}}_{0,n}$ since it can be regarded as an AD counterpart of the $R_{0,n}$ fixture. This quiver description implies that the Schur index of $\CT^{(n)}_{0,n,0}$ can also be expressed as
\begin{align}
 \CI_{\CT^{(n)}_{0,n,0}}(q;{\bf y},c) &=
 \int_{SU(2)}d\mu({\bf z})\, \CI_{\AD_{3}}(q;{\bf
 z},c^{n})\CI_\text{vec}^{SU(2)}(q;{\bf z})
 \CI_{R^{2,\text{AD}}_{0,n}}(q;{\bf z},{\bf y})~,
\label{eq:Tn0n0}
\end{align}
where $\CI_{R^{2,\text{AD}}_{0,n}}$ is the Schur index of the $R^{2,\text{AD}}_{0,n}$ theory. Note that previously this index was obtained only for the special case $n=3$ \cite{Buican:2017fiq}, while here we describe it for all odd $n\ge3$.\footnote{The identification of the flavor $U(1)$ fugacity in $\mathcal{I}_{\AD_3}(q;{\bf z},c^n)$ can be understood as follows. From the quiver description in Fig.~\ref{fig:quiver3}, we see that $\mathcal{T}^{(n)}_{0,n,0}$ has two baryonic Higgs branch operators of dimension $n$. These operators are charge conjugate to each other and contribute $q^{\frac{n}{2}}c^{\pm n}$ to the Schur index. In the dual description shown in Fig.~\ref{fig:quiver-n0n0-2}, these operators are realized as the product of a flavor $SU(3)$ moment map in $\AD_3$ and a Higgs branch operator in $R^{2,\text{AD}}_{0,n}$. Indeed, from the 3D mirror of $R^{2,\text{AD}}_{0,n}$ discussed in appendix \ref{app:monopoles}, we see that $R^{(2,\text{AD})}_{0,n}$ has a Higgs branch operator of dimension $(n-2)$ in the ${\bf 2}\otimes {\bf 1}$ representation of the flavor $SU(2)\times SU(n)$ symmetry (this operator corresponds to a mirror monopole of scaling dimension $(n-2)/2$). Let us denote it by $\mathcal{O}^a$ with $a=1,2$ being the $SU(2)$ index. Let us also denote by $\mathcal{O}_\pm^a$ two flavor $SU(3)$ moment maps in the doublet of $SU(2)\subset SU(3)$, where the subscript stands for the charge under $U(1)\subset SU(3)$. Then we see that $\epsilon_{ab}\mathcal{O}_\pm^a \mathcal{O}^b$ can be identified as the baryonic Higgs branch operators mentioned above. This discussion implies that the flavor $U(1)$ fugacity in $\mathcal{I}_{\AD_3}$ is $c^n$.}

By substituting \eqref{eq:Ln0n0} and \eqref{eq:Tn0n0} into \eqref{eq:connection2} and using the identities \eqref{eq:AD2} and \eqref{eq:rewrite-vec}, we obtain
\begin{align}
0 &= \int_{SU(2)}d\mu({\bf z})\, \CI_\text{fund}(q^2;{\bf
 z},c^n)\,\CI^{SU(2)}_\text{vec}(q^2;{\bf
 z})
\nonumber\\
&\qquad \times \left\{ \CI_{R_{0,n}}(q^2;{\bf z},q,{\bf
 y},{\bf y}^*) -
 P.E.\left[\frac{q}{1-q^2}\left(1-\chi_\text{adj}^{SU(2)}({\bf
 z})\right)\right]\CI_{R_{0,n}^{2,\text{AD}}}(q;{\bf z},{\bf
 y})\right\}~.
\label{eq:integral}
\end{align}
This equation is solved by
\begin{align}
 \CI_{R_{0,n}^{2,\text{AD}}}(q;{\bf z}, {\bf y}) &=
 P.E.\left[\frac{q}{1-q^2}\left(-1+\chi_\text{adj}^{SU(2)}({\bf
 z})\right)\right]\CI_{R_{0,n}}(q^2;{\bf z},q,{\bf y},{\bf
 y}^*)~.
\label{eq:RAD0n-1}
\end{align}
Indeed, there exists an inversion formula \cite{Gadde:2010te} that extracts the integrand of \eqref{eq:integral}, which implies that \eqref{eq:RAD0n-1} is the unique solution to \eqref{eq:integral}. Combining \eqref{eq:RAD0n-1} and \eqref{eq:R0n},
 we obtain the following TQFT expression for the Schur index of $R_{0,n}^{2,\text{AD}}$
\begin{align}
 \CI_{R_{0,n}^{2,\text{AD}}}(q;{\bf z},{\bf y}) &=
 \frac{1}{(zq;q^2)(z^{-1}q;q^2)}\sum_{R:\text{ irrep of
 }\mathfrak{su}(n)}\frac{f_R^{Y_2^{(1)}}(q^2;{\bf z},q)
 f_R^{Y_\text{full}}(q^2;{\bf y}) f_R^{Y_\text{full}}(q^2;{\bf
 y}^*)}{C_R(q^2)}~.
\label{eq:ADR0n}
\end{align}
Note that, even though the flavor $U(1)\subset U(2)$ fugacity $r$ of $f_R^{Y_2^{(1)}}(q^2;{\bf z},r)$ is set to $q$, one can show that \eqref{eq:ADR0n} only has integer and half-integer powers of $q$ as it should. Moreover, one can check that for $n>3$ the index does not have an $\CO(q^{1\over2})$ term and so the theory does not have free hypermultiplets. In appendix \ref{monopoles} we find another proof of this fact by bounding monopole operator dimensions in the 3D mirror.\footnote{Due to the non-trivial quiver topology of the 3D mirror that will be discussed further in the next section, this computation is non-trivial and does not follow directly from the results in \cite{Gaiotto:2008ak}.}

For $n=3$, one can perform a stronger consistency check of the above result. Indeed, the $R_{0,3}^{2,\text{AD}}$ theory was carefully studied in \cite{Buican:2017fiq}, where it was shown that $R_{0,3}^{2,\text{AD}}$ splits into an exotic AD theory called $\CT_X$ and a decoupled half-hypermultiplet in the fundamental representation of the flavor
$SU(2)$.\footnote{In \cite{Buican:2017fiq}, the $R_{0,3}^{2,\text{AD}}$ theory is denoted as $\CT_{3,\frac{3}{2}}$.} The Schur index of the $\CT_X$ SCFT is then
\begin{align}
 \CI_{\CT_X}(q;{\bf z},{\bf y}) &=
 (zq^{\frac{1}{2}};q)(z^{-1}q^{\frac{1}{2}};q)\CI_{R_{0,3}^{2,\text{AD}}}(q;{\bf
 z},{\bf y})~,
\label{eq:TX}
\end{align}
where the first two factors comprise the Schur index of the free matter fields. One can check, order by order in $q$, that \eqref{eq:TX} with \eqref{eq:ADR0n} substituted in is identical to
the following expression for the index of $\CT_X$ obtained in \cite{Buican:2017fiq}:
\begin{align}\label{TXAKM}
 \CI_{\CT_{X}}(q;{\bf z},{\bf y}) &= \sum_{\lambda =
 0}^\infty
 q^{\frac{3}{2}\lambda}\,P.E.\left[\frac{2q^2}{1-q}+2q-2q^{\lambda+1}\right]\mathrm{ch}_{R_\lambda}^{SU(2)}(q;{\bf
 z})\mathrm{ch}_{R_{\lambda,\lambda}}^{SU(3)}(q;{\bf y})~,
\end{align}
where $\mathrm{ch}_{R}^{SU(N)}(q;{\bf x})$ is the character of a representation $R$ of $\widehat{\mathfrak{su}(N)}_{-N}$, and $R_{\lambda}$ and $R_{\lambda,\lambda}$ are the highest weight representations of $\widehat{\mathfrak{su}(2)}_{-2}$ and $\widehat{\mathfrak{su}(3)}_{-3}$ corresponding to the Dynkin labels $(-2-\lambda,
\lambda)$ and $(-3-2\lambda,\lambda,\lambda)$, respectively.

Let us further analyze the two equivalent expressions in \eqref{eq:TX}, with \eqref{eq:ADR0n} substituted in,  and \eqref{TXAKM}. Note that these two expressions have very different origins. Indeed, the expression in \eqref{TXAKM} is written in terms of affine Kac-Moody representations\footnote{This expansion is natural considering that the Schur index is related to the vacuum character of the corresponding 2D chiral algebra under the 4D/2D map of \cite{Beem:2013sza}.} while \eqref{eq:ADR0n} is closely related to the correlator of a TQFT on a sphere with three regular punctures. Moreover, \eqref{TXAKM} takes the form of a sum over a full set of $SU(2)$ representations (with the $SU(3)$ representations restricted in terms of the $SU(2)$ data), while \eqref{eq:ADR0n} takes the form of a sum over a full set of $SU(3)$ representations (here the $SU(2)$ data is fixed in terms of the larger $SU(3)$ data). In spite of these differences, the two formulas both take the form of a product of group theoretical factors of $SU(2)\times SU(3)$. Indeed, the second $SU(3)$ wave function in \eqref{eq:ADR0n} is dependent on the first $SU(3)$ wave function since their fugacities are complex conjugates of each other (therefore, in some sense, both expressions involve restrictions on $SU(3)$ data). In Sec. \ref{Wfn3D}, we will reinterpret this dependence of the wave functions in terms of the topology of the corresponding 3D mirrors of the $R_{0,n}^{2,AD}$ SCFTs.

\subsection{S-duality of $\CT^{(1)}_{1,n,1} = (A_{2n-1},A_{2n-1})$ theory}
Next let us consider the $\CT^{(1)}_{1,n,1}$ theories for positive integer $n\geq 2$. Taking $\ell=n_1=n_2=1$, the TQFT expression \eqref{eq:qYM1} reduces to 
\begin{align}
 \CI_{\CT^{(1)}_{1,n,1}}(q;\vec{a},c,\vec{b}) &=
 \frac{1}{(q;q^2)^{2n-2}}\sum_{R:\text{ irreps of }\mathfrak{su(n)}}
 \frac{\left(\prod_{i=0}^{n-1}f_R^{Y_\text{simple}}(q^2;e_i)\right)\left(\prod_{j=0}^{n-1}f_R^{Y_\text{simple}}(q^2;f_j)\right)}{\big(C_R(q^2)\big)^{2n-2}}~,
\label{eq:TQFT-T1n1}
\end{align}
where $e_i$ and $f_i$ are determined by \eqref{eq:def-ef0} and \eqref{eq:def-efi}.  Note that this index is invariant under the $S_{2n}$ that permutes $e_0,\cdots,e_{n-1}$ and $f_0,\cdots,f_{n-1}$. These permutations are realized by transforming the flavor fugacities as in \eqref{eq:S-gen-1}  and \eqref{eq:S-gen-2}, but now for $i=0,\cdots,n-1$. In particular, the permutation symmetry is \lq\lq accidentally" enhanced in this case from $S_{2(n-1)}$ to $S_{2n}$.

This $S_{2n}$ invariance can be interpreted as reflecting the S-duality invariance of the theories. Indeed, it has been argued in \cite{Buican:2014hfa, Xie:2016uqq} that the $\CT^{(1)}_{1,n,1}$ theories are identical to the so-called $(A_{2n-1},A_{2n-1})$ SCFTs \cite{Cecotti:2010fi}, whose S-duality group acts on the flavor fugacities through $S_{2n}$ \cite{Buican:2017uka}. Our formula \eqref{eq:TQFT-T1n1} clarifies how this $S_{2n}$ acts on the $(2n-1)$ flavor fugacities, $(\vec{a},c,\vec{b})$, of $\CT^{(1)}_{1,n,1}$.

\begin{figure}
 \begin{center}
\begin{tikzpicture}[place/.style={circle,draw=blue!50,fill=blue!20,thick,inner sep=0pt,minimum size=8mm},transition2/.style={rectangle,draw=black!50,fill=red!20,thick,inner sep=0pt,minimum size=8mm},transition3/.style={rectangle,draw=black!70,thick,inner sep=0pt,minimum size=8mm},auto]

\node[transition2] (1) at (0,.0) {\;$T^{2,\text{AD}}_{(m_1,m_2,m_3)}$\;};
\node[place](2) at (0,1.1) {\small\;$m_2$\;} edge (1);
\node[place](3) at (1.85,0) {\small\;$m_3$\;} edge (1);
\node[place](4) at (-1.85,0) {\small\;$m_1$\;} edge (1);
\node[transition2] (5) at (-3.2,0) {\small\;$\CT^{(1)}_{1,m_1}$\;} edge (4);
\node[transition2] (6) at (3.2,0) {\small\;$\CT^{(1)}_{1,m_3}$\;} edge (3);
\node[transition2] (7) at (0,2.2) {\small\;$\CT^{(1)}_{1,m_2}$\;} edge (2);

\node[transition2] (1) at (-9,.0) {\;$T_{(m_1,m_2,m_3)}$\;};
\node[place](2) at (-9,1.1) {\small\;$m_2$\;} edge (1);
\node[place](3) at (-9+1.85,0) {\small\;$m_3$\;} edge (1);
\node[place](4) at (-9-1.85,0) {\small\;$m_1$\;} edge (1);
\node[transition3] (5) at (-9-3.2,0) {\small\;$\mathcal{L}^{(1)}_{1,m_1}$\;} edge (4);
\node[transition3] (6) at (-9+3.2,0) {\small\;$\mathcal{L}^{(1)}_{1,m_3}$\;} edge (3);
\node[transition3] (7) at (-9,2.2) {\small\;$\mathcal{L}^{(1)}_{1,m_2}$\;} edge (2);

  \end{tikzpicture}
\caption{The S-dual descriptions of $\mathcal{L}^{(1)}_{1,n,1}$ 
  and $\CT^{(1)}_{1,n,1}$ corresponding to $(m_1,m_2,m_3)$ such that $m_1+m_2+m_3 = 2n$ and $1<m_i<n$. Here a circle with $m_i$ inside stands for an $SU(m_i)$ gauge group.}
\label{fig:T1n1-quiver1}
 \end{center}
\end{figure}
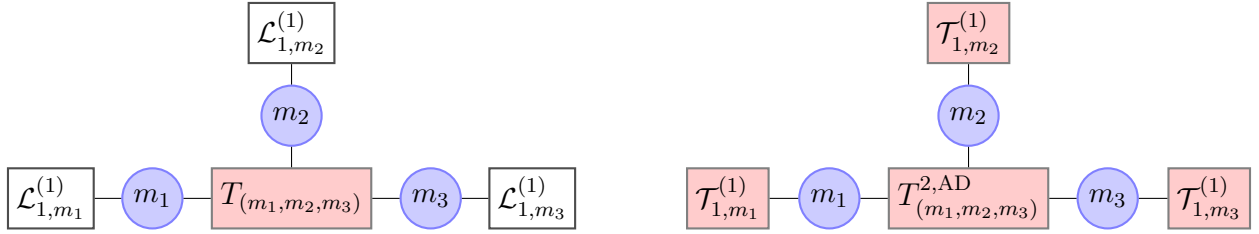

As in the case of $\CT^{(n)}_{0,n,0}$, other $S$-dual descriptions of our theories lead us to expressions for the Schur indices of a series of exotic type $III$ AD fixtures. Indeed, by applying the technique developed in \cite{Xie:2016uqq}, we see that the  $\CT^{(1)}_{1,n,1}$ SCFTs have an S-dual description for each set $(m_1,m_2,m_3)$ of integers such that $2\leq m_i\leq 2n-4$ and $m_1+m_2+m_3=2n$. We focus on the case in which $2\leq m_i<n$ for all $i=1,2,3$ (we will discuss relaxing the condition that $m_i<n$ below). Then this dual description is characterized by the quiver diagram shown on the right of Fig.~\ref{fig:T1n1-quiver1}. The quiver has three tails corresponding to three $\CT^{(1)}_{1,m_i}$ SCFTs, which are connected to the central node by an $SU(m_i)$ gauge group. The central node corresponds to an isolated type $III$ AD theory labeled by three Young diagrams with $n$ boxes $Y_1 = Y_2 = [m_1,m_2,m_3]$ and $Y_3 = [1,\cdots,1]$, which we denote by $T^{2,\text{AD}}_{(m_1,m_2,m_3)}$. The flavor symmetry of $T^{2,\text{AD}}_{(m_1,m_2,m_3)}$ is generically $U(1)^2\times \prod_{i=1}^3 SU(m_i)$. From this S-dual description of $\CT^{(1)}_{1,n,1}$, we see that its Schur index can also be written as
\begin{align}
 \CI_{\CT^{(1)}_{1,n,1}}(q;\vec{a},c,\vec{b}) &=
 \int\left(\prod_{i=1}^3d\mu({\bf
 z}_i)\,\CI_\text{vec}^{SU(m_i)}(q;{\bf z}_i)\,\CI_{\CT^{(1)}_{1,m_i}}(q;\vec{s}_i,{\bf
 z}_i)\right)\CI_{T^{2,\text{AD}}_{(m_1,m_2,m_3)}}(q;{\bf z}_1,{\bf
 z}_2,{\bf z}_3,t_1,t_2)~,
\label{eq:T1n1-2}
\end{align}
where $\vec{s}_i \equiv (s_{i,1},\cdots,s_{i,m_i-1})$, $t_j$ are some functions of $\vec{a},c$ and $\vec{b}$, and the last factor is the Schur index of $T^{2,\text{AD}}_{(m_1,m_2,m_3)}$. This latter index has not been worked out in the literature before.

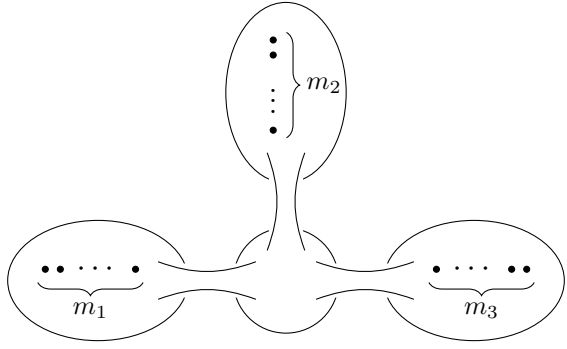
\begin{figure}
 \begin{center}
\begin{tikzpicture}[place/.style={circle,draw=blue!50,fill=blue!20,thick,inner sep=0pt,minimum size=8mm},transition2/.style={rectangle,draw=black!50,fill=red!20,thick,inner sep=0pt,minimum size=8mm},transition3/.style={rectangle,draw=black!70,thick,inner sep=0pt,minimum size=8mm},auto]

\draw (0,0) circle (.7);

\draw (-2.5,0) ellipse (1.2 and .8);
\draw (2.5,0) ellipse (1.2 and .8);
\draw (0,2.5) ellipse (.8 and 1.2);

\draw[fill=white,color=white] (-1.5,.22) rectangle (-.5,-.22);
\draw[fill=white,color=white] (1.5,.22) rectangle (.5,-.22);
\draw[fill=white,color=white] (.22,.5) rectangle (-.22,1.5);

\draw[bend right=20] (-.25,.4) to (-.25,1.7);
\draw[bend left=20] (.25,.4) to (.25,1.7);

\draw[bend right=20] (.4,.25) to (1.7,.25);
\draw[bend left=20] (.4,-.25) to (1.7,-.25);

\draw[bend left=20] (-.4,.25) to (-1.7,.25);
\draw[bend right=20] (-.4,-.25) to (-1.7,-.25);

\draw[fill] (-3.2,.15) circle (.04);
\draw[fill] (-3,.15) circle (.04);
\node at (-2.5,.15) {$\cdots$};
\draw[fill] (-2,.15) circle (.04);

\draw [decorate,decoration={brace,mirror,amplitude=5pt,raise=5pt}]
(-3.3,.15) -- (-1.9,.15) node[midway,below=.3]{\footnotesize $m_1$};

\draw[fill] (3.2,.15) circle (.04);
\draw[fill] (3,.15) circle (.04);
\node at (2.5,.15) {$\cdots$};
\draw[fill] (2,.15) circle (.04);

\draw [decorate,decoration={brace,mirror,amplitude=5pt,raise=5pt}]
(1.9,.15) -- (3.3,.15) node[midway,below=.3]{\footnotesize $m_3$};

\draw[fill] (-.17,3.2) circle (.04);
\draw[fill] (-.17,3) circle (.04);
\node at (-.17,2.5) {$\vdots$};
\draw[fill] (-.17,2) circle (.04);

\draw [decorate,decoration={brace,mirror,amplitude=5pt,raise=5pt}]
(-.17,1.9) -- (-.17,3.3) node[midway,right=.3]{\footnotesize $m_2$};

\end{tikzpicture}
\vskip.3cm
\caption{The decomposition of the punctured sphere corresponding to the $S$-dual description of $\mathcal{L}^{(1)}_{1,n,1}$ shown on the left of Fig.~\ref{fig:T1n1-quiver1}. The $i$-th tail contains $m_i$ (simple) punctures and corresponds to $\mathcal{L}^{(1)}_{1,m_i}$. The three cylinders correspond to $SU(m_i)$ vector multiplets. The middle sphere corresponds to $T_{(m_1,m_2,m_3)}$.}
\label{fig:decomposition2}
 \end{center}
\end{figure}

The Lagrangian counterpart, $\mathcal{L}_{1,n,1}^{(1)}$, has a similar $S$-dual frame described by the quiver diagram on the left of Fig.~\ref{fig:T1n1-quiver1} \cite{Gaiotto:2009we}, where the gauge group is the same but each tail now corresponds to $\mathcal{L}_{1,m_i}^{(1)}$. The central node now stands for the theory obtained by compactifying the 6d (2,0) $A_{n-1}$ theory on a sphere with three regular punctures associated with Young diagrams $Y_{m_1}^{(1)}$ \cite{Gaiotto:2009we, Chacaltana:2010ks}, which we call the $T_{(m_1,m_2,m_3)}$ theory. The flavor symmetry of $T_{(m_1,m_2,m_3)}$ contains $\prod_{i=1}^3 U(m_i)$, and the diagonal $U(1)$ enhances to
$SU(2)$.\footnote{There can be additional enhancements when $n=m_i+1$ for at least one $i$. If this statement holds for all $i$, then we get the usual $E_6$ SCFT (i.e., $T_{(2,2,2)}=T_3$).} This description of $\mathcal{L}^{(1)}_{1,n,1}$ corresponds to a decomposition of the punctured sphere as in Fig.~\ref{fig:decomposition2} (note that the punctures are all simple punctures when $\ell=n_1=n_2=1$). This S-dual description implies that the Schur indices of $\mathcal{L}^{(1)}_{1,n,1}$ have integral expressions similar to \eqref{eq:T1n1-2} but with the indices of $\mathcal{L}_{1,m_i}^{(1)}$ and $T_{(m_1,m_2,m_3)}$ replacing those of $\CT_{1,m_i}^{(1)}$ and $T^{2,\text{AD}}_{(m_1,m_2,m_3)}$. Note that the Schur indices of $T_{(m_1,m_2,m_3)}$ have already been written down in \cite{Gadde:2011ik} as $\sum_{R} (C_R(q))^{-1}\prod_{i=1}^3f_R^{Y_{m_i}^{(1)}}(q;{\bf z}_i,v_i)$, where the sum runs over irreducible representations of $\mathfrak{su}(n)$, and ${\bf z}_i$ and $v_i$ are fugacities for $SU(m_i)\subset U(m_i)$ and $U(1)\subset U(m_i)$, respectively. Using \eqref{eq:connection2}, \eqref{eq:connection1}, and \eqref{eq:rewrite-vec}, one can translate this integral expression for $\CI_{\mathcal{L}_{1,n,1}^{(1)}}$ into the following formula for $\CI_{\CT^{(1)}_{1,n,1}}$:
\begin{align}
 \CI_{\CT^{(1)}_{1,n,1}}(q;\vec{a},c,\vec{b}) &=
 \frac{1}{(q;q^2)}\int\left(\prod_{i=1}^3 d\mu({\bf
 z}_i) \,\CI_\text{vec}^{SU(m_i)}(q;{\bf z}_i)
 \,\CI_{\CT^{(1)}_{1,m_i}}(q;\vec{u}_i,{\bf z}_i)\right)
\nonumber\\
&\qquad \times  P.E.\left[\frac{q}{1-q^2}\sum_{i=1}^3
 \chi_\text{adj}^{SU(m_i)}({\bf z}_i)\right]\sum_{R: \text{ irrep of
 }\mathfrak{su}(n)}\!\!\!\! \frac{\prod_{i=1}^3f_R^{Y_{m_i}^{(1)}}(q^2;{\bf z}_i,q^{\frac{m_i}{2n}}v_i) }{C_R(q^2)}~,
\label{eq:T1n1-3}
\end{align}
where ${\bf z}_i$ is an $SU(m_i)$ fugacity, and $\vec{u}_i \equiv (u_{i,1},\cdots,u_{i,m_i-1})$ and $v_i$ are $U(1)$ fugacities related to $e_k$ and $f_k$ by
\begin{align}
 (u_{i,k})^{k(k+1)} = \frac{(g_{i,k+1})^{kn}}{(g_{i,1}\cdots
 g_{i,k})^n}~,\qquad v_i = q^{-\frac{m_i}{2n}}\prod_{k=1}^{m_i}g_{i,k}~,
\label{eq:st}
\end{align}
with $(g_{1,1},\cdots,g_{1,m_1}, g_{2,1},\cdots,g_{2,m_2},g_{3,1},\cdots,g_{3,m_3}) = (e_0,\cdots,e_{n-1},f_{0},\cdots,f_{n-1})$. From \eqref{eq:def-ef0} and \eqref{eq:def-efi}, we see that $u_{i,k}$ and $v_{i}$ are functions only of flavor fugacities $\vec{a},\vec{b}$ and $c$ and are therefore independent of $q$. Note also that, since $v_1v_2v_3 = 1$,  only two of the $v_i$ are independent.

We now see that the two expressions \eqref{eq:T1n1-2} and \eqref{eq:T1n1-3} are consistent if $\vec{s}_i = \vec{u}_i,\, t_i = v_i$ and the Schur index of $T^{2,\text{AD}}_{(m_1,m_2,m_3)}$ is given by
\begin{align}\label{TADind}
&\CI_{T^{2,\text{AD}}_{(m_1,m_2,m_3)}}(q;{\bf z}_1,{\bf
 z}_2,{\bf z}_3,t_1,t_2) 
\nonumber\\
&\qquad = P.E.\left[\frac{q}{1-q^2}\left(1+\sum_{i=1}^3
 \chi_\text{adj}^{SU(m_i)}({\bf z}_i)\right)\right]
\sum_{R: \text{ irrep of
 }\mathfrak{su}(n)}\!\!\!\!
 \frac{\prod_{i=1}^3f_R^{Y_{m_i}^{(1)}}(q^2;{\bf
 z}_i,q^{\frac{m_i}{2n}}t_i) }{C_R(q^2)}~,
\end{align}
with $t_3 \equiv {1\over t_1t_2}$ (as in the case of the
$R^{2,AD}_{0,n}$ SCFTs, this fugacity dependence will have consequences
for the corresponding 3D mirrors to be discussed in the next
section). While we don't have a full proof that this is the only
expression consistent with \eqref{eq:T1n1-2} and \eqref{eq:T1n1-3}, we
see that it gives a physically meaningful result, since there are only
integer and half-integer powers of $q$ (which is necessary for the
quantity to be a Schur index of an $\CN=2$ SCFT), and it has the
expected $S_3$ symmetry acting on the ${\bf z}_i$ and
$t_i$.

Finally, let us note that the expression in \eqref{eq:T1n1-3} assumes that $m_i<n$ (at least for the corresponding 4D regular puncture theory to make sense). Indeed, for $m_i\ge n$, we would end up with a Young diagram with $m_i$ columns of height one and one column of non-positive height, $n-m_i\le 0$.\footnote{If $m_i>n$ for some $i\in \{1,2,3\}$, the decomposition of the punctured sphere shown in Fig.~\ref{fig:decomposition2} leads to a different S-dual description of $\mathcal{L}^{(1)}_{1,n,1}$ from the one described by the left quiver of Fig.~\ref{fig:T1n1-quiver1}. In particular, the central three-punctured sphere corresponds to a different fixture from $T_{(m_1,m_2,m_3)}$. It would be interesting to find an AD analog of this class $\mathcal{S}$ fixture.}

On the other hand, the expression in \eqref{eq:T1n1-2} may in principle make sense for $m_i\ge n$. It would be interesting to understand if we can analytically continue the expression in \eqref{eq:T1n1-3} to the regime of $m_i\ge n$ and understand the corresponding regular puncture theory, $T_{(m_1,m_2,m_3)}$, as a non-unitary 4D theory (perhaps generalizing the discussion in \cite{Anninos:2011ui,Hertog:2017ymy,Hertog:2019uhy,Vafa:2014iua,Dijkgraaf:2016lym}).

\newsec{Wave function relations and topology of 3D mirrors}\label{Wfn3D}
In this section, we interpret the TQFT formulas \eqref{eq:ADR0n} and \eqref{TADind} for the Schur indices of the $R_{0,n}^{2,\rm AD}$ and $T_{(m_1,m_2,m_3)}^{2,\rm AD}$ SCFTs in terms of the corresponding 3D mirrors given in Fig. \ref{R0nADquiver} and Fig. \ref{quiverTnAD} respectively. In the following subsection, we argue that this discussion implies the existence of RG flows with accidental SUSY enhancement to thirty-two (Poincar\'e plus special) supercharges.

\begin{figure}
\begin{center}
\includegraphics[height=1.5in,width=2.0in,angle=0]{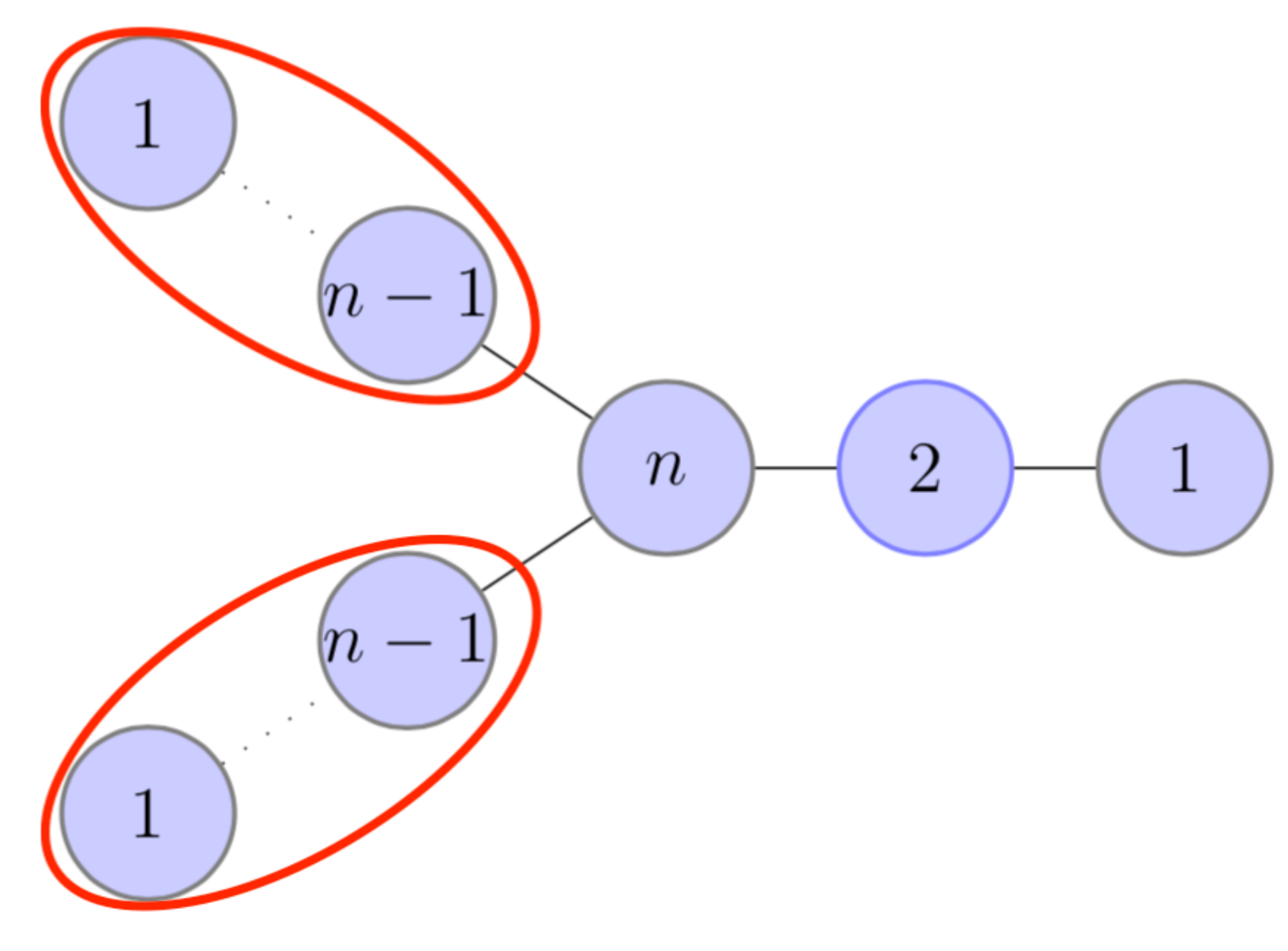}
\caption{The 3D mirror of the $S^1$ reduction of the $R_{0,n}$ SCFT. Nodes labeled by \lq\lq$N$" represent $U(N)$ gauge nodes and lines between nodes denote bifundamental hypermultiplets (an overall decoupled $U(1)$ is removed). The $R_{0,n}$ index has a TQFT expression with three independent wave functions corresponding to the three quiver tails in the above diagram of the 3D mirror. The two quiver tails circled in red generate monopoles which are responsible for the $SU(n)^2\subset SU(2n)\times SU(2)$ flavor symmetry of the theory (the $SU(2)\subset SU(2n)\times SU(2)$ factor comes from the third tail, and the balanced central node is responsible for the $U(1)\times SU(n)^2\to SU(2n)\subset SU(2n)\times SU(2)$ enhancement). When we perform the transformation that takes us from the Schur index of $R_{0,n}$ to that of  $R_{0,n}^{2,\rm AD}$, the two $SU(n)$ tails fuse to form a single $SU(n)$ line of nodes as in Fig. \ref{R0nADquiver}.}
\label{R0nquiver}
\end{center}
\end{figure}

We begin by discussing the TQFT formula for the $R_{0,n}^{2,\rm AD}$ index, which we reproduce below for ease of reference
\begin{align}
 \CI_{R_{0,n}^{2,\text{AD}}}(q;{\bf z}, {\bf y}) &=
 P.E.\left[\frac{q}{1-q^2}\left(-1+\chi_\text{adj}^{SU(2)}({\bf
 z})\right)\right]\CI_{R_{0,n}}(q^2;{\bf z},q,{\bf y},{\bf
 y}^*)~.
\label{RADind2a}
\end{align}
Using the expression for $\CI_{R_{0,n}}$ \eqref{eq:R0n} we then have
\begin{align}
 \CI_{R_{0,n}^{2,\text{AD}}}(q;{\bf z},{\bf y}) &=
 \frac{1}{(zq;q^2)(z^{-1}q;q^2)}\sum_{R:\text{ irrep of
 }\mathfrak{su}(n)}\frac{f_R^{Y_2^{(1)}}(q^2;{\bf z},q)
 f_R^{Y_\text{full}}(q^2;{\bf y}) f_R^{Y_\text{full}}(q^2;{\bf
 y}^*)}{C_R(q^2)}~.
\label{RADind2}
\end{align}
Let us pay special attention to the transformation on the flavor fugacities when we go from the TQFT expression for $R_{0,n}$ to that for $R_{0,n}^{2,AD}$. At the level of flavor symmetries, recall that $R_{0,n}$ has a $G_{R_{0,n}}=SU(2)\times SU(2n)$ flavor symmetry (which is enhanced to $E_6$ for $n=3$) \cite{Chacaltana:2010ks}. On the other hand, $R_{0,n}^{2,AD}$ has flavor symmetry $G_{R_{0,n}^{2,AD}}=SU(2)\times SU(n)$ (which is enhanced to $SU(2)^2\times SU(3)$ for $n=3$).

\begin{figure}
\begin{center}
\includegraphics[height=2.1in,width=2.5in,angle=0]{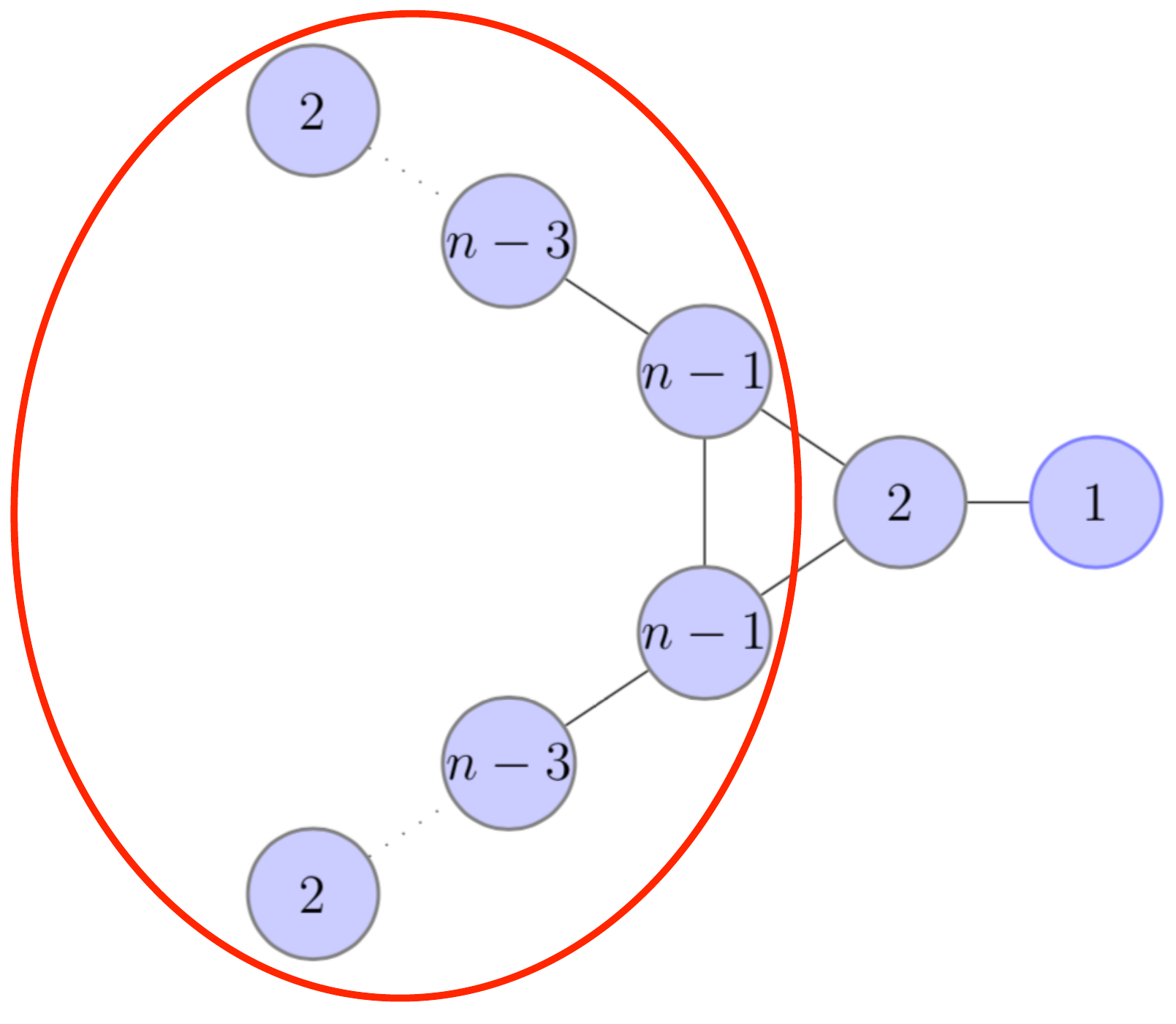}
\caption{The 3D mirror of the $S^1$ reduction of the $R_{0,n}^{2,AD}$ theory. Nodes and lines are defined as in Fig. \ref{R0nquiver}. The two quiver tails corresponding to the TQFT wave functions with conjugate fugacities are fused together to give one linear set of nodes generating an $SU(n)$ symmetry. The corresponding nodes are inside the red oval. The remaining quiver tail gives an $SU(2)$ symmetry and is symmetrically fused with the $SU(n)$ nodes via an unbalanced $SU(2)$ node. As a result, the quiver contains a closed non-abelian loop, and the theory can flow to an interacting $\CN=8$ SCFT via the procedure described in the text.}
\label{R0nADquiver}
\end{center}
\end{figure}

The TQFT expression in \eqref{eq:R0n} makes manifest a $U(2)\times SU(n)^2\subset G_{R_{0,n}}$ flavor subgroup via the wave function with $U(2)$ symmetry, $f_R^{Y^{(1)}_2}(q,{\bf z},r)$, and the two wave functions with $SU(N)$ symmetry, $f_{R}^{Y_{\rm full}}(q,{\bf w_1})$ and $f_{R}^{Y_{\rm full}}(q,{\bf w_2^*})$. These wave functions, and the
flavor symmetries they describe, are related to punctures in the $A_{n-1}$ $(2,0)$ theory. The punctures appear in the 3D mirrors of the $S^1$ compactifications of our 4D theories via the presence of certain quiver tails radiating off a central $SU(n)$ node as in Fig. \ref{R0nquiver} \cite{Chacaltana:2010ks,Benini:2010uu}. In particular, the two tails with gauge groups $U(n-1)\times\cdots \times U(1)$ correspond to punctures described by the $f_{R}^{Y_{\rm full}}$ wave functions, while the tail with gauge group $U(2)\times U(1)$ corresponds to the puncture described by $f_{R}^{Y_2^{(1)}}$. Indeed, by the linear quiver rules given in \cite{Gaiotto:2008ak}, the dimension one monopole operators with fluxes supported on, say, one of the $U(n-1)\times\cdots\times U(1)$ tails give rise to multiplets containing the additional symmetry currents that enhance the corresponding $U(1)^{n-1}$ topological symmetry to $SU(n)$.\footnote{Recall that any 3D $U(n_c)$ gauge group has a corresponding topological symmetry current, $j_{\mu}=\epsilon_{\mu\nu\rho}F^{\nu\rho}$, where $F^{\nu\rho}$ is the field strength corresponding to the trace part of $U(n_c)$. Note that this is a global flavor symmetry acting on the Coulomb branch. In the direct reduction (i.e., the mirror of the mirror quivers we are discussing), the topological symmetry (along with any additional enhanced symmetry via monopole operators) acts on the Higgs branch and descends from the usual 4D flavor symmetry.} This statement follows from the fact that the corresponding line of nodes is \lq\lq balanced," i.e., each $U(n_c)$ node has $n_f=2n_c$ flavors. A similar phenomenon occurs in the other $U(n-1)\times\cdots\times U(1)$ tail and the $U(2)\times U(1)$ tail, thereby giving rise to the $U(2)\times SU(n)^2\subset G_{R_{0,n}}$ non-abelian symmetry (the $U(1)\times SU(n)^2\to SU(2n)$ enhancement occurs because of monopole operators with flux through the central $U(n)$ node).

Given this discussion and the relations between \eqref{eq:R0n} and \eqref{RADind2}, let us give an explanation for the form of the quiver tails for the 3D mirror of $R_{0,n}^{2, AD}$ shown in Fig. \ref{R0nADquiver}. First, note that the two independent $SU(n)$ TQFT $R_{0,n}$ wave functions in \eqref{eq:R0n} are no longer independent in \eqref{RADind2}. Indeed, we must set $w_1=w_2=y$ (in addition to taking $q\to q^2$) and so there is just one independent set of $SU(n)$ fugacities. Since the two $SU(n)$ wave functions are no longer independent, it is natural that in going from Fig. \ref{R0nquiver} to Fig. \ref{R0nADquiver} we should fuse the two previously independent quiver tails into a single tail giving rise to a single $SU(n)$ symmetry.\footnote{In fact, since the wave functions have conjugate fugacities, it is tempting to write $f_R^{Y_{\rm full}}(q^2, y^*)=f_{\bar R}^{Y_{\rm full}}(q^2, y)$, where $\bar R$ is the $SU(n)$ representation conjugate to $R$. We may then write the product of $SU(n)$ wave functions in \eqref{RADind2} as
\begin{equation}
f_R^{Y_{\rm full}}(q^2, {\bf y})f_R^{Y_{\rm full}}(q^2, {\bf y^*})=f_R^{Y_{\rm full}}(q^2, {\bf y})f_{\bar R}^{Y_{\rm full}}(q^2, {\bf y})=\CI_V^{-{1\over2}}(q^2,{\bf y})\sum_{R'\in R\otimes \bar R}f_{R'}^{Y_{\rm full}}(q^2,{\bf y})~,
\end{equation}
where $\CI_V$ is the Schur index of the $SU(n)$ vector multiplet. The appearance of a single wave function suggests that the $SU(n)$ symmetry should be associated with a single line of nodes in the 3D mirror. Moreover, the additional inverse factor of $\CI_V^{-{1\over2}}$ reminds us that this symmetry was associated with two punctures in the original regular puncture theory. At the level of the mirror quiver, this factor reflects the fact that the ranks of the gauge groups in the red oval increase by two between successive nodes in the tails.
} Indeed, note that the line of nodes in the red oval in Fig. \ref{R0nADquiver} have a bifundamental connecting the two previously independent tails and consist of $n-1$ total balanced nodes. By the rules of \cite{Gaiotto:2008ak}, this line of nodes gives rise to the $SU(n)\subset G_{R^{2,AD}_{0,n}}$ symmetry.

Since the two previously independent $SU(n)$ wave functions are now related by complex conjugation of fugacities, non-chirality demands that that their corresponding line of nodes connects to the quiver tail corresponding to the $SU(2)$ wave function in a symmetric fashion. Indeed, the loop of nodes appearing in Fig. \ref{R0nADquiver} is precisely such a symmetric connection. The shortening of the remaining tail reduces the $U(2)$ global symmetry factor to $SU(2)$ and also ensures that the line of nodes generating the $SU(n)$ symmetry are indeed balanced. Note that this loop topology of the $R_{0,n}^{2,AD}$ quiver will be important in arguing for flows to theories with thirty-two (Poincar\'e plus special) supercharges in the next section.

\begin{figure}
\begin{center}
\vskip .5cm
\begin{tikzpicture}[place/.style={circle,draw=blue!50,fill=blue!20,thick,inner sep=0pt,minimum size=10mm},transition2/.style={rectangle,draw=black!50,fill=blue!20,thick,inner sep=0pt,minimum size=10mm},transition3/.style={rectangle,draw=black!50,fill=blue!20,thick,inner sep=0pt,minimum size=10mm},transition4/.style={rectangle,draw=black!50,fill=blue!20,thick,inner sep=0pt,minimum size=10mm},transition5/.style={rectangle,draw=black!50,fill=blue!20,thick,inner sep=0pt,minimum size=10mm},transition6/.style={rectangle,draw=black!50,fill=blue!20,thick,inner sep=0pt,minimum size=10mm},transition7/.style={rectangle,draw=black!50,fill=blue!20,thick,inner sep=0pt,minimum size=10mm},auto]
\node[place] (2) at (-5,0) [shape=circle] {$m_3$} edge [-] node[auto]{} (2);
\node[transition2] (3) at (-3.5,0) [shape=circle]  {$1$} edge [loosely dotted] node[auto]{} (2);
\node[transition3] (n) at (-6.5,0) [shape=circle]  {$n$} edge [-] node[auto]{} (2);
\node[transition4] (n-1) at (-8,1) [shape=circle]  {$m_1$} edge [-] node[auto]{} (n);
\node[transition5] (n-1a) at (-8,-1) [shape=circle]  {$m_2$} edge [-] node[auto]{} (n);
\node[transition6] (1) at (-9.5,-2) [shape=circle]  {$1$} edge [thick, loosely dotted] node[auto]{} (n-1a);
\node[transition7] (1a) at (-9.5,2) [shape=circle]  {$1$} edge [thick, loosely dotted] node[auto]{} (n-1);
\end{tikzpicture}
\caption{The 3D mirror of the $S^1$ reduction of the $T_{(m_1,m_2,m_3)}$ theory. Nodes and lines are defined as in Fig. \ref{R0nquiver}. The three quiver tails correspond to TQFT wave functions carrying $U(m_i)$ global symmetry.}
\label{quiverTn}
\end{center}
\end{figure}
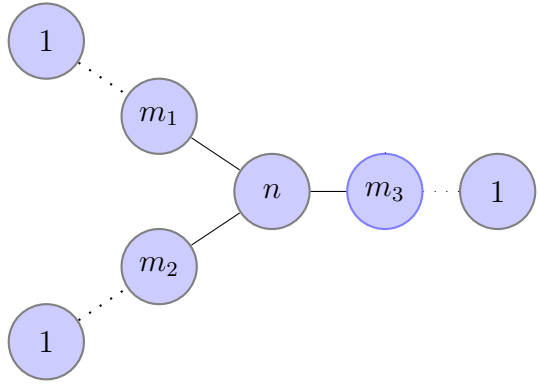

Next let us discuss the case of $T_{(m_1,m_2,m_3)}^{2, AD}$. For ease of reference, we again write the TQFT expression for the Schur indices of these theories originally appearing in \eqref{TADind}
\begin{align}\label{TADind2}
&\CI_{T^{2,\text{AD}}_{(m_1,m_2,m_3)}}(q;{\bf z}_1,{\bf
 z}_2,{\bf z}_3,t_1,t_2) 
\nonumber\\
&\qquad = P.E.\left[\frac{q}{1-q^2}\left(1+\sum_{i=1}^3
 \chi_\text{adj}^{SU(m_i)}({\bf z}_i)\right)\right]
\sum_{R: \text{ irrep of
 }\mathfrak{su}(n)}\!\!\!\!
 \frac{\prod_{i=1}^3f_R^{Y_{m_i}^{(1)}}(q^2;{\bf
 z}_i,q^{\frac{m_i}{2n}}t_i) }{C_R(q^2)}~,
\end{align}
where the $U(1)$ fugacities are constrained to satisfy $t_3={1\over t_1t_2}$ (i.e., there is only a $U(1)^2$ abelian symmetry), and ${\bf z_i}$ are $SU(m_i)$ fugacities. In the case of the $T_{(m_1,m_2,m_3)}$ theory, we have a generic global symmetry group $G_{T_{(m_1,m_2,m_3)}}\supset U(m_1)\times U(m_2)\times U(m_3)$ (the diagonal $U(1)$ enhances to $SU(2)$), while for the $T_{(m_1,m_2,m_3)}^{2,AD}$ theory, we have $G_{T_{(m_1,m_2,m_3)}^{2,AD}}=SU(m_1)\times SU(m_2)\times SU(m_3)\times U(1)^2$.

The correspondence between TQFT wave functions and quiver tails is clear in the case of the 3D mirror of the reduction of $T_{(m_1,m_2,m_3)}$ in Fig. \ref{quiverTn}: each $f_R^{Y_{m_i}^{(1)}}$ wave function corresponds to an independent $U(m_i)\times\cdots\times U(1)$ quiver tail of balanced nodes which, by the rules of \cite{Gaiotto:2008ak} gives rise to monopole operators leading to the $U(1)^{m_i}\to U(m_i)$ flavor enhancement (again, this statement holds assuming generic $m_i$ such that $m_1+m_2+m_3=2n$ and $m_i<n$).

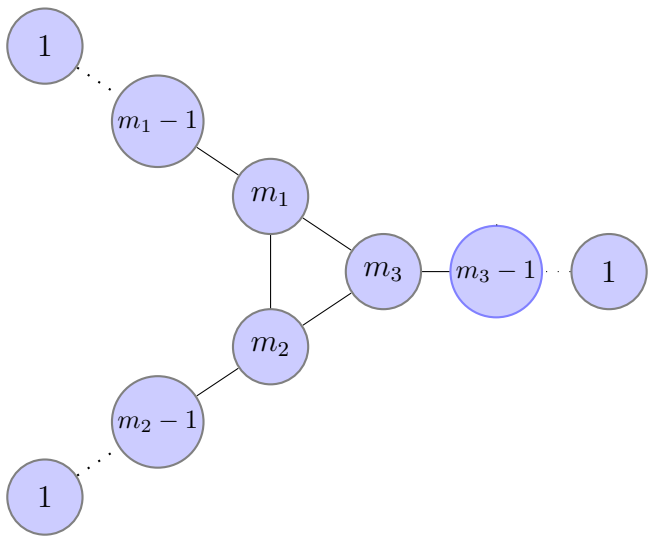
\begin{figure}[h]
\begin{center}
\vskip .5cm
\begin{tikzpicture}[place/.style={circle,draw=blue!50,fill=blue!20,thick,inner sep=0pt,minimum size=10mm},transition2/.style={rectangle,draw=black!50,fill=blue!20,thick,inner sep=0pt,minimum size=10mm},transition3/.style={rectangle,draw=black!50,fill=blue!20,thick,inner sep=0pt,minimum size=10mm},transition4/.style={rectangle,draw=black!50,fill=blue!20,thick,inner sep=0pt,minimum size=10mm},transition5/.style={rectangle,draw=black!50,fill=blue!20,thick,inner sep=0pt,minimum size=10mm},transition6/.style={rectangle,draw=black!50,fill=blue!20,thick,inner sep=0pt,minimum size=10mm},transition7/.style={rectangle,draw=black!50,fill=blue!20,thick,inner sep=0pt,minimum size=10mm},transition8/.style={rectangle,draw=black!50,fill=blue!20,thick,inner sep=0pt,minimum size=10mm},transition9/.style={rectangle,draw=black!50,fill=blue!20,thick,inner sep=0pt,minimum size=10mm},auto]
\draw (-8,1) -- (-8,-1);
\node[place] (2) at (-5,0) [shape=circle] {\footnotesize \,$m_3-1$\,} edge [-] node[auto]{} (2);
\node[transition2] (3) at (-3.5,0) [shape=circle]  {$1$} edge [loosely dotted] node[auto]{} (2);
\node[transition3] (n) at (-6.5,0) [shape=circle]  {$m_3$} edge [-] node[auto]{} (2);
\node[transition4] (n-1) at (-8,1) [shape=circle]  {$m_1$} edge [-] node[auto]{} (n);
\node[transition5] (n-1a) at (-8,-1) [shape=circle]  {$m_2$} edge [-] node[auto]{} (n);
\node[transition6] (1) at (-9.5,-2) [shape=circle]  {\footnotesize \,$m_2-1$\,} edge [-] node[auto]{} (n-1a);
\node[transition7] (1a) at (-9.5,2) [shape=circle]  {\footnotesize \,$m_1-1$\,} edge [-] node[auto]{} (n-1);
\node[transition8] (1b) at (-11,3) [shape=circle]  {$1$} edge [thick, loosely dotted] node[auto]{} (1a);
\node[transition9] (1c) at (-11,-3) [shape=circle]  {$1$} edge [thick, loosely dotted] node[auto]{} (1);
\end{tikzpicture}
\caption{The 3D mirror of the $S^1$ reduction of the $T_{(m_1,m_2,m_3)}^{2,AD}$ SCFT. Nodes and lines are defined as in Fig. \ref{R0nquiver}. The three $U(m_i)\times\cdots\times U(1)$ quiver tails generate independent $SU(m_i)$ fugacities and correspond to the independent $SU(m_i)$ parts of the three TQFT wave functions. On the other hand, the $U(1)$ parts of the three TQFT wave functions are symmetrically dependent. This dependence is reflected in the quiver by the loop of three $U(m_i)$ nodes.}
\label{quiverTnAD}
\end{center}
\end{figure}

On the other hand, the $T_{(m_1,m_2,m_3)}^{2,AD}$ theory no longer has independent wave functions carrying $U(m_i)$ flavor symmetry since the $t_i$ fugacities in \eqref{TADind2} are constrained so that $t_3={1\over t_1t_2}$. Indeed, only the $SU(m_i)$ parts of the wave functions are still independent. We can then give an argument in favor of the 3D mirror of the $S^1$ reduction of $T_n^{2,AD}$ shown in Fig. \ref{quiverTnAD}. The point is that the three balanced $U(m_i-1)\times\cdots U(1)$ tails correspond to the independent $SU(m_i)$ parts of the TQFT wave functions, while the loop of three $U(m_i)$ nodes appears because of the constraint on the $U(1)$ parts of the TQFTs wave functions. Again, this difference in the topology of the AD mirror relative to the $T_{(m_1,m_2,m_3)}$ mirror gives rise to the RG flows to theories with thirty-two supercharges that will be discussed further in the next section.

\newsec{Flows to thirty-two supercharges}\label{flows32}
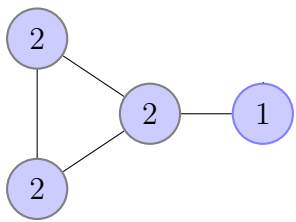
\begin{figure}
\begin{center}
\vskip .5cm
\begin{tikzpicture}[place/.style={circle,draw=blue!50,fill=blue!20,thick,inner sep=0pt,minimum size=8mm},transition2/.style={rectangle,draw=black!50,fill=blue!20,thick,inner sep=0pt,minimum size=8mm},transition3/.style={rectangle,draw=black!50,fill=blue!20,thick,inner sep=0pt,minimum size=8mm},transition4/.style={rectangle,draw=black!50,fill=blue!20,thick,inner sep=0pt,minimum size=8mm},transition5/.style={rectangle,draw=black!50,fill=blue!20,thick,inner sep=0pt,minimum size=8mm},transition6/.style={rectangle,draw=black!50,fill=blue!20,thick,inner sep=0pt,minimum size=8mm},transition7/.style={rectangle,draw=black!50,fill=blue!20,thick,inner sep=0pt,minimum size=8mm},transition8/.style={rectangle,draw=black!50,fill=blue!20,thick,inner sep=0pt,minimum size=8mm},transition9/.style={rectangle,draw=black!50,fill=blue!20,thick,inner sep=0pt,minimum size=8mm},auto]
\draw (-8,1) -- (-8,-1);
\node[place] (2) at (-5,0) [shape=circle] {$1$} edge [-] node[auto]{} (2);
\node[transition3] (n) at (-6.5,0) [shape=circle]  {$2$} edge [-] node[auto]{} (2);
\node[transition4] (n-1) at (-8,1) [shape=circle]  {$2$} edge [-] node[auto]{} (n);
\node[transition5] (n-1a) at (-8,-1) [shape=circle]  {$2$} edge [-] node[auto]{} (n);
\end{tikzpicture}
\caption{The mirror quiver obtained after performing the Coulomb branch flow from Fig. \ref{R0nADquiver} described in the main text. Nodes and lines are defined as in Fig. \ref{R0nquiver}.}
\label{Cflow}
\end{center}
\end{figure}

\begin{figure}
\begin{center}
\vskip .5cm
\begin{tikzpicture}[place/.style={circle,draw=blue!50,fill=blue!20,thick,inner sep=0pt,minimum size=8mm},transition2/.style={circle,draw=black!50,fill=blue!20,thick,inner sep=0pt,minimum size=8mm},auto]
\draw [black] (3,.2) arc [radius=0.5, start angle=20, end angle= 340];
\node[place] (2) at (3,0) [shape=circle] {$2$} edge [-] node[auto]{} (2);
\node[transition2] (3) at (4.3,0)  {$1$} edge [-] node[auto]{} (2);
\end{tikzpicture}
\caption{The mirror quiver obtained after performing the Higgs branch flow from Fig. \ref{Cflow} described in the main text (we drop decoupled hypermultiplets). The circular line is an adjoint hypermultiplet, and the remaining lines and nodes are defined as in Fig. \ref{R0nquiver}. This theory flows to 3D $\CN=8$ in the IR.}
\label{Hflow}
\end{center}
\end{figure}
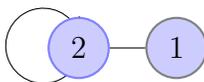

As alluded to in the previous section and also in the introduction, one important characteristic of the isolated AD fixtures we are discussing is that, unlike the regular puncture theories they are related to, the AD theories have RG flows (triggered by vevs and genuinely relevant deformations) with accidental SUSY enhancement to interacting theories with thirty-two (Poincar\'e plus special) supercharges (thereby generalizing the examples in \cite{Buican:2018ddk}). In the next section, we will argue that such flows are in fact generic in the landscape of AD theories with known 3D mirrors. Note that these flows will proceed via reduction to 3D and via flowing onto the moduli spaces of the resulting theories. We briefly discuss the possibility of uplifting this discussion to 4D at the end of this subsection while postponing a more detailed analysis for future work.

To first understand why the RG flows to interacting theories with thirty-two supercharges occur in the $R_{0,n}^{2,AD}$ and $T_{(m_1,m_2,m_3)}^{2,AD}$ theories discussed above, it is sufficient to compactify these theories on $S^1$ and consider the corresponding 3D mirrors. Let us start with the mirror in Fig. \ref{R0nADquiver}. Flowing to generic points on the Coulomb branch of the two lines of nodes with gauge groups $U(2)\times U(4)\times\cdots\times U(n-3)$ and also onto points of the Coulomb branch of $U(n-1)\times U(n-1)$ with symmetry breaking pattern $U(n-1)\times U(n-1)\to U(2)\times U(2)\times U(1)^{2(n-3)}$, we obtain the quiver in Fig. \ref{Cflow}, where we have dropped decoupled $U(1)$ factors. This is the mirror of the lowest rank theory studied in \cite{Buican:2018ddk}, which we know from that reference flows to $\CN=8$ via mass terms in the direct reduction. However, it will be useful for our more general discussion below to analyze a purely moduli space flow to $\CN=8$ in the mirror theory itself.\footnote{The mirror analog of the flow in \cite{Buican:2018ddk} proceeds by turning on Fayet-Iliopoulos terms.} To that end, consider turning on Higgs branch vevs
\begin{equation}\label{Higgsvev}
\langle Q_1\tilde Q_1\rangle=\langle Q_2\tilde Q_2\rangle=\langle Q_3\tilde Q_3\rangle\ne0~,
\end{equation}
where the $Q_i, \tilde Q_i$ pairs correspond to the three edges in the loop of Fig. \ref{Cflow} so that we break $U(2)^3\to U(2)_{\rm diag}$ leaving the quiver in Fig. \ref{Hflow} after dropping decoupled fields.\footnote{In the direct reduction, this maneuver corresponds to turning on vevs for the overall $U(1)\subset U(2)$ vector multiplet primary.} In terms of the squark fields, we may imagine turning on vevs 
\begin{equation}\label{Sqvevs}
\langle Q_i\rangle=\langle\tilde Q_i\rangle=v\mathds{1}_{2\times2}\ne0~, 
\end{equation}
for $i=1,2,3$. This latter theory flows directly to $\CN=8$ in the IR. Therefore, we see that through a combination of Coulomb and Higgs branch flows in the mirror theory, we flow to an interacting 3D $\CN=8$ SCFT.

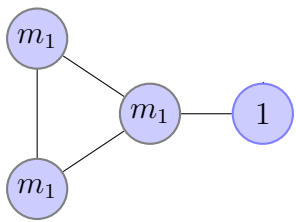
\begin{figure}
\begin{center}
\vskip .5cm
\begin{tikzpicture}[place/.style={circle,draw=blue!50,fill=blue!20,thick,inner sep=0pt,minimum size=8mm},transition2/.style={rectangle,draw=black!50,fill=blue!20,thick,inner sep=0pt,minimum size=8mm},transition3/.style={rectangle,draw=black!50,fill=blue!20,thick,inner sep=0pt,minimum size=8mm},transition4/.style={rectangle,draw=black!50,fill=blue!20,thick,inner sep=0pt,minimum size=8mm},transition5/.style={rectangle,draw=black!50,fill=blue!20,thick,inner sep=0pt,minimum size=8mm},transition6/.style={rectangle,draw=black!50,fill=blue!20,thick,inner sep=0pt,minimum size=8mm},transition7/.style={rectangle,draw=black!50,fill=blue!20,thick,inner sep=0pt,minimum size=8mm},transition8/.style={rectangle,draw=black!50,fill=blue!20,thick,inner sep=0pt,minimum size=8mm},transition9/.style={rectangle,draw=black!50,fill=blue!20,thick,inner sep=0pt,minimum size=8mm},auto]
\draw (-8,1) -- (-8,-1);
\node[place] (2) at (-5,0) [shape=circle] {$1$} edge [-] node[auto]{} (2);
\node[transition3] (n) at (-6.5,0) [shape=circle]  {$m_1$} edge [-] node[auto]{} (2);
\node[transition4] (n-1) at (-8,1) [shape=circle]  {$m_1$} edge [-] node[auto]{} (n);
\node[transition5] (n-1a) at (-8,-1) [shape=circle]  {$m_1$} edge [-] node[auto]{} (n);
\end{tikzpicture}
\caption{The mirror quiver obtained after performing the Coulomb branch flow from Fig. \ref{quiverTnAD} described in the main text. Nodes and lines are defined as in Fig. \ref{R0nquiver}.}
\label{Cflow2}
\end{center}
\end{figure}

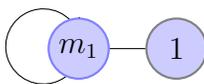
\begin{figure}
\begin{center}
\vskip .5cm
\begin{tikzpicture}[place/.style={circle,draw=blue!50,fill=blue!20,thick,inner sep=0pt,minimum size=8mm},transition2/.style={circle,draw=black!50,fill=blue!20,thick,inner sep=0pt,minimum size=8mm},auto]
\draw [black] (3,.2) arc [radius=0.5, start angle=20, end angle= 340];
\node[place] (2) at (3,0) [shape=circle] {$m_1$} edge [-] node[auto]{} (2);
\node[transition2] (3) at (4.3,0)  {$1$} edge [-] node[auto]{} (2);
\end{tikzpicture}
\caption{The mirror quiver obtained after performing the Higgs branch flow from Fig. \ref{Cflow2} described in the main text (we drop decoupled hypermultiplets). The circular line is an adjoint hypermultiplet, and the remaining lines and nodes are defined as in Fig. \ref{R0nquiver}. This theory flows to 3D $\CN=8$ in the IR.}
\label{Hflow2}
\end{center}
\end{figure}

Now consider the $T_{(m_1,m_2,m_3)}^{2,AD}$ SCFTs. Since the $T_{(m_1,m_2,m_3)}^{2,AD}$ mirror clearly flows to the $R_{0,n}^{2,AD}$ mirror via excursions on the Coulomb branch, we can appeal to the above discussion and conclude that there are flows from $T_{(m_1,m_2,m_3)}^{2,AD}$ to theories with thirty-two supercharges. On the other hand, the $T_{(m_1,m_2,m_3)}^{2,AD}$ theories also admit other flows to a richer set of $\CN=8$ theories, which we now describe.

Without loss of generailty, we will assume that $m_3\ge m_2\ge m_1$. Now, let us flow to points on the Coulomb branches of the two quiver tails of length $m_i$ with $i=3,2$ such that the corresponding gauge groups break as $U(m_i)\times U(m_i-1)\times\cdots\times U(1)\to U(m_1)\times U(1)^{m_i-m_1}\times U(1)^{m_i-1}\times\cdots\times U(1)$ (where the ellipses on the RHS of the breaking contain only abelian gauge groups). Simultaneously, we flow to a generic point on a Coulomb sub-branch in the third tail specified by $SU(m_1-1)\times U(m_1-2)\times\cdots\times U(1)$ and obtain the theory in Fig. \ref{Cflow2}. We can then turn on vevs as in \eqref{Higgsvev} where the $Q_i, \tilde Q_i$ pairs are now bifundamentals of $U(m_1)\times U(m_1)$.\footnote{In analogy with the previous case, we turn on vevs $\langle Q_i\rangle=\langle\tilde Q_i\rangle=v\mathds{1}_{m_1\times m_1}\ne0$ for all $i=1,2,3$.}  This procedure produces the interacting $\CN=8$ theory described in Fig. \ref{Hflow2}.

Note that in all RG flows described in this subsection, we flow on both the Coulomb and Higgs branches of the 3D mirror. Therefore, by mirror symmetry, in order to reach the $\CN=8$ fixed points, we flow on both the Higgs and Coulomb branches of the direct $S^1$ reductions of our 4D SCFTs. It would be interesting to understand if these flows uplift to 4D flows along the Higgs and Coulomb branches of our 4D theories (i.e., if the corresponding 4D RG flows commute with the $S^1$ reduction as in \cite{Buican:2018ddk}).

If the flows do uplift, then it would also be interesting to understand if the 3D $\CN=8$ fixed points map to $\CN=4$ theories in 4D. In principle, if the flows are well behaved enough, then the detailed properties of these possible $\CN=4$ fixed points---e.g., if they are of Super-Yang Mills (SYM) type or not---can be studied.

\subsec{Universality of flows to interacting SCFTs with thirty-two supercharges}
In this section, we briefly state and prove a theorem governing how universally we may expect the existence of RG flows to interacting theories with thirty-two (Poincar\'e plus special) supercharges. This discussion is motivated by our TQFT formulae for the Schur indices of the $R^{2,AD}_{0,n}$ and $T_{(m_1,m_2,m_3)}^{2,AD}$ theories and our reinterpretation of these formulae as leading to closed loops of non-abelian nodes in the corresponding 3D mirrors. Indeed, we saw that the existence of such closed loops generically led to RG flows ending on interacting SCFTs with thirty-two supercharges.

Combined with the infinite class of examples in \cite{Buican:2018ddk}, it is then tempting to wonder whether such flows are generic in the class of (untwisted) type $III$ theories (and therefore, perhaps, in the space of $\CN=2$ theories coming from compactifications of the $(2,0)$ theory on surfaces with untwisted punctures). In fact, it is straightforward to show this is the case, if we assume the classification of such theories given in \cite{Xie:2017vaf,Xie:2017aqx}. In this classification, the space of type $III$ theories is specified by $N\ge2$ Young diagrams (the theories discussed above have $N=3$). The $N=2$ theories cannot flow to theories with thirty-two supercharges (we do not consider turning on additional gauge couplings in the UV), and so we focus on the more generic theories with $N\ge3$.\footnote{Interestingly if one adds a regular singularity one finds, among the $N=2$ theories, 3D mirrors equivalent to the star-shaped quivers found in the case of some theories with regular punctures (and no irregular punctures).} The Young diagrams in question take the form \cite{Xie:2017vaf,Xie:2017aqx}
\begin{equation}\label{ydiag}
Y_1=Y_{2}=\cdots=Y_{N-1}=[h_1,h_2,\cdots,h_p]~, \ \ \ Y_N=[a_{1,1},\cdots,a_{1,n_1},a_{2,1}\cdots,a_{2,n_2}\cdots a_{p,n_p}]~,
\end{equation}
where the column heights $h_i$ and $a_{i,b}$ are non-decreasing (from left to right) positive integers satisfying
\begin{equation}
\sum_{b=1}^{n_b}a_{i,b}=h_i~.
\end{equation}
The above Young diagrams correspond to the degeneracy of the eigenvalues of the singular terms in the Higgs field one obtains in the Hitchin system describing the type $III$ compactification \cite{Xie:2012hs} (although note that in our conventions $Y_1$ corresponds to the {\it most} singular piece). At the level of the 3D mirror, the quiver consists of a core with gauge group
\begin{equation}\label{coreGp}
G=U(h_1)\times U(h_2)\times\cdots\times U(h_p)~,
\end{equation}
and $N-2$ bifundamentals between each node.\footnote{In the case of the $R_{0,n}^{2,AD}$ and $T_{(m_1,m_2,m_3)}^{2,AD}$ theories, the cores are the triangular loops in Fig. \ref{R0nADquiver} and Fig. \ref{quiverTnAD} respectively.} The final Young diagram, $Y_N$, describes the quiver tails. For example, if the column of height $h_b$ is broken up into $[a_{b,1},\cdots,a_{b, n_b}]$, we attach a tail to $U(h_b)$ with gauge group
\begin{equation}
G_{b}^{\rm tail}=U(h_b-a_{b,1})\times U(h_b-a_{b,1}-a_{b,2})\times\cdots\times U(h_b-a_{b,1}-\cdots-a_{b,n_{b-1}})~,
\end{equation}
and bifundamentals between each corresponding node (and also a single bifundamental between the $U(h_b-a_{b,1})$ and $U(h_b)$ node). One repeats this procedure for all $b\in\left\{1,\cdots,p\right\}$. Given this setup and assumptions, we can prove the following theorem on the universality of non-perturbative flows from sixteen to thirty-two supercharges:

\bigskip
\noindent
{\bf Theorem:} If the quantities $h_3$ and $n_1$ in \eqref{ydiag} satisfy $h_3, n_1>1$, the corresponding type $III$ SCFT flows, up to free decoupled factors, to an interacting theory with thirty-two (Poincar\'e plus special) supercharges upon compactification to 3D, flowing to certain points on the moduli space of the theory, and, for $N>3$, turning on mass terms in the 3D mirror.\footnote{The same caveats described at the end of the previous section apply in lifting these flows to 4D.}

\bigskip
\noindent
{\bf Proof:} We would like to reduce the 3D mirror to the diagram in Fig. \ref{Cflow2} with $m_1=h_3>1$. To accomplish this task, we can first move along the Coulomb branch to reduce our theory to a diagram similar to the one in Fig. \ref{Cflow2}, but containing $N-2$ bifundamentals between each node. To get to this diagram, first go to generic points on the Coulomb branches of the subset of the core nodes (see \eqref{coreGp}) characterized by $U(h_4)\times\cdots\times U(h_p)\subset G$ and to generic points on the Coulomb branches of all their tails (if any exist). Next, we go to generic points on the Coulomb branches of the tails of the $U(h_2)\times U(h_3)$ nodes to remove them as well. Then, we go to generic points on the Coulomb branch of the $U(h_1-a_{1,1}-1)\times\cdots\times U(h_1-a_{1,1}-\cdots-a_{1,n_1})$ part of the $U(h_1)$ quiver tail. This procedure leaves us (up to decoupled $U(1)$ factors, which we drop) with a $U(h_1)\times U(h_2)\times U(h_3)$ group of core nodes connected by $N-2$ bifundamentals between each node and a $U(1)$ node connected to $U(h_1)$ via a fundamental. To proceed, we now go to a point on the $U(h_1)\times U(h_2)$ Coulomb branch that breaks the gauge symmetry as $U(h_1)\times U(h_2)\to U(h_3)^2\times U(1)^{h_1-h_3}\times U(1)^{h_2-h_3}$. Up to decoupled $U(1)$'s, we have a diagram equivalent to that in Fig. \ref{Cflow2} with $m_1=h_3$ except for the fact that there are $N-2$ bifundamentals between each non-abelian node. We may add mass terms to remove $N-3$ of the bifundamentals between each node to end up with a diagram identical to the one in Fig. \ref{Cflow2}. Combined with the Higgs branch flow described below Fig. \ref{Cflow2}, we flow to an interacting $\CN=8$ theory. Therefore, if we are willing to go on the Coulomb and Higgs branches of the 3D mirror and, at the same time, add mass terms for some of the bifundamentals between the remaining non-abelian nodes, we flow to a theory with thirty-two supercharges.\footnote{Note that adding a regular singularity to the above set of theories does not change the above proof: we can decouple the additional nodes associated with this singularity via flowing to generic points on the corresponding Coulomb branches.} {\bf q.e.d.}

\newsec{Conclusions}
In this paper we found various new relations between theories with non-integer scaling dimension $\CN=2$ chiral operators (i.e., AD theories) and those with purely integer dimensional $\CN=2$ chiral operators (the regular puncture class $\CS$ theories). The latter theories have TQFT index expressions that are typically simpler (and more uniformly presented) than those of the former. The additional complication in the TQFT expressions for the case of AD theories (e.g., see \cite{Buican:2017uka,Song:2015wta}) is related to the fact that the corresponding singularities in the compactification from 6D to 4D generally contain more data. However, we saw that we can, in some sense, encode this additional data by taking TQFT data for regular puncture theories (which only have integer dimension $\CN=2$ chiral operators) and demanding interdependence of the different TQFT wave functions through intricate fugacity relations. This fugacity interdependence has important physical consequences: a large class of AD theories flow to interacting IR SCFTs with thirty-two (Poincar\'e plus special) supercharges via flows of the type discussed in Sec. \ref{flows32}. Using these index relations, we also found expressions for the Schur indices of various classes of exotic type $III$ AD theories.

Clearly, there is a lot more to be said. We conclude with some open problems (and potential solutions):
\begin{itemize}
\item{It would be interesting to understand if the RG flows we discussed above can be lifted to 4D (for some flows, we know this is the case; e.g., see \cite{Buican:2018ddk}). If so, then it would be particularly intriguing to try to compute the indices of some of the resulting IR theories and see if they are $\CN=4$ theories or not. If they are $\CN=4$ theories, then it would be interesting to understand if they are Lagrangian (SYM theories) or not.}
\item{One way to address the above point would be to try to construct better-behaved RG flows in the class described in Sec. \ref{flows32}. This might involve better understanding the role that monopole operators can play in the corresponding mirror RG flows. Alternatively, this might involve a better understanding of non-abelian mirror symmetry.}
\item{Another approach to the problem in the first bullet point might be as follows. The authors of \cite{Razamat:2019vfd} find $\CN=1$ Lagrangians for certain class $\CS$ regular puncture theories by considering excursions along $\CN=1$ conformal manifolds that include these $\CN=2$ SCFTs as special points. In their discussion, the authors find $\CN=1$ Lagrangians on certain conformal manifolds containing $\CN=2$ SCFTs that have both dimension three Higgs branch and dimension three Coulomb branch operators. Some of the theories discussed in the present article satisfy this condition. Moreover, given the similarity of the Schur indices of our theories to those in the regular puncture class $\CS$ case, it would be interesting to see if one can find $\CN=1$ Lagrangians for some of the $R_{0,n}^{2,AD}$ and $\CT_{(m_1,m_2,m_3)}^{2,AD}$ theories in this manner. Having an $\CN=1$ Lagrangian or, at the very least, an $\CN=1$ conformal manifold might in turn make it easier to study flows to $\CN=4$.}
\item{The ubiquity of RG flows to interacting theories with thirty-two supercharges emanating from compactifications of the 6D $(2,0)$ theory on Riemann surfaces with irregular punctures strongly suggests the existence of another way of understanding these theories via D3 branes probing type IIB / F-theory backgrounds far beyond what has been explored in the literature.}
\item{It would be interesting to understand the most general class of $\CN=2$ SCFTs with non-integer dimensional $\CN=2$ chiral operators (i.e., Coulomb branch operators) that are involved in RG flows with SUSY enhancement either as UV or IR end points.}
\item{We had to rescale fugacities as $q\to q^2$ in order to find a match between the indices of the AD theories and those of the regular puncture theories. In the process, we had to consider going from the $A_{n-1}$ to the $A_{2n-1}$ 6D $(2,0)$ parent theories. It would be interesting to understand why this is the case and also to see if more general $q\to q^m$ rescalings are meaningful.}
\item{Finally, we saw that there is a close relation between regular puncture class $\CS$ fixtures and our AD fixtures. It would be interesting to understand if to each class $\CS$ fixture there exists an AD counterpart and, if so, how many such counterparts exist. In addition, we saw that in our class of theories, the AD fixtures with interacting regular puncture relatives admitted RG flows to interacting thirty-two supercharge theories. On the other hand, AD fixtures with free class $\CS$ relatives did not admit such flows (even though the corresponding AD theories are strongly interacting). It would be interesting to understand if this story is completely general in the space of theories of class $\CS$.}
\end{itemize}

\medskip
\ack{It is a pleasure to acknowledge P.~Argyres, T.~Creutzig, S.~Giacomelli, Z.~Laczko, and M.~Martone for interesting discussions and correspondence. We also thank Z.~Laczko for collaboration on closely related topics. M.~B. thanks the Centro de Ciencias de Benasque Pedro Pascual and organizers for a wonderful working environment during the excellent workshop \lq\lq Gauge theories, supergravity, and superstrings." T.~N. is grateful to the organizers of the conferences ``Infinite dimensional algebras, geometry and integrable systems'' at RIMS and ``Vertex algebras, factorization algebras and applications'' at Kavli IPMU where he had many useful discussions. M.~B.'s research is partially supported by the Royal Society under the grant, \lq\lq New Constraints and Phenomena in Quantum Field Theory." T.~N.'s research is partially supported by JSPS Grant-in-Aid for Early-Career Scientists 18K13547.}

\newpage
\begin{appendices}

\section{Useful identities}

In this appendix, we derive useful identities for index contributions from a vector multiplet and a bifundamental hypermultiplet. The index contribution from an $SU(n)$ vector multiplet is given by
\begin{align}
 \mathcal{I}^{SU(N)}_\text{vec}(q;{\bf z}) &=
 P.E.\left[-\frac{2q}{1-q}\chi_\text{adj}^{SU(N)}({\bf z})\right]~.
\end{align}
Using  $q/(1-q) = q/(1-q^2) + q^2/(1-q^2)$, we find the
following identity
\begin{align}
 \CI^{SU(N)}_\text{vec}(q;{\bf z}) =
 \CI^{SU(N)}_\text{vec}(q^2;{\bf z}) \times
 P.E.\left[-\frac{2q}{1-q^2}\chi_\text{adj}^{SU(N)}({\bf z})\right]~.
\label{eq:rewrite-vec}
\end{align}
Similarly, for the Schur index of a bifundamental hypermultiplet of
$SU(N)\times SU(M)$
\begin{align}
 \CI_\text{bifund}^{N \times M}(q;{\bf y},{\bf z},a) &=
 P.E.\left[\frac{q^{\frac{1}{2}}}{1-q}\left(a
 \chi_\text{fund}^{SU(N)}({\bf y})\chi_\text{afund}^{SU(M)}({\bf z}) +
 a^{-1}\chi_\text{afund}^{SU(N)}({\bf y}) \chi_\text{fund}^{SU(M)}({\bf z})\right)\right]~,
\end{align}
we can show the identity
\begin{align}
 \CI_\text{bifund}^{N \times M}(q;{\bf y},{\bf z},a) &=
 \CI_\text{bifund}^{N \times M}(q^2;{\bf y},{\bf
 z},aq^{\frac{1}{2}})\CI_\text{bifund}^{N \times M}(q^2;{\bf
 y},{\bf z}, aq^{-\frac{1}{2}})~,
\label{eq:rewrite-bfund}
\end{align}
using $q^{\frac{1}{2}}a^{\pm1}/(1-q) = q(q^{-\frac{1}{2}}+q^{\frac{1}{2}})a^{\pm 1}/(1-q^2)$.

\section{$\mathrm{III}_{2\times [n-1,n-1,2],[2,\cdots,2,1,1]}$ theory}
\label{app:Tn0n0}

In this appendix, we argue that theory described by the right quiver in Fig.~\ref{fig:quiver-n0n0-2} is equivalent to $\mathcal{T}^{(n)}_{0,n,0}$. To that end, first note that the former theory is equivalent to  the type $III$ AD theory associated with three Young diagrams, $Y_1=Y_2=[n-1,n-1,1,1]$ and $Y_3=[2,\cdots,2,1,1]$, in the language of \cite{Xie:2012hs}. Indeed, the prescription of \cite{Xie:2016uqq} suggests that this type $III$ theory has a weak coupling description corresponding to the splitting of $2n$ boxes in $Y_1 = [n-1,n-1,1,1]$ into the two groups, $[1,1]$ and $[n-1,n-1]$.\footnote{Here, the idea of \cite{Xie:2016uqq} is that there exists an S-dual frame  for each splitting of boxes in $Y_1$ into two groups.} From the 3d mirror analysis, we see that the sector corresponding to $[1,1]$ is $D_2(SU(3)) = \AD_3$, the one corresponding to $[n-1,n-1]$ is $R_{0,n}^{2,\text{AD}}$, and an $SU(2)$ vector multiplet is coupled to them.\footnote{Recall that $R_{0,n}^{2,\text{AD}}$ is the type $III$ AD theory associated with $Y_1 = Y_2 = [n-1,n-1,2]$ and $Y_3= [2,\cdots,2,1,1]$.} Therefore, all we have to show here is that this type $III$ AD theory is equivalent to $\mathcal{T}^{(n)}_{0,n,0}$.

\begin{figure}
 \begin{center}
\begin{tikzpicture}[place/.style={circle,draw=blue!50,fill=blue!20,thick,inner sep=0pt,minimum size=8mm},transition2/.style={rectangle,draw=black!50,fill=red!20,thick,inner sep=0pt,minimum size=8mm},transition3/.style={rectangle,draw=black!70,thick,inner sep=0pt,minimum size=8mm},auto]

 \node[place] (1) at (0,0) [shape=circle] {\footnotesize\;$n$\;};
 \node[transition2] (2) at (2.7,0) {\;$\mathrm{IV}_{2\times [n-1,1],[2,\cdots,2,1]}^\text{full}$\;} edge (1); 
 \node[transition2] (3) at (-2.7,0) {\;$\mathrm{IV}_{2\times
 [n-1,1],[2,\cdots,2,1]}^\text{full}$\;} edge (1); 
\node[transition3] at (0,1.2) {\;$1$\;} edge (1);

  \end{tikzpicture}
\vskip .5cm
\caption{Another weak coupling description of the type $III$ AD theory associated with the Young diagrams $Y_1 = Y_2 = [n-1,n-1,2]$ and $Y_3 = [2,\cdots,2,1,1]$. The left and right boxes each stand for one copy of the type $IV$ AD theory described in the main text, while the top box stands for a fundamental hypermultiplet. We argue that this quiver theory is identical to $\mathcal{T}^{(n)}_{0,n,0}$.}
\label{fig:IVIV}
 \end{center}
\end{figure}
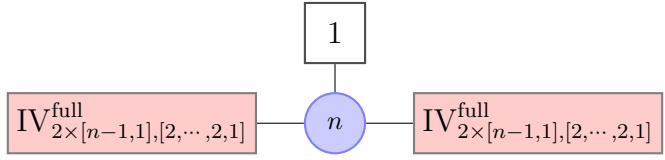

To see the equivalence of the above-mentioned type $III$ theory and $\mathcal{T}^{(n)}_{0,n,0}$, let us consider a weak coupling description of the type $III$ theory corresponding to the splitting of $2n$ boxes in $Y_1$ into $[n-1,1]$ and $[n-1,1]$. From the prescription of \cite{Xie:2016uqq} and the spectrum of $\mathcal{N}=2$ chiral operators, we see that the sector corresponding to each $[n-1,1]$ is the type $IV$ AD theory (in the language of \cite{Xie:2012hs}) associated with an irregular puncture labeled by three Young diagrams $Y_1= Y_2 = [n-1,1]$ and $Y_3=[2,\cdots,2,1]$, and a full (and therefore regular) puncture.\footnote{A type $IV$ theory is obtained by compactifying the 6d (2,0) $A_{n-1}$ theory on sphere with an irregular puncture and a regular puncture. These punctures are characterized by the singularity of an $\mathfrak{sl}(n)$-valued meromorphic $(1,0)$-form, $\varphi$, around them. Suppose that a regular puncture is at $z=0$. Then $\varphi$ behaves near $z=0$ as $\varphi \sim (\frac{M}{z} + \cdots)dz$ with $M\in \mathfrak{sl}(n)$, up to conjugation. When the regular puncture is a full puncture, the eigenvalues of $M$ are all different. When an irregular puncture associated with $Y_1,Y_2$ and $Y_3$ are at $z=0$, $\varphi$ behaves as $\varphi \sim  \left(\frac{M_1}{z^3}+\frac{M_2}{z^2}+\frac{M_3}{z} + \cdots\right)dz$ up to conjugation, where $M_1,M_2,M_3\in \mathfrak{sl}(n)$ and the eigenvalues of  $M_i$ are such that the ordered list of the numbers of equal eigenvalues is identical to $Y_i$.}
We also see that an $SU(n)$ vector multiplet is coupled to these type $IV$ AD theories as well as an extra fundamental hypermultiplet. Therefore, this weak coupling description corresponds to the quiver diagram in Fig.~\ref{fig:IVIV}.

Hence, all we need to show is the equivalence of $\mathcal{T}^{(n)}_{0,n,0}$ and the theory described by the quiver in Fig.~\ref{fig:IVIV}. Note that, for this purpose, it is sufficient to show that the type $IV$ theory involved in the quiver is equivalent to the $\mathcal{T}^{(\ell)}_{0,n}= \AD_n$ with $\frac{n-1}{2}$ extra fundamental hypermultiplets of $SU(n)$.\footnote{Recall here that $n$ is odd, and therefore $\frac{n-1}{2}$ is an integer.}  In the rest of this appendix, we show that the Seiberg-Witten (SW) curves of these two theories are indeed identical, which strongly suggests the equivalence of these two theories.

\subsection{Curve of type $IV$ theory}
Let us first write down the SW curve of the above-mentioned type IV theory. Since the theory is obtained by compactifying the 6d (2,0) $A_{n-1}$ theory on a sphere with one irregular puncture and a regular puncture, its SW curve is
\begin{align}
 \det (xdz - \varphi) = 0~,
\label{eq:SW-curve}
\end{align}
where $xdz$ is the SW 1-form and $\varphi=\varphi_z dz$ is a meromorphic $(1,0)$-form valued in $\mathfrak{sl}(n)$. We take a holomorphic coordinate, $z$, on the sphere so that the irregular puncture is at $z=\infty$ and the full puncture is at $z=0$. The Young diagrams characterizing the irregular puncture, $Y_1=Y_2=[n-1,1]$ and $Y_3 = [2,\cdots,2,1]$, imply that $\varphi$ behaves near $z=\infty$ as
\begin{align}
 \varphi \sim dz\left(
T_1 z + T_2 + \frac{T_3}{z} + \cdots~,
\right)
\label{eq:z=infty}
\end{align}
where, up to conjugations, $T_1 = \mathrm{diag}(a,\cdots,a,-(n-1)a),\, T_2 = \mathrm{diag}(b,\cdots,b,-(n-1)b)$ and $T_3= \mathrm{diag}(m_1,m_1,m_2,m_2,\cdots,m_{\frac{n-1}{2}},m_{\frac{n-1}{2}},-2\sum_{i=1}^{\frac{n-1}{2}}m_i)$. On the other hand, near $z=0$, $\varphi$ behaves as
\begin{align}
 \varphi \sim dz\left(\frac{M}{z}+\cdots\right)~,
\end{align}
where $M =\mathrm{diag}(M_1,\cdots,M_{n})$ such that $\sum_{i=1}^n M_i = 0$. By a change of coordinates that preserves the SW 1-form, the first two matrices can be mapped to $T_1=\mathrm{diag}(0,\cdots,0,-1)$ and $T_2=\mathrm{diag}(0,\cdots,0,-\tilde{b})$. Here, $m_i$ and $M_i$ are identified as mass parameters, and $\tilde{b}$ is identified as a relevant coupling of the type IV theory.

While the masses and couplings of the 4D theory are encoded in the singular terms described above, the vacuum expectation values (vevs) of Coulomb branch operators are encoded in less singular terms. To write down the most general expression for the curve including these vevs, let us consider the first correction, $U/z^2$, to the terms in the bracket of \eqref{eq:z=infty}, where we parameterize $U$ as $U = \mathrm{diag}(u_1+v_1,u_1-v_1,\cdots,u_{n-1}+v_{n-1},u_{n-1}-v_{n-1},-2\sum_{i=1}^{\frac{n-1}{2}}u_i)$. The parameters $u_i$ and $v_i$ are not fixed by the boundary conditions, but they are partially restricted so that $\det (x -\varphi_z)$ has only integer powers of $x$ and $z$. This condition implies that the most general expression for the curve $0 = \det(x-\varphi_z)$ is 
\begin{align}
0 &= x^n + x^{n-1}(z + \tilde{b}) + \sum_{i=2}^{n}
 x^{n-i}\left((z+\tilde{b})\frac{t_{i-1}}{z^{i-1}} + \frac{w_{i-1}}{z^{i-1}}
 + \frac{s_i}{z^i}\right)~.
\label{eq:curve1}
\end{align}
where $s_i,t_i$ and $w_i$ are combinations of the parameters such that $\prod_{i=1}^{n}\left(x-\frac{M_i}{z}\right) = x^n + \sum_{i=2}^{n} s_i  \frac{x^{n-i}}{z^i},\;\prod_{i=1}^{\frac{n-1}{2}}\left(x- \frac{m_i}{z}\right)^2 = x^{n-1} + \sum_{i=2}^{n} t_{i-1}\frac{x^{n-i}}{z^{i-1}}$ and 
\begin{align}
\frac{1}{z}
 \sum_{i=1}^{\frac{n-1}{2}}u_i\prod_{j\neq
 i} \left(x-\frac{m_i}{z}\right) \prod_{k=1}^{\frac{n-1}{2}}\left(x-\frac{m_k}{z}\right)&= \sum_{i=2}^n w_{i-1}\frac{x^{n-i}}{z^{i-1}}~.
\end{align}
Note that the curve \eqref{eq:curve1} can be rewritten as
\begin{align}
 0 &= \prod_{i=1}^{n}\left(x-\frac{M_i}{z}\right) +
 z\prod_{i=1}^{\frac{n-1}{2}}\left(x-\frac{m_i}{z}\right)^2
 + \left(\tilde{b}
 x^{\frac{n-1}{2}}+\sum_{i=1}^{\frac{n-1}{2}}\tilde{u}_i
 \frac{x^{\frac{n-1}{2}-i}}{z^i}\right)\prod_{i=1}^{\frac{n-1}{2}}\left(x-\frac{m_i}{z}\right)~,
\label{eq:curve2}
\end{align}
where $\tilde{u}_i$ are defined by
\begin{align}
 \tilde{b}\prod_{i=1}^{\frac{n-1}{2}}\left(x-\frac{m_i}{z}\right) +
 \frac{1}{z}\sum_{i=1}^{\frac{n-1}{2}}u_i\prod_{j\neq
 i}\left(x-\frac{m_i}{z}\right) = \tilde{b}x^{\frac{n-1}{2}} +
 \sum_{i=1}^{\frac{n-1}{2}} \tilde{u}_i\frac{x^{\frac{n-1}{2}-i}}{z^i}~.
\end{align}

\subsection{Curve of $\mathcal{T}^{(n)}_{0,n}$ with $\frac{n-1}{2}$ fundamental hypers}

Let us now turn to the SW curve of the $\AD_n$ theory with $\frac{n-1}{2}$ extra fundamental hypermultiplets of $SU(n)$. Our strategy is the same as in Appendix B of \cite{Buican:2014hfa}, i.e., we start with the curve of $\AD_n$, weakly gauge its $SU(n)$ flavor symmetry, introduce $\frac{n-1}{2}$ extra fundamental hypermultiplets of $SU(n)$, and then turn off the $SU(n)$ gauge coupling. The SW curve of $\AD_n = D_2(SU(n))$ is \cite{Cecotti:2012jx}
\begin{align}
 0 &= t^2 + t\sum_{i=0}^{\frac{n-1}{2}}U_iw^i + \prod_{i=1}^{n}\left(w-M_i\right)~,
\end{align}
where $M_i$ are the mass parameters associated with the $SU(n)$ flavor symmetry and therefore subject to $\sum_{i=1}^{N}M_i = 0$, $U_0$ is the relevant coupling of dimension $\frac{1}{2}$, and $U_i$ for $i\geq 1$ are the vevs of Coulomb branch operators. The SW 1-form is given by $\lambda = w\frac{dt}{t}$~. When we weakly gauge the $SU(n)$ flavor symmetry, the curve becomes
\begin{align}
 0 &= t^2 + t\sum_{i=0}^{\frac{n-1}{2}}U_iw^i +
 \prod_{i=1}^{n}\left(w-M_i\right) + \frac{\Lambda^{\frac{3n}{2}}}{t}~,
\end{align}
where $\Lambda$ is the corresponding dynamical scale, and $M_i$ is identified with the vevs of the Coulomb branch operators arising from the $SU(n)$ vector multiplet. When we introduce $\frac{n-1}{2}$ extra fundamental hypermultiplets of $SU(n)$, the curve becomes
\begin{align}
 0 &= t^2 + t\sum_{i=0}^{\frac{n-1}{2}}U_iw^{\frac{n-1}{2}-i} +
 \prod_{i=1}^{n}\left(w-M_i\right) +
 \frac{\Lambda^{n+\frac{1}{2}}}{t}\prod_{i=1}^{\frac{n-1}{2}}(w-m_i)~.
\end{align}
In terms of $z\equiv t/\prod_{i=1}^{\frac{n-1}{2}}(w-m_i)$ and $x
\equiv w/z$, the curve is
\begin{align}
 0 &= z\prod_{i=1}^{\frac{n-1}{2}}\left(x-\frac{m_i}{z}\right)^2 + \left(
 \sum_{i=0}^{\frac{n-1}{2}} U_i \frac{x^{\frac{n-1}{2}-i}}{z^i}\right)\prod_{i=1}^{\frac{n-1}{2}}\left(x-\frac{m_i}{z}\right)+
 \prod_{i=1}^n \left(x-\frac{M_i}{z}\right) + \frac{\Lambda^{n+\frac{1}{2}}}{z^{n+1}}~,
\end{align}
and the 1-form is $\lambda = xdz$ up to exact terms. We finally turn off the $SU(n)$ gauge coupling by setting $\Lambda = 0$. We then see that the resulting curve is precisely identical to the curve  in \eqref{eq:curve2}, where $U_0$ is identified as $\tilde{b}$ and $U_i$ for $i\geq 1$ are identified as $\tilde{u}_i$. This strongly suggests that the type $IV$ theory discussed in the previous sub-section is identical to the $\AD_n$ theory with $\frac{n-1}{2}$ extra decoupled hyper multiplets of $SU(n)$. The last identification then implies the equivalence of $\mathcal{T}^{(n)}_{0,n,0}$ and the theory described by the quiver in Fig.~\ref{fig:IVIV}.

\section{Monopole dimension bounds}\label{monopoles}
\label{app:monopoles}

In this appendix, we argue that the dimensions of monopole operators in the 3D mirror SCFTs associated with the $R_{0,n}^{2,AD}$ theories, $\Delta(\CO_i)$, satisfy the following bounds
\begin{equation}\label{bound}
\Delta\ge \begin{cases}
{1\over2}~, \ \ \ n=3\\
1~, \ \ \ n>3\ (n\ {\rm odd})~.
\end{cases}
\end{equation}
This result is in agreement with our 4D index analysis in the main text. Indeed, we argued that the $R_{0,n}^{2,AD}$ SCFT only has a decoupled free field sector for $n=3$. Note that the linear quiver discussion in \cite{Gaiotto:2008ak} does not directly apply here since, as discussed around Fig. \ref{R0nADquiver}, the mirror quiver contains a closed loop of nodes. Indeed, the fact that the $n=3$ case has free hypermultiplets even though it is \lq\lq good" by the naive application of the criteria of \cite{Gaiotto:2008ak} motivates us to examine the case for general $n$ more carefully.

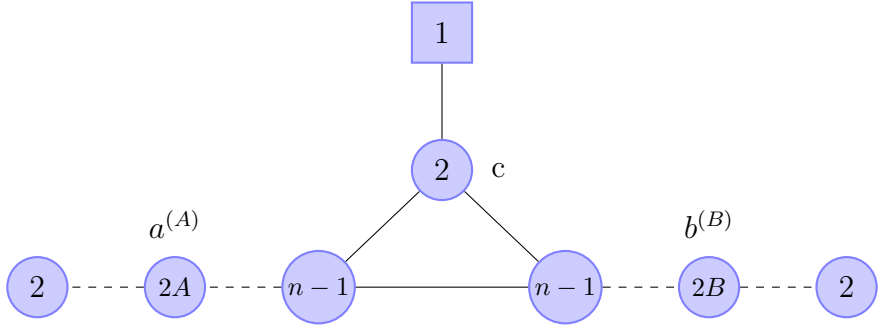
\begin{figure}
	\begin{center}
		\vskip .5cm
		\begin{tikzpicture}[round/.style={circle,draw=blue!50,fill=blue!20,thick,inner sep=0pt,minimum size=8mm},square/.style={rectangle,draw=blue!50,fill=blue!20,thick,inner sep=0pt,minimum size=8mm},auto]
		\node[square] (top) {1};
		\node (d) [right=1mm of top] {};
		\node[round] (middle) [below=of top] {2};
		\node (c) [right=1mm of middle] {c};
		\node (null) [below=of middle] {};
		\node[round] (L1) [left=of null] {\footnotesize\,$n-1$\,};
		\node[round] (L2) [left=of L1] {\footnotesize\,$2A$\,};
		\node (a) [above=1mm of L2] {$a^{(A)}$};
		\node[round] (L3) [left=of L2] {2};
		\node[round] (R1) [right=of null] {\footnotesize\,$n-1$\,};
		\node[round] (R2) [right=of R1] {\footnotesize\,$2B$\,};
		\node (b) [above=1mm of R2] {$b^{(B)}$};
		\node[round] (R3) [right=of R2] {2};
		\draw (top)--(middle);
		\draw (middle)--(L1);
		\draw (middle)--(R1);
		\draw (L1)--(R1);
		\draw[dashed] (L1)--(L2);
		\draw[dashed] (L2)--(L3);
		\draw[dashed] (R1)--(R2);
		\draw[dashed] (R2)--(R3);
		\end{tikzpicture}
		\caption{We reproduce the quiver from Fig. \ref{R0nADquiver} but rotated and with labels $a^{(A)}\in\mathbb{Z}^{2A}, b^{(B)}\in\mathbb{Z}^{2B}, c\in\mathbb{Z}^2$ denoting magnetic charges through the corresponding gauge nodes (the nodes to the left of the central $U(2)$ node have fluxes labeled by \lq\lq$a$," while those to the right have fluxes labeled by \lq\lq $b$").}
		\label{Basic}
	\end{center}

\end{figure}

While the bound for $n=3$ follows from the mirror symmetry discussion in \cite{Buican:2017fiq,Buican:2018ddk} (and also the analysis in \cite{Buican:2014hfa}), we will prove the result in this case and also for all $n>3$ directly via an analytic monopole analysis in the mirror. To that end, the quantity we wish to bound is
\begin{eqnarray}\label{qty}
\Delta&=&-\left(\sum_{A=1}^{n-1\over2}\sum_{i_A<j_A}|a_{i_A}^{(A)}-a_{j_A}^{(A)}|+\sum_{B=1}^{n-1\over2}\sum_{i_B<j_B}|b_{i_B}^{(B)}-b_{j_B}^{(B)}|+|c_1-c_2|\right)\ \ \ \ \ \ \ \ \ \ \ \ \ \ \ \ \cr&+&{1\over2}\left(\sum_{A=1}^{n-3\over2}\sum_{i_A,j_{A+1}}|a^{(A)}_{i_A}-a^{(A+1)}_{j_{A+1}}|+\sum_{B=1}^{n-3\over2}\sum_{i_B,j_{B+1}}|b^{(B)}_{i_B}-b^{(B+1)}_{j_{B+1}}|\right)+{1\over2}\left(|c_1|+|c_2|\right)\cr&+&{1\over2}\left(\sum_{i,j}|a^{\left({n-1\over2}\right)}_{i}-b^{\left({n-1\over2}\right)}_{j}|+\sum_{i,j}|c_i-a^{\left(n-1\over2\right)}_{j}|+\sum_{i,j}|c_i-b^{\left(n-1\over2\right)}_{j}|\right)~,
\end{eqnarray}
where $i_A, j_A\in\left\{1,\cdots,2A\right\}$, $i_B, j_B\in\left\{1,\cdots,2B\right\}$, and $a^{(A)}\in\mathbb{Z}^{2A}, b^{(b)}\in\mathbb{Z}^{2B}, c\in\mathbb{Z}^2$ label the magnetic flux through each gauge node in the quiver (note that we have dropped subscripts denoting the particular entry in the flux vector)---see Fig. \ref{Basic}. Note that the negative contributions in \eqref{qty} arise from the gauge nodes while the positive contributions arise from the (bi)fundamentals.

The main strategy in proving \eqref{bound} is repeated use of the triangle inequality to cancel four positive matter contributions to $\Delta$ against single gauge contributions (we perform the cancelation between lines and the nodes that they end on). We will start from the leftmost $U(2)$ node in Fig. \ref{Basic} and then inductively argue that we can cancel all the negative contributions from all the nodes in the left tail up to and including negative contributions from the $U(n-3)$ node that neighbors the left $U(n-1)$ node. By $\mathbb{Z}_2$ symmetry, the corresponding negative contributions from the $U(2)$ to $U(n-3)$ nodes from the right tail will also be cancelled by corresponding matter contributions. We then move on to consider the core of the quiver and prove \eqref{bound}.

Before continuing, let us note that we may always use Weyl transformations at each gauge node to arrange that
\begin{equation}
a^{(\alpha)}_1\ge a_2^{(\alpha)}\ge\cdots\ge a_{2\alpha}^{(\alpha)}~, \ \ \ b^{(\beta)}_1\ge b_2^{(\beta)}\ge\cdots\ge b_{2\beta}^{(\beta)}~, \ \ \ c_1\ge c_2~,
\end{equation}
for all $\alpha,\beta\in\left\{1,2,\cdots,{n-1\over2}\right\}$. This maneuver has the effect of removing absolute values from gauge node contributions in \eqref{qty}. We may then write the contributions from the $U(2A)$ node as
\begin{equation}\label{countneg}
\Delta\supset-\sum_{i=1}^{A}(2(A-i)+1)(a^{(A)}_i-a^{(A)}_{2A+1-i})
\end{equation}
Note that there are $A^2=\sum_{i=1}^A(2(A-i)+1)$ such contributions in total.

\subsec{Inductive proof of the canceling of negative contributions from the quiver tails}
Let us begin by focusing on the left quiver tail in Fig. \ref{Basic}. We start with the somewhat special $U(2)$ contributions to $\Delta$ and the contributions of the corresponding eight hypermultiplets in the bifundamental of $U(2)\times U(4)$
\begin{equation}
\Delta\supset\Delta_2=-(a_1^{(1)}-a_2^{(1)})+{1\over2}\sum_{i,j}|a_i^{(1)}-a_j^{(2)}|~.
\end{equation}
We can cancel the negative contributions from $U(2)$ against four hypermultiplet contributions by using the triangle inequality twice
\begin{equation}\label{tri1}
-(a_1^{(1)}-a_2^{(1)})+{1\over2}\left(|a_1^{(1)}-a_2^{(2)}|+|a_1^{(1)}-a_3^{(2)}|+|a_2^{(1)}-a_2^{(2)}|+|a_2^{(1)}-a_3^{(2)}|\right)\ge0
\end{equation}
This procedure leaves a surplus of four matter contributions we can use to cancel contributions from the adjoining $U(4)$ node. Moreover, since we have not used matter contributions involving $a^{(2)}_{1,4}$, we can use this surplus to cancel one of the most negative terms from $U(4)$ (i.e., one proportional to $a^{(2)}_1-a^{(2)}_4$).

Let us now discuss the $U(4)$ node and adjoining matter contributions more carefully. Since this computation contains contributions from matter fields to the left and right of the gauge node, we can use this discussion to build a base case for an inductive proof of the positivity of contributions to $\Delta$ from the left quiver tail. To that end, consider the contributions
\begin{eqnarray}
\Delta&\supset&\Delta_4=-\sum_{i=1}^{2}(2(2-i)+1)(a^{(2)}_i-a^{(2)}_{5-i})+{1\over2}\Big(|a_1^{(1)}-a_1^{(2)}|+|a_1^{(1)}-a_4^{(2)}|+\cr&+&|a_2^{(1)}-a_1^{(2)}|+|a_2^{(1)}-a_4^{(2)}|\Big)+{1\over2}\sum_{k,\ell}|a_k^{(2)}-a_{\ell}^{(3)}|~.
\end{eqnarray}
We may use the surplus contributions in the second term above to cancel one of the contributions from the $U(4)$ gauge node so that
\begin{equation}
\Delta_4\ge-(2(a_1^{(2)}-a_4^{(2)})+(a_2^{(2)}-a_3^{(2)}))+{1\over2}\sum_{k,\ell}|a_k^{(2)}-a_{\ell}^{(3)}|~.
\end{equation}
Let us now use twelve of the twenty-four $U(4)\times U(6)$ hypermultiplets to cancel the remaining three negative $U(4)$ contributions. To see how this cancelation is done, it is useful to visualize the hypermultiplet contributions via a $4\times6$ matrix with a \lq\lq $1$" indicating an unused matter contribution and a \lq\lq $0$" indicating a used matter contribution. We start with
\begin{equation}
{\bf L_{4,6}}=\begin{pmatrix}
1&1&1&1&1&1\\
1&1&1&1&1&1\\
1&1&1&1&1&1\\
1&1&1&1&1&1\\
\end{pmatrix}~.
\end{equation}
Our strategy is to leave as surplus the first and last columns while using the remainder of the first and last rows (eight terms in all) to cancel the two $U(4)$ contributions proportional to $a_1^{(2)}-a_4^{(2)}$ (this is done via four applications of the triangle inequality). In other words, we have
\begin{equation}
{\bf L_{4,6}}\to\begin{pmatrix}
1&0&0&0&0&1\\
1&1&1&1&1&1\\
1&1&1&1&1&1\\
1&0&0&0&0&1\\
\end{pmatrix}~,
\end{equation}
which leads to the bound
\begin{eqnarray}
-2(a_1^{(2)}-a_4^{(2)})&+&{1\over2}\Big([|a^{(2)}_1-a^{(3)}_2|+|a^{(2)}_1-a^{(3)}_5|+|a^{(2)}_4-a^{(3)}_2|+|a^{(2)}_4-a^{(3)}_5|]\ \ \ \ \ \ \ \ \ \cr&+&[|a^{(2)}_1-a^{(3)}_3|+|a^{(2)}_1-a^{(3)}_4|+|a^{(2)}_4-a^{(3)}_3|+|a^{(2)}_4-a^{(3)}_4|]\Big)\ge0
\end{eqnarray}
We cancel the remaining negative contribution from $U(4)$ by using the middle four entries of ${\bf L_{4,6}}$ so that
\begin{equation}\label{L46final}
{\bf L_{4,6}}\to\begin{pmatrix}
1&0&0&0&0&1\\
1&1&0&0&1&1\\
1&1&0&0&1&1\\
1&0&0&0&0&1\\
\end{pmatrix}~.
\end{equation}
Indeed, we see that 
\begin{equation}
-(a_2^{(2)}-a_3^{(2)})+{1\over2}[|a_2^{(2)}-a^{(3)}_3|+|a_2^{(2)}-a^{(3)}_4|+|a_3^{(2)}-a^{(3)}_3|+|a_3^{(2)}-a^{(3)}_4|]\ge0~.
\end{equation}
This procedure leaves a surplus of 12 hypermultiplets we can use to cancel negative contributions from $U(6)$.

Now that we have shown how the negative $U(2)\times U(4)$ contributions in the left quiver tail are cancelled, we can move on to the induction hypothesis in our proof. We assume that all the negative contributions in $U(1)\times\cdots\times U(2A)$ have been canceled. In particular, the $A^2$ negative $U(2A)$ contributions (see the discussion below \eqref{countneg}) have been canceled as follows: ${A(A-1)\over2}$ of them from $U(2(A-1))\times U(2A)$ bifundamentals and ${A(A+1)\over2}$ of them from $U(2A)\times U(2(A+1))$ bifundamentals.

Let us understand these statements in more detail. In particular, we should first focus on the ${\bf L_{2(A-1),2A}}$ generalization of \eqref{L46final} we get after finishing the cancelation of terms in $U(2(A-1))$. This matrix has its first column filled with $1$'s. The next column has all $1$'s except in the first and last row which are $0$. For $2\le p\le A$, the $p^{\rm th}$ column consists of zeros in positions $i$ such that $1\le i\le p-1$ and $2A-p\le i\le2(A-1)$ with $1$'s everywhere else. This discussion specifies half the matrix. The remaining half is set by demanding that ${\bf L_{2(A-1),2A}}$ is symmetric under reflections through a line running between columns $A$ and $A+1$, i.e.
\begin{equation}\label{red1}
{\bf L_{2(A-1),2A}}\to\begin{pmatrix}
1&0&0&\cdots&\cdots&0&0&1\\
1&1&0&\cdots&\cdots&0&1&1\\
\vdots&\vdots&\vdots&\ddots&\ddots&\vdots&\vdots&\vdots\\
1&1&0&\cdots&\cdots&0&1&1\\
1&0&0&\cdots&\cdots&0&0&1\\
\end{pmatrix} ~. 
\end{equation}
By using the $2A(A-1)$ hypermultiplet contributions corresponding to the $1$'s in \eqref{red1}, we assume we cancel ${A(A-1)\over2}$ of the negative $U(2A)$ contributions via repeated applications of the triangle innequality.

Next we move to ${\bf L_{2A,2(A+1)}}$. This is a $2A\times2(A+1)$ matrix full of $1$'s. Now, as in the $U(4)$ case, we leave the first column alone. In the $p^{\rm th}$ column, with $2\le p\le A+1$, we set to zero all rows $i$ such that $1\le i\le p-1$ and $2(A+1)-p\le i\le 2A$. This procedure again specifies the left half of the matrix. The right half is fixed by requiring the matrix to be symmetric under reflection through a line running between columns $A+1$ and $A+2$, i.e.
\begin{equation}\label{indproof}
{\bf L_{2A,2(A+1)}}\to\begin{pmatrix}
1&0&0&\cdots&\cdots&0&0&1\\
1&1&0&\cdots&\cdots&0&1&1\\
\vdots&\vdots&\vdots&\ddots&\ddots&\vdots&\vdots&\vdots\\
1&1&0&\cdots&\cdots&0&1&1\\
1&0&0&\cdots&\cdots&0&0&1\\
\end{pmatrix}~.  
\end{equation}
We have therefore set to zero $2A(A+1)$ entries, and we assume we can use the corresponding hypermultiplet contributions to cancel the remaining ${A(A+1)\over2}$ negative contributions in $U(2A)$ via repeated use of the triangle inequality.

Given these assumptions, we now show that we can cancel the negative contributions in $U(2(A+1))$ and complete our inductive proof. The negative contributions in this case take the form
\begin{equation}
\Delta\supset-\sum_{i=1}^{A+1}(2(A+1-i)+1)(a^{(A+1)}_i-a^{(A+1)}_{2(A+1)+1-i})
\end{equation}
Let us now focus on the matter contributions from the first and last columns in \eqref{indproof}. We have
\begin{eqnarray}
{1\over2}&\Big(&[|a_1^{(A)}-a_1^{(A+1)}|+|a_1^{(A)}-a_{2(A+1)}^{(A+1)}|+|a_{2A}^{(A)}-a_1^{(A+1)}|+|a_{2A}^{(A)}-a_{2(A+1)}^{(A+1)}|]\ \ \ \ \ \ \ \ \ \ \ \cr&+&[|a_2^{(A)}-a_1^{(A+1)}|+|a_2^{(A)}-a_{2(A+1)}^{(A+1)}|+|a_{2A-1}^{(A)}-a_1^{(A+1)}|+|a_{2A-1}^{(A)}-a_{2(A+1)}^{(A+1)}|]\cr&+&\cdots+[|a_A^{(A)}-a_1^{(A+1)}|+|a_A^{(A)}-a_{2(A+1)}^{(A+1)}|+|a_{A+1}^{(A)}-a_1^{(A+1)}|+|a_{A+1}^{(A)}-a_{2(A+1)}^{(A+1)}|]\Big)\cr&\ge&A(a_1^{(A+1)}-a_{2(A+1)}^{(A+1)})~,
\end{eqnarray}
where, in the last line, we have repeatedly used the triangle inequality. Working inward, a similar computation shows that the contributions from columns $p$ and $2(A+1)-p+1$ are bounded from below by $(A+1-p)(a_p^{(A+1)}-a_{2(A+1)-p+1}^{(A+1)})$. Therefore, after using the $2A(A+1)$ $1$'s in \eqref{indproof}, we have the following remaining negative contributions from $U(2(A+1))$
\begin{equation}\label{firstc}
\Delta\supset-\sum_{i=1}^{A+1}(A+2-i)(a^{(A+1)}_i-a^{(A+1)}_{2(A+1)+1-i})
\end{equation}

To cancel the remaining negative terms, we must use the $U(2(A+1))\times U(2(A+2))$ bifundamental contributions captured by ${\bf L_{2(A+1),2(A+2)}}$. In particular, this latter matrix has $1$'s in all $2(A+1)\times2(A+2)$ entries. Let us use entries $2$ through $2A+3$ of the first and last rows to cancel the $-(A+1)(a^{(A+1)}_1-a^{(A+1)}_{2(A+1)})$ contribution in \eqref{firstc}. Indeed, we see
\begin{eqnarray}
{1\over2}&\Big(&[|a_1^{(A+1)}-a_2^{(A+2)}|+|a_1^{(A+1)}-a_{2A+3}^{(A+2)}|+|a_{2(A+1)}^{(A+1)}-a_2^{(A+2)}|+|a_{2(A+1)}^{(A+1)}-a_{2A+3}^{(A+2)}|]\ \ \ \ \ \ \ \ \ \ \ \cr&+&[|a_1^{(A+1)}-a_3^{(A+2)}|+|a_1^{(A+1)}-a_{2(A+1)}^{(A+2)}|+|a_{2(A+1)}^{(A+1)}-a_3^{(A+2)}|+|a_{2(A+1)}^{(A+1)}-a_{2(A+1)}^{(A+2)}|]\cr&+&\cdots+[|a_1^{(A+1)}-a_{A+2}^{(A+2)}|+|a_1^{(A+1)}-a_{A+3}^{(A+2)}|+|a_{2(A+1)}^{(A+1)}-a_{A+2}^{(A+2)}|+|a_{2(A+1)}^{(A+1)}-a_{A+3}^{(A+2)}|]\Big)\cr&\ge&(A+1)(a_1^{(A+1)}-a_{2(A+1)}^{(A+1)})~,
\end{eqnarray}
where we have repeatedly used the triangle inequality. Proceeding in a similar fashion with rows $p$ and $2(A+1)-p+1$ (but now using entries $p+1$ through $2(A+2)-p$ of each row), we find that each contribution is bounded from below by $(A+2-p)(a^{(A+1)}_p-a^{(A+1)}_{2(A+1)+1-p})$.

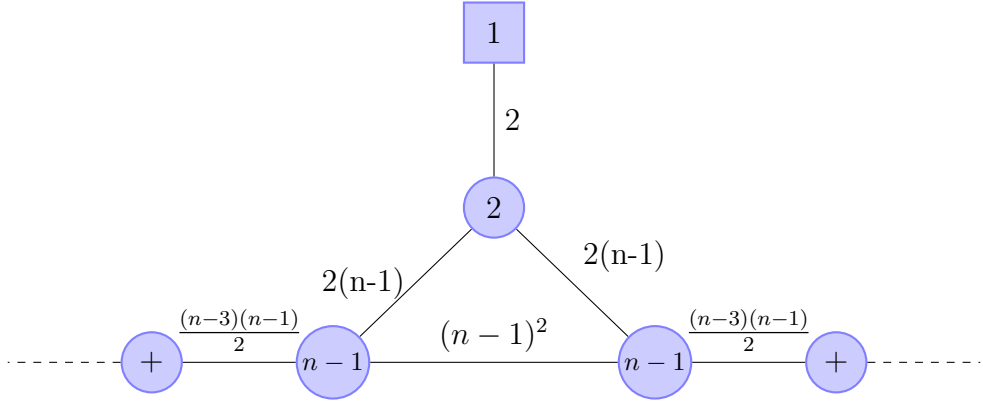
\begin{figure}
	\begin{center}
		\vskip .5cm
		\begin{tikzpicture}[round/.style={circle,draw=blue!50,fill=blue!20,thick,inner sep=0pt,minimum size=8mm},square/.style={rectangle,draw=blue!50,fill=blue!20,thick,inner sep=0pt,minimum size=8mm},auto,node distance=1.5cm]
		\node[square] (top) {1};
		\node[round] (middle) [below=of top] {2};
		\node (null) [below=of middle] {};
		\node[round] (L1) [left=of null] {\footnotesize\,$n-1$\,};
		\node[round] (L2) [left=of L1] {+};
		\node (L3) [left=of L2] {};
		\node[round] (R1) [right=of null] {\footnotesize\,$n-1$\,};
		\node[round] (R2) [right=of R1] {+};
		\node (R3) [right=of R2] {};
		\draw (top)--(middle) node[midway] {2};
		\draw (middle)--(L1)node[midway,left] {2(n-1)};
		\draw (middle)--(R1)node[midway] {2(n-1)};
		\draw (L1)--(R1)node[midway] {$(n-1)^2$};
		\draw (L1)--(L2)node[midway,above] {${(n-3)(n-1)\over2}$};
		\draw[dashed] (L2)--(L3);
		\draw (R1)--(R2)node[midway] {${(n-3)(n-1)\over2}$};
		\draw[dashed] (R2)--(R3);
		\end{tikzpicture}
		\caption{After cancelling the negative contributions from the left and right quiver tails, we put a \lq\lq$+$" in each corresponding node. We are left over with ${(n-3)(n-1)\over2}$ surplus contributions to $\Delta$ from bifundamentals of $U(n-3)\times U(n-1)$ in each quiver tail, and this has been encoded in the corresponding numbers on the tail links emanating from the $U(n-1)$ nodes. The remaining numbers associated with the core links indicate the total number of unused (bi)fundamental contributions to $\Delta$.}
		\label{Reduced}
	\end{center}
\end{figure}

Therefore, we have succeeded in cancelling all the negative contributions of $U(2(A+1))$. Note that, after canceling the $U(2(A+1))$ contributions, we have $2(A+1)(A+2)$ contributions from bifundamentals of $U(2(A+1))\times U(2(A+2))$ left over as surplus. By $\mathbb{Z}_2$ symmetry, we have now proven that all the non-core nodes of the quiver have their negative contributions to $\Delta$ canceled, and we are left over with ${(n-3)(n-1)\over2}$ bifundamental contributions of $U(n-3)\times U(n-1)$ in both gauge tails of Fig. \ref{Basic}. In particular, we have shown
\begin{eqnarray}\label{qtybd1}
\Delta&\ge&-\left(\sum_{i<j}|a_{i}^{\left({n-1\over2}\right)}-a_{j}^{\left({n-1\over2}\right)}|+\sum_{i<j}|b_{i}^{\left({n-1\over2}\right)}-b_{j}^{\left({n-1\over2}\right)}|+|c_1-c_2|\right)\ \ \ \ \ \ \ \ \ \ \ \ \ \ \ \ \cr&+&{1\over2}\left(\sum_{i,j\in\CS_a}|a^{\left({n-3\over2}\right)}_{i}-a^{\left({n-1\over2}\right)}_{j}|+\sum_{i,j\in \CS_b}|b^{\left({n-3\over2}\right)}_{i}-b^{\left({n-1\over2}\right)}_{j}|\right)+{1\over2}\left(|c_1|+|c_2|\right)\cr&+&{1\over2}\left(\sum_{i,j}|a^{\left({n-1\over2}\right)}_{i}-b^{\left({n-1\over2}\right)}_{j}|+\sum_{i,j}|c_i-a^{\left(n-1\over2\right)}_{j}|+\sum_{i,j}|c_i-b^{\left(n-1\over2\right)}_{j}|\right)~,
\end{eqnarray}
where the first line contains the only remaining negative contributions (i.e., those from the core $U(n-1)\times U(n-1)\times U(2)$ nodes of the quiver), the first two sums in the second line are restricted to lie in the sets $\CS_{a,b}$ that run over the surplus $U(n-3)\times U(n-1)$ nodes in the left and right tails respectively (the \lq\lq $a$" and \lq\lq $b$" subscripts distinguish these tails), and the final line contains bifundamentals from the core of the quiver. This discussion is summarized in Fig. \ref{Reduced}.

\subsec{Analyzing the quiver core and proving \eqref{bound}}
To complete our proof, we now proceed to the quiver core in Fig. \ref{Reduced}. In particular, let us begin by canceling some of the negative contributions to $\Delta$ from the left $U(n-1)$ node
\begin{equation}
\Delta\supset-\sum_{i=1}^{n-1\over2}\left(n-2i\right)\left(a^{\left({n-1\over2}\right)}_i-a^{\left(n-1\over2\right)}_{n-i}\right)~.
\end{equation}
First we use the remaining ${(n-3)(n-1)\over2}$ bifundamental contributions of $U(n-3)\times U(n-1)$ as in the discussion above \eqref{firstc} to cancel some of the $U(n-1)$ contributions and obtain
\begin{equation}\label{nnred}
\Delta\supset-\sum_{i=1}^{n-1\over2}\left({n+1\over2}-i\right)\left(a^{\left({n-1\over2}\right)}_i-a^{\left(n-1\over2\right)}_{n-i}\right)~.
\end{equation}
Without loss of generality, we may also choose to use the $2(n-1)$ bifundamentals of $U(2)\times U(n-1)$ to cancel more of these negative contributions.\footnote{This choice of cancellation will make some of the later inequalities we derive look less manifestly $\mathbb{Z}_2$ symmetric, but this choice does not affect the final result.} Indeed, repeated use of the triangle inequality shows that
\begin{equation}
{1\over2}\sum_i\left(|c_1-a_i^{\left({n-1\over2}\right)}|+|c_2-a_i^{\left({n-1\over2}\right)}|\right)\ge\sum_{i=1}^{n-1\over2}\left(a_i^{\left(n-1\over2\right)}-a_{n-i}^{\left(n-1\over2\right)}\right)~.
\end{equation}
As a result, we have that the remaining negative contributions from $U(n-1)$ are 
\begin{equation}\label{remneg}
\Delta\supset-\sum_{i=1}^{n-3\over2}\left({n-1\over2}-i\right)\left(a^{\left({n-1\over2}\right)}_i-a^{\left(n-1\over2\right)}_{n-i}\right)~.
\end{equation}
Let us now use some of the $U(n-1)\times U(n-1)$ bifundamentals to cancel the remaining negative contributions in \eqref{remneg}. To that end, consider using entries $2$ through $n-2$ in the first and last rows of ${\bf L_{n-1,n-1}}$. We have
\begin{eqnarray}
{1\over2}&\Big(&[|a_1^{\left(n-1\over2\right)}-b_2^{\left(n-1\over2\right)}|+|a_1^{\left(n-1\over2\right)}-b_{n-2}^{\left(n-1\over2\right)}|+|a_{n-1}^{\left(n-1\over2\right)}-b_2^{\left(n-1\over2\right)}|+|a_{n-1}^{\left(n-1\over2\right)}-b_{n-2}^{\left(n-1\over2\right)}|]\ \ \ \ \ \ \ \ \ \ \ \cr&+&[|a_1^{\left(n-1\over2\right)}-b_3^{\left(n-1\over2\right)}|+|a_1^{\left(n-1\over2\right)}-b_{n-3}^{\left(n-1\over2\right)}|+|a_{n-1}^{\left(n-1\over2\right)}-b_3^{\left(n-1\over2\right)}|+|a_{n-1}^{\left(n-1\over2\right)}-b_{n-3}^{\left(n-1\over2\right)}|]\cr&+&\cdots+[|a_1^{\left(n-1\over2\right)}-b_{n-1\over2}^{\left(n-1\over2\right)}|+|a_1^{\left(n-1\over2\right)}-b_{n+1\over2}^{\left(n-1\over2\right)}|+|a_{n-1}^{\left(n-1\over2\right)}-b_{n-1\over2}^{\left(n-1\over2\right)}|+|a_{n-1}^{\left(n-1\over2\right)}-b_{n+1\over2}^{\left(n-1\over2\right)}|]\Big)\cr&\ge&\left({n-1\over2}-1\right)\left(a_1^{\left(n-1\over2\right)}-a_{n-1}^{\left(n-1\over2\right)}\right)
\end{eqnarray}
Similarly, we see that the contributions from rows $p\ge2$ and $n-p$ are bounded from above by $\left({n-1\over2}-p\right)\left(a_p^{\left(n-1\over2\right)}-a_{n-p}^{\left(n-1\over2\right)}\right)$. As a result, we have countered all negative contributions from the left $U(n-1)$ node.

We must still counter the negative contributions from the remaining $U(n-1)\times U(2)$ nodes with contributions from $2(n-1)$ bifundamentals of $U(n-1)\times U(2)$, ${n(n-1)\over2}$ bifundamentals of $U(n-1)\times U(n-1)$, and $(n-3)(n-1)\over2$ bifundamentals of $U(n-1)\times U(n-3)$ (from the right quiver tail in Fig. \ref{Reduced}). Proceeding in analogy with the discussion for the other $U(n-1)$ node in \eqref{nnred}, we use the remaining $U(n-1)\times U(n-3)$ bifundamentals to get rid of some of the $U(n-1)$ contributions. We are left with 
\begin{equation}\label{nnred2}
\Delta\supset-\sum_{i=1}^{n-1\over2}\left({n+1\over2}-i\right)\left(b^{\left({n-1\over2}\right)}_i-b^{\left(n-1\over2\right)}_{n-i}\right)~.
\end{equation}
Now we may use the remaining contributions from the $U(n-1)\times U(n-1)$ bifundamentals to cancel the negative contribution in \eqref{nnred2}.\footnote{Note that we have more such bifundamentals left over than we used in the cancelation of the contributions from the left $U(n-1)$ node since we chose to use the left $U(2)\times U(n-1)$ bifundamentals in the cancelation of the contributions from the left $U(n-1)$ node.}

We start with the first and last columns of $1$'s remaining in ${\bf L_{n-1,n-1}}$ and find the following bound via repeated uses of the triangle inequality
\begin{eqnarray}
{1\over2}&\Big(&[|a_1^{\left(n-1\over2\right)}-b_1^{\left(n-1\over2\right)}|+|a_1^{\left(n-1\over2\right)}-b_{n-1}^{\left(n-1\over2\right)}|+|a_{n-1}^{\left(n-1\over2\right)}-b_1^{\left(n-1\over2\right)}|+|a_{n-1}^{\left(n-1\over2\right)}-b_{n-1}^{\left(n-1\over2\right)}|]\ \ \ \ \ \ \ \ \ \ \ \cr&+&[|a_2^{\left(n-1\over2\right)}-b_1^{\left(n-1\over2\right)}|+|a_2^{\left(n-1\over2\right)}-b_{n-1}^{\left(n-1\over2\right)}|+|a_{n-2}^{\left(n-1\over2\right)}-b_1^{\left(n-1\over2\right)}|+|a_{n-2}^{\left(n-1\over2\right)}-b_{n-1}^{\left(n-1\over2\right)}|]\cr&+&\cdots+[|a_{n-1\over2}^{\left(n-1\over2\right)}-b_1^{\left(n-1\over2\right)}|+|a_{n-1\over2}^{\left(n-1\over2\right)}-b_{n-1}^{\left(n-1\over2\right)}|+|a_{n+1\over2}^{\left(n-1\over2\right)}-b_{1}^{\left(n-1\over2\right)}|+|a_{n+1\over2}^{\left(n-1\over2\right)}-b_{n-1}^{\left(n-1\over2\right)}|]\Big)\cr&\ge&\left({n-1\over2}\right)\left(b_1^{\left(n-1\over2\right)}-b_{n-1}^{\left(n-1\over2\right)}\right)
\end{eqnarray}
Similarly, we find that the remaining contributions from columns $p\ge2$ and $n-p$ can be bounded from above as $\left({n+1\over2}-p\right)\left(b_p^{\left(n-1\over2\right)}-b_{n-p}^{\left(n-1\over2\right)}\right)$. Therefore, we cancel all the remaining negative contributions in \eqref{nnred2}. 

We are left with one final source of negative contributions, those from the top $U(2)$ node
\begin{equation}
\Delta\supset-(c_1-c_2)~.
\end{equation}
However, we still have all $2(n-1)$ bifundamentals of the right $U(2)\times U(n-1)$ left to cancel them. This is more than enough since
\begin{equation}
{1\over2}\left(|b^{\left(n-1\over2\right)}_{n-1\over2}-c_1|+|b^{\left(n-1\over2\right)}_{n-1\over2}-c_2|+|b^{\left(n-1\over2\right)}_{n+1\over2}-c_1|+|b^{\left(n-1\over2\right)}_{n+1\over2}-c_2|\right)\ge c_1-c_2~.
\end{equation}
As a result, we have proven that
\begin{equation}
\Delta\ge{1\over2}\left(|c_1|+|c_2|\right)+{1\over2}\sum_{j\ne{n\pm1\over2},i}|c_i-b_j^{\left(n-1\over2\right)}|~.
\end{equation}
While our choice of cancelation below \eqref{nnred} has the effect of making this inequality less manifestly $\mathbb{Z}_2$ symmetric (the contributions of the \lq\lq$a$" side of the quiver have already been taken into account in the above bound), this choice does not affect our conclusions.

To prove \eqref{bound}, we need only consider a few simple cases. For $c_1=c_2=0$, we know that all monopole operators have $\Delta\ge1$ by \cite{Gaiotto:2008ak} since the quiver effectively reduces to a linear quiver and all nodes are \lq\lq good." Moreover, if $|c_i|\ge2$ for either $i=1$ or $i=2$, then clearly $\Delta\ge1$. Similar statements hold if $|c_1|=|c_2|=1$. Therefore, we need only consider the case where, without loss of generality, $|c_1|=1$ and $c_2=0$. We then have
\begin{equation}\label{final}
\Delta\ge{1\over2}+{1\over2}\sum_{j\ne{n\pm1\over2}}|b_j^{\left(n-1\over2\right)}|+{1\over2}\sum_{j\ne{n\pm1\over2}}|c_1-b_j^{\left(n-1\over2\right)}|
\end{equation}
For $n=3$, this bound reduces to \eqref{bound} since the second and third terms are trivial. For $n>3$, if we choose any of the $b_j^{\left({n-1\over2}\right)}\ne0$, then $\Delta\ge1$ due to contributions from the second term in \eqref{final}. However, if we set all $b_j^{\left({n-1\over2}\right)}=0$, then the third term leads to $\Delta\ge1$. Therefore, we have proven \eqref{bound}.

\end{appendices}

\newpage
\bibliography{chetdocbib}


\end{document}